\documentclass{SciPost}

\hypersetup{
    colorlinks,
    linkcolor={red!50!black},
    citecolor={blue!50!black},
    urlcolor={blue!80!black}
}
\usepackage{braket}
\usepackage{slashed}
\usepackage{color}
\usepackage{dsfont}
\usepackage{graphicx}
\usepackage{marginnote}
\usepackage{setspace}
\usepackage{ulem}
\usepackage{amsmath,amsthm,amsfonts,amssymb,amscd}
\usepackage{graphicx}
\usepackage{enumerate}
\usepackage{hyperref}
\usepackage{amsbsy}
\usepackage{tikz-cd}
\usepackage{tensor}
\usepackage{comment}

\usepackage{todonotes}
\usepackage[arrowdel]{physics}
\usepackage{mathtools}
\usepackage{graphicx} 
\usepackage{subcaption} 
\usepackage[font=normalsize,labelfont=bf,
   format=plain]{caption}
\usepackage{tikz-cd}

\newcommand{\Diff}{\operatorname{Diff}_{+}(S^1)}
\newcommand{\SL}{\mathrm{SL}(2,\mathbb{R})}
\newcommand{\PSL}{\mathrm{PSL}(2,\mathbb{R})}

\newcommand{\wb}{\overline{w}}
\newcommand{\te}{\tau}

\newcommand{\xib}{\overline{\xi}}

\newcommand{\partialb}{\overline{\partial}}

\newcommand{\gb}{\overline{g}}
\newcommand{\ep}{\varepsilon}
\newcommand{\V}[1]{\mathcal{V}_{#1}}
\newcommand{\Rb}{\mathbb{R}}
\newcommand{\Mc}{\mathcal{M}}

\newcommand{\mop}{\pmb{\tau}}

\newcommand*\pFq[2]{{}_{#1}F_{#2}\genfrac[]{0pt}{}}
\newcommand{\mn}{{\mu\nu}}
\newcommand{\ali}[1]{\begin{align} #1 \end{align}}
\newcommand{\ra}{\rightarrow}
\newcommand{\tuv}{\tau_{\text{uv}}}

\newcommand\kwout{\bgroup\markoverwith{\textcolor{olive}{\rule[0.5ex]{2pt}{1pt}}}\ULon}
\graphicspath{{figures/}}
\usepackage{graphicx}

\usepackage[bitstream-charter]{mathdesign}
\DeclareSymbolFont{usualmathcal}{OMS}{cmsy}{m}{n}
\DeclareSymbolFontAlphabet{\mathcal}{usualmathcal}

\fancypagestyle{SPstyle}{
\fancyhf{}

\fancyfoot[C]{\textbf{\thepage}}
}

\begin{document}

\pagestyle{SPstyle}

\begin{center}{\Large \textbf{\color{scipostdeepblue}{
The chiral SYK model in three-dimensional holography\\
}}}\end{center}

\begin{center}\textbf{
Alexander Altland\textsuperscript{1 $\star$},
Dmitry Bagrets\textsuperscript{1,2 $\dagger$},
Nele Callebaut\textsuperscript{1 $\ddagger$}, \\ and
Konstantin Weisenberger\textsuperscript{1 $\circ$}
}\end{center}

\begin{center}
{\bf 1} Institute for Theoretical Physics, University of Cologne, Zülpicher Str. 77a, \\50937 Cologne, Germany
\\
{\bf 2} Peter Gr\"unberg Institute, Quantum Computing Analytics (PGI-12), \\Forschungszentrum J\"ulich, 52425 J\"ulich, Germany
\\[\baselineskip]
$\star$ \href{mailto:alexal@thp.uni-koeln.de}{\small alexal@thp.uni-koeln.de} \\
$\dagger$ \href{mailto:d.bagrets@fz-juelich.de}{\small d.bagrets@fz-juelich.de} \\ 
$\ddagger$ \href{nele.callebaut@thp.uni-koeln.de}{\small nele.callebaut@thp.uni-koeln.de} \\ 
$\circ$ \href{mailto:kweisenb@uni-koeln.de}{\small kweisenb@uni-koeln.de}
\end{center}

\section*{\color{scipostdeepblue}{Abstract}} \textbf{ A celebrated realization
of the holographic principle posits an approximate duality between the
$(0+1)$-dimensional quantum mechanical SYK model and two-dimensional
Jackiw-Teitelboim gravity, mediated by the Schwarzian action as an effective low
energy theory common to both systems. We here propose a generalization of this
correspondence to one dimension higher. Starting from different microscopic
realizations of effectively chiral $(1+1)$-dimensional generalizations of the
SYK model, we derive a reduction to the Alekseev-Shatashvilli (AS)-action, a
minimal extension of the Schwarzian action which has been proposed as the
effective boundary action of three-dimensional gravity. In the bulk, we show how
the same action describes fluctuations around the Euclidean BTZ black hole
configuration, the dominant stationary solution of three-dimensional gravity. 
These two constructions allow us to match bulk and boundary coupling constants,
and to compute observables. Specifically, we apply semiclassical techniques
inspired by condensed matter physics to the computation of out-of-time-order
correlation functions (OTOCs), demonstrating maximal chaos in the chiral SYK
chain and its gravity dual.}

\vspace{\baselineskip}

\vspace{10pt}
\noindent\rule{\textwidth}{1pt}
\tableofcontents
\noindent\rule{\textwidth}{1pt}
\vspace{10pt}


\section{Introduction and summary of results}
\label{sec:Intro} 
Holography posits a duality  between gravitational bulk theories and quantum
boundary theories in one dimension lower \cite{1993thooft, Susskind:1994vu, Maldacena1999}. While this principle is formulated at
a great level of generality, only relatively few concretely worked out realizations
exist. Alongside Maldacena's  duality between four-dimensional super Yang-Mills theory and five-dimensional 
AdS quantum gravity 
\cite{Maldacena1999, Gubser1998, Witten1998, AHARONY2000183},
the holographic correspondence linking two-dimensional Jackiw-Teitelboim gravity \cite{1983Teitelboim, JACKIW1985343} to
the  quantum mechanics of the SYK model describing $N$ randomly interacting Majorana fermions \cite{1993Sachdev, Kitaev:2018} has been another case
study attracting a great deal of attention \cite{Almheiri2015, Maldacena2016, Kitaev:2018, Maldacena2016Remarks, Bagrets2016, Bagrets2017, Polchinski2016, Mertens162}. This low-dimensional manifestation
of holography is not only simpler  than the classic example but also
conceptually different: While the latter involves
two precisely defined theories, the two-dimensional duality relates an
\textit{ensemble} of random boundary theories to a gravitational bulk describing
ensemble correlations \cite{Saad2019}. This statistical correspondence is formulated at a high
level of  concreteness. In fact, there exist \textit{two} 
bridges between bulk and boundary, describing parametrically distinct regimes.
The first addresses fine-grained structures of microscopic spectra or,
equivalently, late times of the order of the inverse of the many-body level
spacing, exponential in the number of microscopic constituents of the boundary
theory, $N$ \cite{Cotler2017, SSS2018, Saad2019, Altland2018, Altland21}. 
The second  focuses on the complementary regime of early times, polynomial in
$N$. It is based on a reduction of the SYK model and the JT partition sum to
Liouville quantum mechanics or the Schwarzian action as a common effective
theory in the relevant time window \cite{Bagrets2016,Maldacena2016,Mertens162,Jensen:2016pah}.

What both correspondences just outlined have in common is that they rely on a
high level of microscopic control individually for bulk and boundary theory. On
its basis, we understand that the correspondence between the SYK ensemble and of
the JT partition sum is, in fact, not perfect but limited to effective theories
with parametrically defined scopes, as indicated above. For example, at finite
temperature $T=\beta^{-1}$, Liouville quantum mechanics is described by the
imaginary time action
\begin{align}
    \label{eq:SchwarzianAction}
    S[f]=-M\int_0^\beta d\tau \{f,\tau\},\qquad \{f,\tau\}=- \frac{1}{2}\left( \frac{f''}{f'} \right)^2+ \left( \frac{f''}{f'} \right)', 
\end{align}
where $f'=d_\tau f(\tau)\ge 0$, and $M\sim N$ is a coupling constant of
dimensionality `time'. In the context of the SYK model, this action describes
the invariance of the theory under reparametrizations of  time, where
$f:S^1\to S^1,\tau\mapsto f(\tau)$, is a diffeomorphism of the imaginary time
circle onto itself, $f\in \Diff$ \cite{Kitaev:2018}. In  JT gravity, on the other hand, $S$
quantifies  the action cost associated to fluctuations of a one-dimensional
boundary of two-dimensional space, as described by a geometric deformation
$f(\tau)$ \cite{Maldacena2016}. The holographic correspondence unfolds via the approximate reduction
of boundary and bulk theory to this common effective action in the
`semiclassical' regime of time scales $\tau\sim N$. 

The success of such concepts has sparked a surge of activity aiming for
generalizations to higher dimensions. An important step in this direction was
achieved by Cotler and Jensen, who proposed the Alekseev-Shatashvili (AS) action to assume
the role of Liouville quantum mechanics as a boundary theory for
AdS$_3$ gravity. This theory, a $(1+1)d$ extension of the Schwarzian theory, is defined by the action  \cite{Alekseev1989, Alekseev1990}
\begin{equation}
    \label{eq:S_graviton}
    S_\pm [f] =  \frac{C}{24\pi} \int\limits_0^\beta d\tau \int\limits_0^{L} 
    d x \left(  \frac{f''\partial_\pm f'}{ f'^2} - \frac{ 4\pi^2 }{\beta^2} f' \partial_\pm  f\right),
     \qquad \partial_\pm = \frac 12 (u^{-1}\partial_\tau \mp  i \partial_x), 
    \end{equation}
where $f(x,\tau)$ now is a reparametrization field depending on space and time,
    subject to periodic boundary conditions in both directions, and $f' \equiv \partial_\tau f> 0$. The action
    depends on two coupling constants, the dimensionless $C \sim N \gg 1$, and
    the velocity scale $u$. Finally, the sign `$\pm$' indicates a sense
    of  chirality to be discussed below.

    In the limit of highly anisotropic boundaries, $\beta\gg L/u$, nonvanishing
 spatial derivatives get frozen out, $\partial_x f(x,\tau)=0$, and $S_\pm[f]$
 reduces to the Schwarzian action. Much as  the latter describes black hole
 states in JT-gravity, the spatiotemporal fluctuations of the AS theory describe
 black hole states in  AdS$_3$\cite{Cotler2018, Mertens2022}, the so-called BTZ
 solutions \cite{Banados1992, Banados1993}. The latter dominate the
 gravitational path integral at high temperatures, $ \beta\ll L/u$,
 indicating that the AS action  is the relevant semiclassical theory for
 this\footnote{By analogy with the two-dimensional correspondence, the
 description of long imaginary time scales $\sim \exp(N)$, calls for a different
 boundary theory.}  regime.

The purpose of the present paper is to add concreteness to the holographic
principle revolving around the AS action. Our first contribution is the
definition of a microscopic boundary theory, i.e. a theory assuming the role of
the SYK model in one-dimension lower. The second is an explicit derivation of the AS
theory as effective theory of fluctuations around BTZ states in the
three-dimensional bulk. These derivations will get us in a position to relate bulk and boundary on a
microscopic level, matching coupling constants. Building on this correspondence,
we will introduce streamlined approaches to the computation of correlation
functions, specifically out-of-time-order correlation functions (OTOCs). On this
basis, we will discuss the concept of `maximal chaos' on both sides of the
holographic correspondence.

\subsection{Boundary perspective}

\begin{figure}
    \centering
    \includegraphics[width=0.45\linewidth]{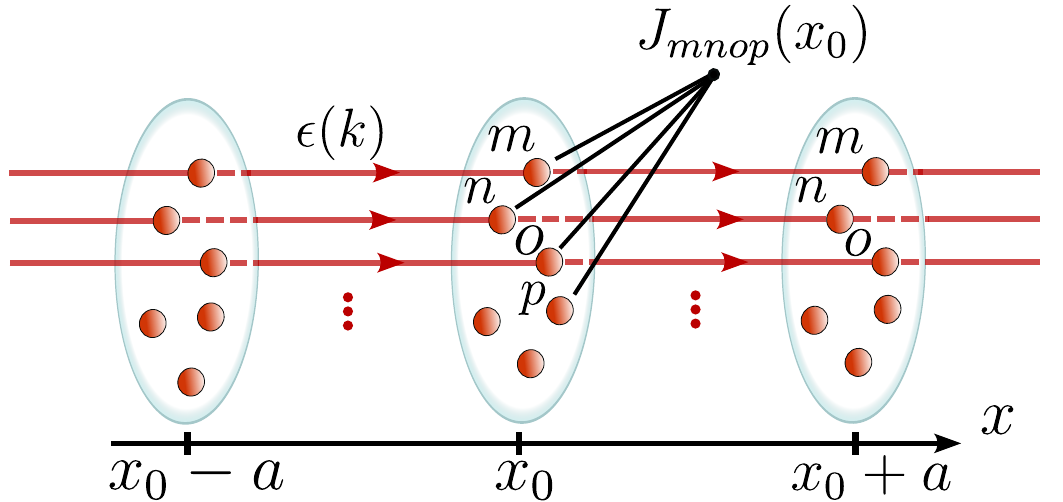}
    \caption{Sketch of the chiral SYK model: A chain  of SYK grains, each
    containing $N$ Majorana fermions with the well-known grain-local Majorana
    interaction $J_{mnop}(x_0)$.  Across grains, each species of Majoranas is
    coupled independently by a kinetic term with chiral dispersion
    $\epsilon(k)$. }
    \label{fig:chain}
\end{figure}

What is missing so far in the discussion of the three-dimensional holographic
correspondence is a microscopic boundary theory assuming the role of the SYK
model in one dimension lower. In this paper we propose such a model system,
building on a combination of four principles, exhibited by various
`realistic' classes of condensed matter systems: a chiral, approximately linear single
particle dispersion, strong local interactions, static randomness, and the
violation of particle number conservation. The first is realized at edge modes
of topological insulators, such as two-dimensional quantum Hall insulators\cite{Altland2023}. In
these systems, a nonvanishing bulk topological invariant requires the presence
of left- or right-propagating gapless boundary modes governed by an
approximately linear dispersion. In the presence of Coulomb interactions --- the
second principle ---  these modes define a universality class known as the
\textit{helical liquid}\cite{wuHelicalLiquidEdge2006}. We consider a helical liquid with $N/2>1$
co-propagating edge modes, corresponding to a bulk with invariant $N/2$. We also
consider the system coupled to a nearby  superconductor `proximitizing' the
system via particle-number non-conserving Andreev scattering operators. Finally,
we assume the presence of impurities rendering the systems single particle
states effectively random (cf. Fig.~\ref{fig:chain}). 

All these principles can be individually realized at topological insulator
surfaces. However, we here make the stronger assumption that for an
appropriate parameter configuration they combine to define a system in the
universality class of the chiral SYK model. To see how this may come about in
principle, consider the system coarse grained into spatial units whose extension
is defined by the range of the two-body interaction. If we now consider the
creation operators of helical edge fermions in a Majorana representation,
symbolically $c=\gamma+i\gamma'$, and represent the interaction operator in a basis
of the random single particle eigenstates defined by the combination of impurity
and Andreev scattering, we obtain the model Hamiltonian 
\begin{equation}
    \label{eq:H_SYK4_D2}
    H = \sum_{ik} \epsilon(k) \gamma^{i \dagger}_k \gamma^i_k + \sum_{ijkl}  \int_0^L dx\, J_{ijkl}(x) \gamma^i(x) \gamma^j(x) \gamma^k(x) \gamma^l(x),
    \end{equation}
for $N=(N/2)\times 2$ Majorana modes $\gamma_i$. Here the coefficients $J_{ijkl}$
inherit the randomness of the single particle states and hence are random
themselves, while the first term accounts for a dispersion with low energy asymptotics
$\epsilon_k \sim k$. (Note that a chiral system in which quantum states
propagate only in one direction retains its chirality in the presence of
randomness, there is no `backscattering' in this case.)  

In Eq.~\eqref{eq:H_SYK4_D2},  $\gamma^i(x)=\frac{1}{\sqrt{L}}\sum_k \gamma^i_k e^{ikx}$, are  Majorana
fermion operators, $\gamma^i(x)^\dagger = \gamma^i(x)$, with anti-commutation relations
$\{\gamma^i(x),\gamma^j(x')\} = \delta(x-x') \delta^{ij}$. Adopting a rationale previously applied in the construction of `effective Hamiltonians' for  quantum dots subject to random scattering\cite{Alhassid99}, we assume the interaction coefficients to be Gaussian distributed as
\begin{equation}
\label{eq:J_corr}
\langle J_{ijki}(x) J_{ijkl}(x') \rangle = \frac{3! J^2}{(k_0 N)^3} \delta (x-x'),
\end{equation}
where $J$ is the characteristic single particle interaction energy, and $k_0$ an effective
momentum cut-off related to the correlation length $a$ of the random interaction, $k_0 = \pi/a$. 
We make the non-trivial assumption that the interaction energy is dominant, in the sense that 
\begin{equation}
\label{eq:Lambda_def}
J \gg \Lambda \equiv |\epsilon(k_0)|
\end{equation} 
for all relevant momentum scales below $k_0$, where the scale $\Lambda$ defines
a maximal kinetic energy.

Before turning to the further discussion of this model, let us mention a few
other spatial extensions of the SYK model.  Ref.~\cite{Lian2019} introduced a
model with  linear  dispersion $\epsilon(k) = v k$ and  random but spatially
uniform interaction coefficients. While this system is chaotic and chiral like
Eq.~\eqref{eq:H_SYK4_D2}, its long range correlations lead to symmetries
different from local reparametrization invariance.  
A  model with long-ranged disorder and linearly dispersive non-chiral  fermions,
($\epsilon(k) = \pm v k$ for right/left movers) was discussed and approximately
solved in~\cite{berkooz2017}. The existence of connections to AdS$_3$ was left
as an open question by the authors. Finally, Refs. \cite{Turiaci2017} and
\cite{Pasterski2022} proposed two-dimensional generalizations of the SYK model
(conceptually, one-dimensional models subject to time dependent interactions).
Compared to the standard `0-dimensional' SYK Hamiltonian, this is a stepup by
two dimensions, with a less direct connection the holographic principle in one
dimension lower. The criteria motivating the present model ansatz
Eq.~\eqref{eq:H_SYK4_D2} include spatial locality of interactions and
statistical correlations,  a symmetry contents identical to that of the AS
action, and principal realizability as a many-body system (The  Hamiltonian
Eq.~\eqref{eq:H_SYK4_D2} may be conceptualized as an effective Hamiltonian
describing electrostatically correlated multi-channel quantum Hall edge modes).

\subsection{Bulk perspective}

In section~\ref{sec:grav}, we will derive the AS action from bulk gravity.  
In essence, this amounts to the  lifting of an analogous construction for JT
gravity, where the Schwarzian action emerged as a boundary-fluctuation action of
two-dimensional black hole solutions. In one dimension higher, alongside
global AdS$_3$,  BTZ black holes provide stationary solutions,
and they reduce to JT black holes upon dimensional reduction \cite{Achucarro:1993fd}.
One expects the same to happen at  
the level of the boundary actions, suggesting that fluctuations around the BTZ
saddle are described by the AS action.  
Indeed, a number of works revealed intimate connections between the BTZ geometry
and the AS action~\eqref{eq:S_graviton}. To categorize these previous
contributions, we temporarily denote an AS action with Schwarzian derivatives
acting in the spatial, imaginary, or a real time direction by AS$_{x,\tau,t}$,
respectively. 

Working within the Chern-Simons representation of three-dimensional gravity (for a review, see section \ref{sec:CSAction}), the seminal
paper  \cite{Cotler2018} identified two chiral copies of the AS$_x$-action as
the theory of fluctuations around global AdS$_3$, the vacuum state of
AdS$_3$-gravity. Modular invariance\footnote{The  AdS/CFT correspondence combined with  modular invariance of  CFT on a torus imply a 
dual relation 
 between the Euclidean global AdS and the BTZ solution of AdS$_3$ gravity. Both
solutions are defined on solid tori, and they map onto each other  by an exchange of
the boundary space and Euclidean time coordinate, cf. section ~\ref{sec:CSAction}.} then  implied that two
chiral copies of the AS$_\tau$-action  provide  
	the theory of fluctuations around  Euclidean BTZ black holes.  
Given that the BTZ solution has one boundary in Euclidean signature and two in
	Lorentzian signature (see e.g.~\cite{Callebaut2023}),
	Refs.~\cite{Cotler2018} and \cite{Henneaux2020} proposed  a quadruple AS$_t$
	description  for the two-sided Lorentzian BTZ. 
	Ref.~\cite{Mertens2022} went one step further to suggest AS$_\tau$  as a
	universal effective quantum theory of AdS$_3$-gravity in the
	high-temperature regime. Other investigations of AdS$_3$ gravity
	employing the AS action include \cite{Chua2023}, and works that focus on the phase space of one-
	\cite{Cotler2018, Kraus2021, Ebert2022}  and two-boundary solutions \cite{
		Henneaux2020, Banerjee2022}, or the  spectral form factor \cite{Cotler2021}.

	In this paper we will add a missing element to this web of constructions,
	namely a direct derivation of the AS$_\tau$ action as the theory describing
	fluctuations around the BTZ black hole in Euclidean signature. This
	derivation, as opposed to a more indirect one based on modular invariance arguments, 
    will be
	instrumental for our explicit  comparison of bulk and boundary. Starting
	from Chern-Simons theory defined on a manifold topologically equivalent to a
	solid torus, we will derive the theory of fluctuations in a framework
	exchanging the role of space and time relative to the expansion around the
	AdS$_3$ vacuum~\cite{Cotler2018}. This reassignment plays a crucial role in
	our construction. It leads to a high level of parallelism in the reduction
	of the boundary theory (the chiral SYK chain) and the bulk theory to their
	respective low energy fluctuation actions, and in particular allows us to
	match coupling constants. On this basis, we will then move on to the
	computation of observables.         
	
	The rest of the paper is organized as follows. In the next section, we
process the model Eq.~\eqref{eq:H_SYK4_D2} by an extended version of the
$G\Sigma$-approach to the SYK model~\cite{1993Sachdev}, and derive the AS
action. In section \ref{sec:grav}, we obtain the same action from Euclidean
gravity. Finally, in section \ref{Sec:Liouville} we consider signatures of chaos
described by the Liouville field theory reformulation of the AS functional
integral. We conclude in section \ref{sec:discussion}, and various technical
details are relegated to Appendices.

\section{Derivation of the AS action from the boundary theory}
\label{sec:Chiral-SYK} 
In this section, we derive the semiclassical boundary theory from the
one-dimensional SYK model. The construction proceeds along a sequence of steps,
which are extended versions of those used in one dimension lower en route from
the SYK model to the Schwarzian action: a representation of the disorder
averaged theory in terms of a $G\Sigma$-functional, a mean field treatment of the
latter, an identification of a weakly broken reparametrization symmetry and a
derivation of the corresponding Goldstone mode action, which happens to be of
the AS$_\tau$ form. In course of the mean-field analysis, one realizes that the
behavior of the one-dimensional SYK model based on {\it helical liquids} crosses
over to the 'Fermi-liquid' at temperatures below $T_\Lambda \sim \Lambda^2/J$, where 
$\Lambda$ is a measure for the spatial coupling strength. 
As a consequence, the holographic correspondence of this particular model is limited to a
(parametrically wide) temperature interval, $T_\Lambda < T < J$. This limitation motivates
us to formulate another variant of a chiral SYK model, whose dispersion remains
approximately flat over an extended window around $k=0$. Such an exotic
dispersion relation can be physically realized in so-called soft quantum
Hall edges. We demonstrate that the SYK-like phase for this second model can be
stabilized including the deep infrared limit, leading to a holographic duality
also for the range of temperatures $T_* < T < J$, where the scale $T_* \ll T_\Lambda$ 
is identified in Appendix~\ref{App:Rep_invariance_S_star}.

\subsection{Generalized $G\Sigma$-action and mean-field analysis} \label{subsec:Models}
We begin with a recapitulation of the construction of the Luttinger-Ward functional~\cite{luttingerGroundStateEnergyManyFermion1960} 
for a single SYK cell and the emergent reparametrization invariance of the corresponding stationary phase equations.
In the context of the SYK model, this functional is known as $G\Sigma$-action and assumes the following form:
\begin{align}
	\label{eq:action}
	S[\Sigma,G]=-{N\over 2}\left[ \mathrm{Tr}\ln( \partial_{\tau} +  \Sigma)  +  
	\int d\tau_{1,2}\left(\frac{J^2}{4 } \big[G_{\tau_1\tau_2}\big]^4 +  \Sigma_{\tau_2\tau_1} G_{\tau_1\tau_2}\right)\right]. 
\end{align}
Here, $\tau$ is imaginary time, which in the zero temperature limit considered
presently becomes an unbounded real variable, and the pair of field variables $(G,\Sigma)$ is bilocal in time 
(and in a replica index which we suppress for simplicity, as it will not play a role throughout). 

The factor $N$ upfront invites a stationary phase approach, where variation of
the action in $G$ and $\Sigma$ leads to the stationary phase equations 
$-(\partial_\tau+ \Sigma) G=\mathds{1}$ and $\Sigma_{\tau\tau'}=J^3(G_{\tau\tau'})^3$. 
In the low energy  limit, where $\partial_\tau$ is negligibly small compared to the high energy scale $J$, these possess 
a famously degenerate~\cite{Maldacena2016Remarks, Kitaev:2018} set of solutions. To formulate it, consider a \textit{reparametrization of time}, i.e. a smooth 
and invertible function $\tau \mapsto t=f(\tau)$, with inverse $t\mapsto \tau=F(t)$. One can then show~\cite{Maldacena2016Remarks, Kitaev:2018}, that the configurations
\begin{alignat}{3}
	\label{eq:MeanFieldSingleGrain}
	G_{\tau_1\tau_2}&&\equiv {f'_{1} }^{1/4}\,G^0_{t_1 t_2}\, {f'_{2} }^{1/4},\qquad & G^0_{tt'}=-\frac{1}{(4\pi)^{1/4}J^{1/2}}\frac{\textrm{sgn}(t-t')}{|t-t'|^{1/2}},\cr
    \Sigma_{\tau_1\tau_2}&&\equiv {f'_{1} }^{3/4}\,\Sigma^0_{t_1 t_2}\, {f'_{2} }^{3/4},\qquad &
    \Sigma^0_{tt'}=-\frac{J^{1/2}}{(4\pi)^{3/4}}\frac{\textrm{sgn}(t-t')}{|t-t'|^{3/2}},  
\end{alignat}
with $t_i = f(\tau_i)$ and $f'_i = f'(\tau_i)$ solve the mean-field equations. 
Thinking of imaginary time compactified on a circle, and
$f(\tau)$ as a diffeomorphism from the circle onto itself, this identifies the
diffeomorphism group $\textrm{Diff}(S^1)$ as the symmetry group of the theory.
Included in this group is the three-dimensional subgroup
$\textrm{SL}(2,\Bbb{R})$, represented as 
\begin{align}
	\label{eq:SL2RTransformation}
	f(\tau)= \frac{a \tau+b}{c\tau +d},\qquad ad-bc=1.
\end{align}
This subset of transformations leaves the stationary phase configurations
invariant, i.e. $X_{\tau\tau'}=X^0_{\tau\tau'}$ with $X=G,\Sigma$. In this way, we have
identified the coset space  $\textrm{Diff}(S^1)/\textrm{SL}(2,\Bbb{R})$  as
the Goldstone mode manifold of the single SYK grain. (We note that the
innocently looking $\textrm{SL}(2,\Bbb{R})$ subgroup will play an extremely important role as
a consistency checker in the construction of the full theory below.)

We now generalize the $G \Sigma$--functional~(\ref{eq:action}) to the
(1+1)-dimensional case. Defining the Green's function 
\begin{equation} 
\label{eq:G20}
	G^{xx'}_{\tau\tau'} = - \frac 1 N \sum_i \langle \gamma^i(\tau, x) \gamma^i(\tau', x') \rangle, 
\end{equation}
which is now non-local both in time and space, the functional becomes
\begin{align}
	\label{eq:action1+1}
	S[\Sigma,G]=-{N\over 2}\left [ \mathrm{Tr} \ln( \partial_{\tau}+ \epsilon_{\hat k} +  \Sigma ) + 
    \frac{J^2}{4k_0^3} \int \!\!dx d\tau_{1,2} \big[G_{\tau_1 \tau_2}^{\,\, x x}\big]^4  
     + \mathrm{Tr} \, ( G \Sigma) \right], 
\end{align}
Here, the trace extends over temporal and spatial degrees of freedom with a self-energy  $\Sigma_{\tau \tau'}^{x x'}$ whose matrix structure is defined by that of  $G$.
The  caret notation, $k\to \hat k$, indicates  that space and
position operators are subject to canonical commutation relations, $[\hat k,\hat x]=-i$. 
Variation of the action~(\ref{eq:action1+1}) leads to the extended set of equations,
\begin{equation}
	\label{eq:MF}
	(i\epsilon - \epsilon_k - \Sigma_\epsilon)\, G_{\epsilon, k} = 1, \qquad \Sigma_{\tau_1 \tau_2} = (J^2/ k_0^{3})\, [G_{\tau_1 \tau_2}^{\,xx}]^3,
\end{equation}
with a saddle point self-energy $\Sigma_{\tau_1 \tau_2}^{x_1 x_2} = \delta_{x_1 x_2}\Sigma_{\tau_1 \tau_2}$, local in space\footnote{To shorten formulae we will often abbreviate expressions for the $\delta$-function as $\delta_{x_1 x_2} = \delta(x_1-x_2)$.}. Similarly, the  Green's function at coinciding spatial points $G_\epsilon\equiv G_\epsilon^{xx}$  is given
by the momentum integral,
\begin{equation}
	    \label{eq:MF_GF}
		G_\epsilon=\frac{1}{2\pi} \int\limits_{-k_0}^{k_0} dk\, G_{\epsilon,k} =\frac{1}{2\pi} \int\limits_{-k_0}^{k_0} \frac{dk}{i\epsilon -  \epsilon_k   - \Sigma_\epsilon}, \qquad \epsilon_k = v_0 k,
\end{equation}
and $G_{\tau-\tau'}$ is its temporal Fourier transform from the energy domain.

Before turning to the self-consistent solution of
Eqs.~(\ref{eq:MF}-\ref{eq:MF_GF}), we need to address the regularization of our
model and its associated many-body Hilbert space. Recalling that the parameter $a$ defined below Eq.~\eqref{eq:J_corr}
sets a correlation radius of the random interaction, we introduce
coarse-grained Majorana operators $\gamma^j_n$ in position space at discrete
location $x_n = n a$ as 
\begin{equation}
	\label{eq:gamma_n}
	\gamma^j_n  = \left(\frac{a}{L}\right)^{1/2} \sum_{|k|<k_0} \gamma_k^j e^{i k x_n}, 
	\qquad \frac 12 \{\gamma^i_n, \gamma^j_{n'}\} = \delta_{nn'} \delta^{ij},
\end{equation}
with $L$ being the system size. The Hamiltonian of the SYK-like interaction then becomes a sum over lattice sites, 
\begin{equation}
	H_{\rm SYK} = \frac{1}{4!}\sum_n \sum_{ijkl} J^{n}_{ijkl} \gamma^i_n \gamma^j_n \gamma^k_n \gamma^l_n, \qquad  
	\langle J^{n}_{ijkl} J^{n'}_{ijkl} \rangle = \frac{3! J^2 \delta^{nn'}}{(\pi N)^3},
	\label{eq:H_SYK_chain}
\end{equation}
with lattice two-body matrix elements $J^{n}_{ijkl} = a^{-1} J_{ijkl}(x_n)$. The kinetic energy, as before, is best expressed as a sum over discrete momenta, quantized in units of $2\pi/L$, see Eq.~(\ref{eq:H_SYK4_D2}).
With $N_x = L/a$ representing the number of sites, the dimension of the Hilbert space is then given by ${\cal D} = {2}^{\,N_x N/2}$. Fig.~\ref{fig:chain} illustrates this representation of a model reflecting
its coarse-grained regularization. Within such regularization scheme the Majorana Green's function
at coinciding spatial points, $G_{\,\epsilon}$, is expressed via a momentum integral 
bounded by the UV cut-off $k_0$, see Eq.~(\ref{eq:MF_GF}).  

Turning to the solution of the mean-field equations, we introduce the important energy scale,
\begin{equation}
	T_\Lambda = \frac{(\pi \Lambda)^2}{J}, \qquad T_\Lambda \ll \Lambda \ll J,
\end{equation}
 delineating two distinct regimes: {\it 'weakly dispersive'} ($\epsilon \gg
T_\Lambda$) and {\it 'strongly dispersive'} ($\epsilon \ll T_\Lambda$) (see
Ref.~\cite{Geo:2022}). In the weakly dispersive limit, the kinetic energy is
negligible compared to the self-energy, i.e. $\epsilon_k \ll \Sigma(\epsilon)$,
and both the Green's function and self-energy can, with good accuracy, be
approximated by the conventional SYK form~ (cf. Eq.
(\ref{eq:MeanFieldSingleGrain})), shown here in the energy and time domain for
later reference: 
\begin{equation}
	\label{eq:GS0_SYK_E}
	G^{0}_\epsilon = - \frac{i k_0 {\rm sgn}(\epsilon)}{\sqrt{J |\epsilon|}}, \qquad
	\Sigma^{0}_\epsilon = - \frac{i {\rm sgn}(\epsilon)}{\pi} \sqrt{J |\epsilon|}, \qquad \epsilon \gg T_\Lambda,
\end{equation}
and
\begin{equation}
   \label{eq:GS0_SYK_t}
	G^{0}_{\tau \tau'} = - \frac{ k_0 {\rm sgn}({\tau -\tau'})}{\sqrt{2\pi J |{\tau -\tau'}|}}, \qquad 
	\Sigma^{0}_{\tau \tau'} = - \frac{J^{1/2} {\rm sgn}({\tau -\tau'})}{(2\pi|{\tau -\tau'}|)^{3/2}},  \qquad \tau \ll 1/T_\Lambda.
\end{equation}
In the strongly dispersive limit, the integration range in Eq.~(\ref{eq:MF_GF}) can be extended to infinity and the momentum integral is defined by a pole at $k_* = - \Xi_\epsilon/v_0$, where we have introduced $\Xi_\epsilon = \Sigma_\epsilon - i\epsilon$. This results in a simple, 'Fermi-liquid'-like Green's function,
\begin{equation}
	G^{\rm FL}_\epsilon = -\frac{i {\rm sgn}(\epsilon)}{2 v_0}, \qquad 
	G^{\rm FL}_{\tau-\tau'} = - \frac{1}{2\pi v_0} \times \frac{1}{\tau-\tau'},
\end{equation}
which is independent of the details of $\Xi_\epsilon$. Consequently, the self-energy evaluates to 
$\Sigma_\tau = ({J^2}/{k_0^3}) G^3_\tau \propto \tau^{-3}$, which results in a quadratic energy dependence, $\Sigma_\epsilon \propto i\epsilon^2$. Taking into account all numerical
constants, limiting results for the self-energy read,
\begin{eqnarray}
	\label{eq:Self_E-linear_k}
	\Sigma_\epsilon &=&  - \frac{i {\rm sgn}(\epsilon)}{\pi} \sqrt{J |\epsilon|}, \qquad T_\Lambda \ll\epsilon \ll J,  
	\nonumber \\
	\Sigma_\epsilon &=&  \frac{i \sqrt{J}}{16 \,T_\Lambda^{3/2}} \epsilon^2   {\rm sgn}(\epsilon),  
	\qquad \epsilon \ll T_\Lambda.
\end{eqnarray}
The two asymptotics match at the scale $\epsilon \sim T_\Lambda$, indicating the
self-consistency of the above analysis. In the deep infrared limit, the
self-energy when continued to real energies, defines a damping rate of
Majoranas, $\gamma(\epsilon) \propto (\epsilon/T_\Lambda)^2$, reminiscent of the
Fermi-liquid physics. It is also worth noting here that a crossover from SYK to
the 'Fermi-like' saddle-point described by Eq.~(\ref{eq:Self_E-linear_k}) is
fully analogous to that occurring in granular SYK arrays~\cite{Balents:2017,
Altland:2019}. 

The above considerations imply that stationary solutions for $G$ and $\Sigma$
exhibit an approximate reparametrization invariance only within the time range
$J^{-1} < t < T_\Lambda^{-1}$. At longer times, the 'Fermi-liquid' saddle-point
takes over and the invariance is lost. Since the holographic duality for the SYK model relies heavily on
the reparametrization symmetry, we will restrict our analysis of the linearly dispersive
model~(\ref{eq:H_SYK4_D2}) to temperatures $T_\Lambda <
T < J$, and refer to it as  Model~I throughout. In the next subsection we will introduce the complementary flat-band Model~II, for which the SYK phase remains stable down to the deep infrared limit. 

\subsection{Flat-band model} \label{sec:Flat_band_model}
\begin{figure}[t]
\centering	
		\includegraphics[width=0.45\linewidth]{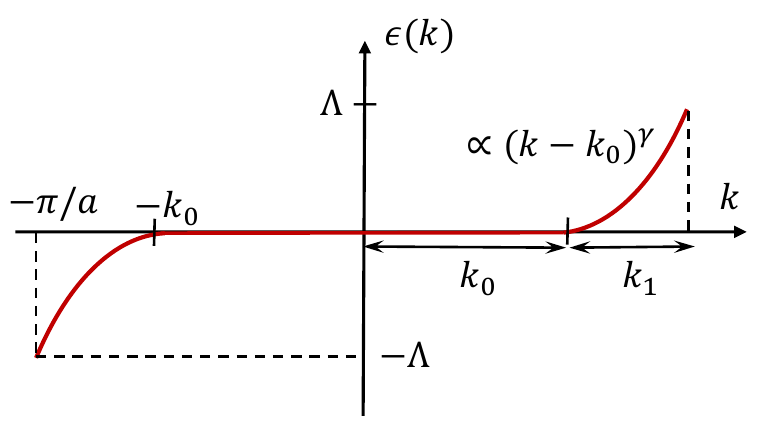}
		\caption{Dispersion relation of Majorana edge modes in the chiral SYK model~(\ref{eq:H_SYK4_D2}) with flat bands (Model II). For the dispersive part one defines $\epsilon_k = \Lambda (\delta k/k_1)^\gamma$, where $\delta k = k - k_0$, so that 
		$\epsilon(\pi/a) = \Lambda \ll  J$ in accordance with Eq.~(\ref{eq:Lambda_def}). 
		}
		\label{fig:Flat_band_model}
\end{figure}

In this section we discuss a variant of the chiral SYK model whose dispersion
relation $\epsilon_k$ is shown in Fig.~\ref{fig:Flat_band_model}, where $k_0 <
\pi/a$ defines the width of an (approximate) flat region and  $\gamma>1$ is the
dispersive exponent for large momenta $>k_0$.  We assume the parameter $k_1/k_0
\ll 1$ to be small. In the following, we will  demonstrate that the SYK phase
remains stable for this model. Although this spectrum of chiral boundary
fermions may seem exotic, it emerges, for example, at the 'soft' edges of
gate-confined IQHE samples. Specifically, within the self-consistent
electrostatic framework developed by Chklovskii et
al.~\cite{Chklovskii:1992,Chklovskii:1993}, one finds the exponent $\gamma=3/2$
(see Appendix~\ref{app:Soft_IQHE_edge}).

The set of self-consistent mean-field equations (\ref{eq:MF}-\ref{eq:MF_GF}) for
Model~II can be solved neglecting the  $i\epsilon$--term. Referring for details of this
procedure to Appendix~\ref{app:mean-field}, the scale $T_\Lambda$ plays the same role  as in 
{Model I}. Specifically, in the weakly dispersive limit ($\epsilon \gg
T_\Lambda$), one arrives at the SYK-like solutions Eqs.~\eqref{eq:GS0_SYK_E} and \eqref{eq:GS0_SYK_t}. 
(One may make these solutions more accurate by substitution  $k_0 \to k_0 + k_1$ and $J^2 \to J^2 (1 + k_1/k_0)^3$, however this refinement will be irrelevant throughout.)

In the strongly dispersive regime (deep IR limit), the Green's function and
self-energy receive perturbative corrections of order $O(k_1/k_0)$ to the SYK
solutions. Specifically, in the energy domain (at $\epsilon \ll T_\Lambda$), they
read
\begin{equation}
	\label{eq:S_E}
	\frac{\delta\Sigma_\epsilon}{\Sigma^{0}_\epsilon} =  \sigma_1(\gamma) \left(\frac{k_1}{k_0}\right)  \left|\frac{\epsilon J}{\Lambda^2}\right|^{1/2\gamma} \!\!\!+
	{\cal O}(k_1^2/k_0^2), \qquad
	\frac{\delta G_\epsilon}{G^{0}_\epsilon} =  g_1(\gamma) \left(\frac{k_1}{k_0}\right)  \left|\frac{\epsilon J}{\Lambda^2}\right|^{1/2\gamma} + \dots,
\end{equation}
where for details of the derivation and the explicit definitions of functions
$\sigma_1(\gamma)$ and $g_1(\gamma)$ we refer the reader to
Appendix~\ref{app:mean-field}. The main point here is that the kinetic energy
term $\epsilon_k$  is an {\it irrelevant} perturbation, including in the deep IR
limit. It produces a correction to the leading SYK self-energy
$\Sigma^{0}_\epsilon$ which is parametrically small in $\epsilon$. Physically,
the flat-band dispersion of the kinetic energy enhances electron correlations,
thereby stabilizing the SYK phase. This is different from  Model I, whose linear
dispersion $\epsilon_k$ qualitatively changes  the nature of the saddle-point at low energies. 

In the Fourier conjugate time domain, at large times $\tau \gg 1/T_\Lambda$, we obtain to the same accuracy 
\begin{equation}
	\label{eq:GS_t}
	\frac{\delta \Sigma_{\,\tau}}{\Sigma^{0}_\tau} = \tilde \sigma_1(\gamma) \left(\frac{k_1}{k_0}\right) \left|\frac{J}{\Lambda^2\tau}\right|^{1/2\gamma} + \dots, \qquad
	\frac{\delta G^{xx}_{\,\tau}}{G^{0}_\tau} = \tilde g_1(\gamma) \left(\frac{k_1}{k_0}\right) \left|\frac{J}{\Lambda^2\tau}\right|^{1/2\gamma} + \dots,
\end{equation}
where the numerical coefficients $\tilde g_1(\gamma)$ and  $\sigma_1(\gamma)$
are discussed in Appendix~\ref{eq:Table}. The result for the Green's function
 affords a clear physical interpretation: The scaling dimension of Majoranas
in the SYK model is $1/4$. The first-order correction to the bare SYK result in
Eq.~(\ref{eq:GS_t}) arises from the small fraction of mobile Majoranas, whose
scaling dimension crosses over to $\frac 1 2 \Delta_1= 1/4 + {1}/{(4 \gamma)}$
at times longer than $T_\Lambda^{-1}$. 

To better illustrate the last point, one can introduce a subset of fermion-like 
operators that describe only mobile Majoranas, 
\begin{equation}
\label{eq:lambda_Majoranas_def}
    \lambda_l^j = \left(\frac{a_1}{L}\right)^{1/2} \sum_{|k|=k_0}^{\pi/a} \gamma_k^j e^{i k x_l}, 
    \qquad \frac 12 \{\lambda^i_l, \lambda^j_{l'}\} = \delta_{ll'} \delta^{ij},
    \qquad a_1 = \pi/k_1,
\end{equation}
and are defined at the coarse-grained positions $x_l = l a_1$. Note, that the summation above goes
only over a narrow strip of momenta where the kinetic energy $\epsilon_k$ is nonvanishing.
By introducing a two-point function,
$g_{\tau\tau'}^{xx'}$, for these fields in analogy to Eq.~\eqref{eq:G20}, its value at coinciding spatial 
points on a level of the mean-field approximation is given by
\begin{equation}
	    \label{eq:MF_g_mobile}
        g^{xx}_\epsilon= - \frac{1}{2\pi} \int_{\epsilon_k \neq 0} 
       \frac{dk}{\epsilon_k  +  \Sigma_\epsilon}.
\end{equation}
In the weakly dispersive limit, $\epsilon \gg T_\Lambda$, the kinetic energy is negligible compared to the self-energy.
Therefore, for short times, $\tau \ll T_\Lambda^{-1}$, a temporal behavior of the correlator $ g^{xx}_\tau$
remains of the conventional SYK form. For longer times, $\tau \gg T_\Lambda^{-1}$, corresponding
to the strongly dispersive limit, the correlator $g^{xx}_\tau$ crosses over to
$\delta G^{xx}_{\,\tau}$. This time dependence is summarized as
\begin{equation}
    g^{xx}_\tau \sim - \frac{k_1 {\rm sgn}(\tau)}{\sqrt{2\pi J}}\times\left\{ 
    \begin{array}{lc}
        |\tau|^{-1/2}, &\quad \tau \ll T_\Lambda^{-1}, \\
         {|\tau|^{-\Delta_1}} |T_\Lambda|^{-1/2\gamma}, &\quad \tau \gg T_\Lambda^{-1}, 
    \end{array}\right.
\end{equation}
where few numerical factors are omitted for brevity. The two asymptotics here match at 
the intermediate time scale $\tau \sim T_\Lambda^{-1}$, where a transmutation of the
scaling exponent from $1/2$ to $\Delta_1 = 1/2 + 1/2\gamma$ governing the temporal dependence 
of the correlator occurs. In Sec.~\ref{Sec:Liouville} we will analyze how quantum
fluctuations beyond the mean-field analysis affect the above result.

\subsection{Reparametrization invariance and soft modes}
\label{sec:Rep_Inv_Soft_Modes}
The concept of  reparametrization invariance has been essential in the context of
holographic duality between the SYK model and JT gravity. At the level of the
$G\Sigma$--functional it manifests itself through a degenerate coset
manifold of saddle-points~(\ref{eq:MeanFieldSingleGrain}) when the symmetry breaking time
derivative operator $\partial_\tau$ is neglected. In the chiral SYK model, the kinetic
energy $\epsilon_k$ acts as an additional source of
symmetry breaking. In this subsection, we argue how global
(i.e.~independent of $x$) and {\it adiabatic} reparametrizations of time, $\tau
\to t=f(\tau)$, can be employed to construct a coset manifold of approximate
saddle-point solutions to the (1+1)-dimensional $G\Sigma$--action, despite the
presence of  this term. In this way, we show that reparametrizations
remain valid soft modes in the problem. We use the latter in the next subsection
to derive the AS action from the chiral SYK model.

The reparameterization invariance of the SYK model can be considered as a
generalization of invariance under scaling $t=\lambda t'$, to diffeomorphism
invariance $t=f(\tau)$. We therefore start our analysis of the higher
dimensional model by considering its behavior under scaling. 
Temporarily
focusing on the case of zero temperature, let $G(t;\Lambda)$ and $\Sigma(t;\Lambda)$ be approximate solutions of the
zero temperature saddle-point equations ignoring the time derivative operator $\partial_\tau$ and
evaluated at  coinciding spatial points ($x'=x$). As shown above, this
approximation is always valid for the weakly dispersive limit and it also holds
in the strongly dispersive limit of the 'flat-band' model. Schematically, the corresponding
equations read
\begin{equation}
	\label{eq:Eqs_MF}
	G(t-t';\Lambda) = -\left[\left(\epsilon_{\hat k} + \Sigma \right)^{-1}\right]_{tt'}^{xx}, \qquad  \Sigma(t;\Lambda) = (J^2/k_0^3)\,G^3(t;\Lambda), 
\end{equation}
where the self-energy operator is defined by $\Sigma_{tt'}^{xx'}\equiv
\delta_{xx'}  \Sigma(t-t';\Lambda)$, and we explicitly indicate the dependence
of the solutions on the kinetic energy scale $\Lambda$. The solutions of
equations~(\ref{eq:Eqs_MF}) satisfy the scaling relations
\begin{equation}
	\label{eq:GS_b}
	G(t/\lambda; \Lambda' ) = \lambda^{1/2} G(t; \Lambda'/\lambda^{1/2}), \qquad 
	\Sigma(t/\lambda; \Lambda' ) = \lambda^{3/2} \Sigma(t; \Lambda'/\lambda^{1/2}), 
\end{equation}
with $\lambda \in \mathds{R}^+$. To see that this condition must hold,  consider the action ${\cal S}
\approx - \int dt\, H$ defined by the Hamiltonian~(\ref{eq:H_SYK4_D2}), where we
substitute Majorana operators by Grassmann variables, $\hat \gamma \to \chi$
(all quantum indices are omitted for brevity), and  we have dropped the term $\int dt \chi \dot{\chi}$,  in
accordance with the rationale of equations~(\ref{eq:Eqs_MF}). The action ${\cal
S}$ is invariant under the (global) rescaling,
\begin{equation}
	t' = t/\lambda, \qquad  \chi' = \lambda^{1/4} \chi, \qquad \Lambda' = \lambda^{1/2}\Lambda, \qquad J' = J.
\end{equation}
In particular, the engineering scaling dimension of the kinetic energy is
$[\Lambda] = 1/2$. Considering that the dimension of the Green's function is
$[G] = 2 \times [\chi] = 1/2$, we arrive at the scaling symmetries of the
approximate mean-field solutions~(\ref{eq:GS_b}). As a sanity check, it is
straightforward to verify that the long-time asymptotic of the Green's function
and self-energy~(\ref{eq:GS_t}) in the 'flat-band' model, which are valid in the
strongly dispersive regime, are consistent with this scaling invariance.

On this basis, we  wish to generalize from scaling to reparametrizations
$f(\tau)$ of time. Staying close to the rationale previously applied to the
local SYK model, we require strict invariance of $(G,\Sigma)$ under the subgroup of ${\rm SL}(2,
\mathds{R})$ transformations Eq.~\eqref{eq:SL2RTransformation}. We begin by introducing the ${\rm SL}(2,
\mathds{R})$-invariant ratio
\begin{equation}
	\label{eq:r12}
	r_{\tau_1 \tau_2} = \frac{f_1 - f_2}{\sqrt{f_1' f_2'}},  \qquad f_i = f(\tau_i), \qquad f'_i = f'(\tau_i),
\end{equation}
 and the rescaled kinetic energy cutoff,
\begin{equation}
	\Lambda_{\tau_1 \tau_2} = \frac{\Lambda}{(f'_1 f_2')^{1/4}},
\end{equation}
where $f' \equiv \partial_\tau f$. Next we define the reparameterized fields
\begin{eqnarray}
	\label{eq:S_rescale}
	\widetilde\Sigma_{\tau_1\tau_2} &=& \Sigma(r_{\tau_1 \tau_2} ;\Lambda) \equiv 
	{f'_1}^{3/4}  \Sigma(f_1-f_2; \Lambda_{\tau_1\tau_2} ) {f'_2}^{3/4}, \\ 
	\label{eq:G_rescale}
	\widetilde G_{\tau_1\tau_2} &=& G(r_{\tau_1 \tau_2} ; \Lambda) \equiv 
	{f'_1}^{1/4}  G(f_1-f_2; \Lambda_{\tau_1\tau_2} ) {f'_2}^{1/4},
\end{eqnarray}
as generalizations of the scaling transformations Eqs.~\eqref{eq:GS_b}. The
correspondence between these definitions follows from the identifications  $t
\to f_1 - f_2$ and $\lambda \to \sqrt{f_1' f_2'}$ together with $\Lambda' \to
\Lambda$. We note that the fields $\widetilde G$ and
$\widetilde \Sigma$ remain invariant under the restricted class of
$\textrm{SL}(2,\mathds{R})$ transformations Eqs.~\eqref{eq:SL2RTransformation}. 

It remains to be shown that the reparameterized fields are approximate
solutions of the stationary phase equations Eq.~\eqref{eq:MF}, provided
$(G,\Sigma)$ are. With Eqs.~\eqref{eq:GS_b} and \eqref{eq:S_rescale}, the second
of these,   $ \widetilde \Sigma = (J^2/k_0^3) \widetilde G^3$, is  manifestly
satisfied. However, the first requires substantially more discussion. In the
following, we analyze this equation to  which the pair
$(\widetilde{G},\widetilde{\Sigma})$ is an approximate solution. Our discussion
will also introduce various concepts and definitions which will play a key role
in our subsequent derivation of the fluctuation action. 

We begin by
noting that the self energy $\widetilde \Sigma^{xx}\propto (\widetilde G^{xx})^3$ is
spatially local, and translationally invariant on average. It therefore does not
exhibit coordinate dependence, and we denote it by $\widetilde \Sigma$ throughout.
Turning to the more involved first equation in \eqref{eq:Eqs_MF}, consider a formal solution $\widetilde{\cal{G}}_{\tau_1\tau_2}^{x_1
x_2}$ evaluated for the self energy $\widetilde \Sigma$. Turning to Fourier space, it reads 
\begin{equation}
\label{eq:1st_DE}
- \int d\tau_2 \left( \epsilon_k\delta_{\tau_1\tau_2} + \widetilde\Sigma_{\tau_1\tau_2}\right) 
\widetilde{\cal G}_{\tau_2\tau_3}(k) = \delta_{\tau_1\tau_3}.
\end{equation}

\subsubsection{Linear transformation in time space}
In the following, it will be helpful to think of Eq.~\eqref{eq:1st_DE} as a
`matrix equation', and to consider the transformation $t\to \tau=F(t)$ as a
change of basis. Referring to Appendix~\ref{App:Rep_S} for the detailed formulation of
this picture in the language of linear algebra, we here note that  the
transformation between the two representations acts on general bilocal operators
$O_{\tau_1 \tau_2}$ with scaling dimensions $\Delta$ via a (non-unitary)
linear map ${\cal M}_\Delta$ as
\begin{equation}
\label{eq:Mapping_M}
  {\cal M}_\Delta:\, O_{\tau_1 \tau_2} \quad {\mapsto} \quad \overline{O}_{t_1 t_2} =  {F_1'}^{\Delta/2} O_{\tau_1 \tau_2} {F_2'}^{\Delta/2}, \quad \tau_i = F(t_i),
\end{equation}
where $F$ is the inverse of the map  $f$,   $F(f(\tau))=\tau$, and we have
abbreviated $F_i' = F'(t_i)$. We  aim to apply ${\cal
M}_\Delta$ in Eq.~(\ref{eq:1st_DE}), keeping in mind the scaling dimensions
$\Delta_\Sigma=3/2$ and  $\Delta_G=1/2$. As seen from
(\ref{eq:S_rescale}-\ref{eq:G_rescale}), this transformation effectively
eliminates the conformal factors ${f'}^{\Delta/2}$ and then changes the time
frame from $\tau$ to $t$. To conveniently describe its action, we define  the function,
\begin{equation}
	\label{eq:b_t-def}
	b_t \equiv 1/\sqrt{F'_t} \equiv \sqrt{f'_\tau}\,\Bigl|_{\tau=F(t)}\,,
\end{equation}
which will be used extensively throughout. Note that $F_t'>0$, making the square root well defined. 
It is straightforward to verify that the transformed representation of the  Dyson equation~(\ref{eq:1st_DE}) reads 
\begin{equation}
	\label{eq:1st_DE_bar}
	- \int d t_2 ( {\epsilon_k}\, b_{t_1}^{-1} \delta_{t_1 t_2} + 
	\overline\Sigma_{t_1 t_2})\, 
	\overline{\cal G}_{t_2 t_3}(k) = \delta_{t_1 t_3}, 
	\end{equation}
where the combination of Eqs.~(\ref{eq:S_rescale}) and
~(\ref{eq:Mapping_M}) implies
\begin{equation}
\label{eq:Sigma_bar}
\overline\Sigma_{t_1 t_2} = \Sigma(t_1-t_2; \Lambda_{t_1 t_2}), \qquad 
\Lambda_{t_1 t_2} = \frac{\!\!\Lambda}{(b_{t_1} b_{t_2})^{1/2}}.
\end{equation}
At this point, the rationale behind the  ${\cal M}$~-~transformation becomes
evident: to  leading order $\overline\Sigma_{t_1 t_2}$ coincides with the SYK
self-energy (\ref{eq:GS0_SYK_t}), while the dependence on $\Lambda_{t_1 t_2}$ is
a next-order effect due to the presence of a kinetic energy term $\epsilon_k$. 

\subsubsection{Wigner-Moyal expansion}
One can solve Eq.~\eqref{eq:1st_DE_bar} in an adiabatic approximation, assuming
the field $b_t$ exhibits  'slow' dependence on the relative time scale $t=t_1-t_2$.  This idea
can be formalized using Wigner symbols and the Moyal expansion,
which are widely employed to represent similarly structured equations in
many-body physics~\cite{Rammer:1986, Kamenev:2011}. We define the
Wigner transform $O_{\epsilon}(s)$ of the operator $\overline{O}_{t_1 t_2}$ as
\begin{equation}
\label{eq:Wigner_function_E}
	\overline{O}_{\epsilon}(s) = \int dt\, e^{i\epsilon t} \overline{O}_{s+\frac t 2, \,s-\frac t 2},
\end{equation}
where $s=(t_1 + t_2)/2$ is the 'center of mass' time, while the energy $\epsilon$ is conjugate to  relative time $t$. 
For the self-energy~(\ref{eq:Sigma_bar}), we use the local approximation by substituting
$\Lambda_{t_1 t_2}\to\Lambda/b_s$. This translates into its Wigner symbol 
$\overline{\Sigma}_{\epsilon}(s) \simeq \Sigma(\epsilon; \Lambda/b_s)$, which is
a conventional Fourier transform of the mean-field self-energy $\Sigma(t; \Lambda)$ over time.
Relying on the above approximation, we then introduce the Wigner symbol of an effective Hamiltonian,
\begin{equation}
	\label{eq:h_Wigner_Et}
	h_{\epsilon, k}(s) \equiv {\epsilon_k}/{b_{s}} + \Sigma(\epsilon; \Lambda/b_s),
\end{equation} 
such that the Dyson equation reads: $- 	h_{\epsilon, k}(s) \star \overline{\cal
G}_{\epsilon, k}(s) = 1$, where  '$\star$' denotes the Moyal product\footnote{ If $A_\epsilon(s)$ and $B_\epsilon(s)$ are Wigner
symbols of  $\hat A_{t_1 t_2}$ and $\hat B_{t_1 t_2}$, the Moyal product $A_\epsilon(s)\star B_\epsilon(s)$ is the Wigner symbol of $(A
B)_{t_1 t_2}$. It affords the semiclassical expansion $(A \star
B)_\epsilon(s) = A_\epsilon(s) e^{-\tfrac{i\hbar}{2}(\overleftarrow{\partial_s}
\overrightarrow{\partial_\epsilon} - \overleftarrow{\partial_\epsilon}
\overrightarrow{\partial_s})} B_\epsilon(s) = A_\epsilon(s) B_\epsilon(s) -
\frac{i\hbar}{2} A_\epsilon(s) (\overleftarrow{\partial_s}
\overrightarrow{\partial_\epsilon} - \overleftarrow{\partial_\epsilon}
\overrightarrow{\partial_s}) B_\epsilon(s) + {\cal O}(\hbar^2)$.}. This latter
equation can be solved to leading Moyal  order, yielding
\begin{equation}
\label{eq:bar_G_h}
	 \overline{\cal G}_{\epsilon, k}(s)  = - 1/h_{\epsilon, k}(s) + {\cal O}(\hbar^2),
\end{equation}
where the ${\cal O}(\hbar^2)$--term symbolically denotes possible 2nd order gradient
    corrections\footnote{ Indeed, if $G_\epsilon(s)= -1/h_\epsilon(s)$, then $
    (G\star h)_\epsilon(s) = -1 - \frac {i\hbar}{2} (\partial_s G
    \,\partial_\epsilon h - \partial_\epsilon G\, \partial_s h) + {\cal
    O}(\hbar^2) = -1 + {\cal O}(\hbar^2)$. } of order $b'^2$ and $b b''$, see
    subsection~\ref{App:Gradient_correctiom_Moyal} below for the detailed discussion of this point. We also note
    that the rescaled kinetic energy in the effective Hamiltonian can be written
    as
\begin{equation}
	 {\epsilon_k}/{b_{s}} = (\Lambda/b_s)(k/k_1)^\gamma =  {\epsilon_k}\bigl|_{\Lambda \to \Lambda/b_s}.
\end{equation}
As a result, the propagator at coinciding spatial points  affords the approximate representation
\begin{equation}
	\label{eq:CalG_t1t2_xx}
	\overline{\cal G}^{xx}_{t_1 t_2} = \int_{\epsilon,k} \overline{\cal G}_{\epsilon, k}(s)e^{- i \epsilon (t_1-t_2)} \simeq
	 - \int_{\epsilon,k} \frac{e^{- i \epsilon (t_1-t_2)} }{ {\epsilon_k}/{b_{s}} + \Sigma(\epsilon; \Lambda/b_s)} = G(t_1-t_2; \Lambda/b_s),
\end{equation}
where $s=\frac{t_1+t_2}{2}$ as before, and we have abbreviated
$\int_{\epsilon,k} \equiv \int dk d\epsilon/(2\pi)^2$ for all subsequent
formulae. We finally transform the rescaled kinetic energy $\Lambda/b_s$ back to its
exact value $\Lambda_{t_1 t_2}$ (the accuracy of this substitution is again of
order $b'^2$), then apply the inverse mapping ${\cal M}^{-1}_{1/2}$, cf.
definition (\ref{eq:Mapping_M}), to both sides of Eq.~(\ref{eq:CalG_t1t2_xx}), and use the scaling identities~\eqref{eq:G_rescale} for the
mean-field Green's function $G(t; \Lambda)$ to arrive at the result
\begin{equation}
	\label{eq:TildeGSolution}
	{\cal G}^{xx}_{\tau_1 \tau_2} = \widetilde G_{\tau_1 \tau_2} + {\cal O}(\hbar^2),
\end{equation}
i.e. an equation stating an identification of the solution to the mean field
equations with the reparameterized Green function $\widetilde G$ in the
semiclassical limit of slowly varying reparameterization transformations.
Corrections, symbolically indicated as ${\cal O}(\hbar^2)$, involve higher order
derivative acting on $f$, which are ignored in the present analysis (but
carefully analyzed in Subsection~\ref{App:Gradient_correctiom_Moyal}). 

To summarize, we have  defined the reparameterized saddle-point $(\widetilde G,
\widetilde \Sigma)$ based on the approximate stationary solution of the
$G\Sigma$-functional, excluding the $\partial_\tau$ operator. The $f$-dependent
fields solve the second equation in \eqref{eq:Eqs_MF} exactly, and the first
approximately. These fields are solutions on the full coset manifold ${\rm
Diff}(S^1)/{\rm SL}(2,\mathds{R})$ of reparameterizations provided the latter
fluctuate slowly. In this way, we have identified reparametrizations $f(\tau,x)$
as the relevant  soft modes of the chiral SYK model. 

\subsection{Gradient expansion}
\label{sec:gradients_GSigma}
In this section, we will expand the $G\Sigma$-action in slow spatial and
temporal fluctuations of the reparametrizations $f(\tau,x)$ to derive the
Alekseev-Shatashvilli action~\eqref{eq:S_graviton}. In the process,  we will
determine the coupling constants $C$ and $u$ of the theory to leading
logarithmic accuracy. Before embarking on the actual derivation, let us
summarize our main results.

For  Model I with dispersion $\epsilon_k = v_0 k$, and at intermediate
temperatures, the two coupling constants evaluate to 
\begin{equation}
	\label{eq:Cu_high_T}
	C =  \left(\frac{3\pi}{4}\right) \frac{N \Lambda}{J}  \ln\frac{J}{T_\Lambda}, \qquad u = 2\Lambda/k_0, \qquad
	T_\Lambda \ll T \ll \sqrt{ J \Lambda},
\end{equation}
where our expansion in derivatives requires  $C \gg 1$, which in turn requires
$N \gg 1$. For lower temperatures Model I lacks a holographic correspondence to
AdS$_3$.  For  Model II, the coupling constants  in the temperature regime above
are the same, Eq.~(\ref{eq:Cu_high_T}). For lower temperatures, $C$ does not
change, while $u$ becomes temperature dependent as 
\begin{equation}
\label{eq:u_T}
	u(T) = \frac{u_0}{\ln(J/T)}, \qquad u_0 = \frac{2\Lambda}{k_0} \ln\frac{J}{T_\Lambda}, \qquad T \ll T_\Lambda.
\end{equation}
In the logarithmic  $T$-dependence manifests a temperature scaling, as described by the flow equation 
\begin{equation}
	\label{eq:u_low_T}
	\frac{d \ln u}{dl} = - \frac{u}{u_0}, \qquad  l = \ln(J/T),
\end{equation}
subject to the initial condition $u(T_\Lambda) = 2\Lambda/k_0$ at the boundary to the
intermediate temperature regime. The diminishing of the velocity $u$ at low
temperatures is a consequence of the vanishing group velocity, $v_k = \partial_k
\epsilon_k$, at low momenta.

To set the stage for our subsequent derivations, let us consider the AS action 
at zero temperature,
\begin{equation}
	\label{eq:Schw_1+1D_h}
	S^{\infty}_\pm [f] =  \frac{C}{24\pi} \int\limits_{-\infty}^{+\infty} d\tau \int\limits_0^{L} 
	d x   \frac{f''\partial_\pm f'}{ f'^2},
	\qquad \partial_\pm = \frac 12 (u^{-1}\partial_\tau \pm  i \partial_x), 
\end{equation} 
as the $\beta \to \infty$ limit of the finite temperature action $S_\pm \equiv
S^{\beta}_\pm$, Eq.~(\ref{eq:S_graviton}), where imaginary  time is
compactified to a circle. The finite temperature generalization is obtained from
Eq.~(\ref{eq:Schw_1+1D_h})  by a substitution $f(\tau,x) \to \tan( \pi
f(\tau,x)/\beta)$, i.e.
\begin{equation}
	\label{eq:Actions_Equivalence}
	S^\beta_\pm [f] = S^{\infty}_\pm [\tan( \pi f/\beta)],
\end{equation}
where a restriction $\tau\in[0,\beta]$ is implied on  both sides of the
relation. In the following, we will focus on the derivation of
(\ref{eq:Schw_1+1D_h}), keeping in mind that its finite temperature
generalization may be realized in a secondary step, by application of
Eq.~\eqref{eq:Actions_Equivalence}.  
Our   starting point is the  
$G\Sigma$-functional, 
\begin{equation}
	\label{eq:S_f}
	S[f] = -{N\over 2} \left[
	\mathrm{tr}\ln\left(\partial_\tau + \epsilon_{\hat k} + \widetilde\Sigma\right)  +
	\frac{J^2}{4 k_0^{3}}  \int d^2 \tau dx\, (\widetilde G^{\,\,xx}_{\tau_1\tau_2})^4 + {\rm Tr} ( \widetilde G\widetilde\Sigma ) \right],
\end{equation}
where the self-energy and Green's function are reparametrized according
to~(\ref{eq:S_rescale}) and (\ref{eq:G_rescale}), and  we have promoted
$f(\tau,x)$ to include space dependence. Importantly, the starting
action~\eqref{eq:S_f} is locally ${\rm SL}(2,\mathds{R})$--invariant. This is a
direct consequence of the invariance of the fields $\widetilde G$ and
$\widetilde\Sigma$ under the transformations~\eqref{eq:SL2RTransformation},
where the coefficients $a,\dots,d$ may depend on position, $a=a_x$, etc. (As
discussed in the foregoing section, the invariance of the fields follows from
the definitions~(\ref{eq:S_rescale}-\ref{eq:G_rescale}) indicating that
$\widetilde G$ and $\widetilde\Sigma$ are expressed through the  
${\rm SL}(2,\mathds{R})$--invariant ratio \eqref{eq:r12}.) Consequently, the
gradient expansion of action~\eqref{eq:S_f} must also preserve this symmetry,
which will turn out to be a highly nontrival consistency check. In fact, the
AS action will  emerge as the minimal local functional respecting this coset
symmetry.

We also note that the AS action~(\ref{eq:Schw_1+1D_h}) contains  real and
imaginary parts: $S^\infty_\pm[f] = S_R[f] \\\pm i S_I[f]$, where the real part is
the position dependent Schwarzian (up to a boundary term), while the imaginary
part is the kinetic term containing mixed spatial and time derivatives of
$f(\tau,x)$. In the following, it will be useful to consider these two terms employing the
field  $b$~\eqref{eq:b_t-def} as an independent variable. 
More precisely, we  generalize the definition~(\ref{eq:b_t-def}) to include 
spatial dependence~\footnote{In what follows an abridged notation is used:
$F_{t,x} \equiv F(t,x)$, and the same for all other functions of time and space,
e.g. $b_{t,x} \equiv b(t,x)$ and so on.}, 
\begin{equation}
	\label{eq:b_tx-def}
	b_{t,x} \overset{{\rm def.}}{=} 1/\sqrt{F'_{t,x}} \equiv \sqrt{f'_{\tau,x}}\,\Bigl|_{\tau=F(t,x)},
\end{equation}
where $f'=\partial_\tau f$, $F' = \partial_t F$ and the relation $t=f_x(F_x(t))$
is defined locally. On changing in Eq.~(\ref{eq:Schw_1+1D_h}) the time
integration variable as $t=f(\tau,x)$ at given $x$, it is straightforward to
verify that the real part becomes  Gaussian (the integration measure,
however, is not flat in terms of the field $b$, i.e. the complexity of the theory is now hiding there), while the
imaginary part assumes the form of a 'cubic' vertex:
\begin{equation}
	\label{eq:S_RI_b}
	S^{\rm AS}_\pm[b] = \frac{C}{12\pi u} 
	\int_{t,x} \!\!\! b'^2 \pm \frac{ i C}{24\pi} 
	\int_{t,x} \!\!\! \partial_x F \left( b'' b - b'^2\right), \quad b' \equiv \partial_t b.
\end{equation}
The equivalence between this representation and  Eq. (\ref{eq:Schw_1+1D_h})
follows from a  few integrations by parts in time, assuming that the boundaries
~$\pm\infty$ of the zero temperature time domain $t \in \mathds{R}$ are
identified. In the following, we will use the pair of fields $(b, \partial_x F)$
to carry out the gradient expansion, and to  derive the AS action  as in
Eq.~\eqref{eq:S_RI_b}. Below, we outline the main steps of this procedure,
relegating the majority of technical details to Appendices.

The starting
reparametrized $G\Sigma$--action~\eqref{eq:S_f} requires some care, as its first
term is expressed via a functional determinant of the operator ${\cal D} =
\partial_\tau + \epsilon_{\hat k} + \widetilde\Sigma$.  We have chosen to
regularize it as
\begin{equation}
	\label{eq:S_reg_f}
	S_{\rm reg}[f] = -\frac N 2 \ln {\rm det}\,{\cal D} + \frac N 2 \ln {\rm det}\, \widetilde\Sigma^0  -\frac N 2 I[\widetilde G,\widetilde \Sigma],
\end{equation}
where $\widetilde\Sigma^0$ is the reparametrized self-energy of the original SYK
solution~\eqref{eq:MeanFieldSingleGrain} and $I[\widetilde G,\widetilde \Sigma]$
adds the  remaining two pieces of the $G\Sigma$--action: 
\begin{equation}
	\label{eq:I_G_Sigma}
	I[G,\Sigma] = \frac{J^2}{4k_0^3} \int d^2 \tau dx\, [G^{\,\,xx}_{\tau_1\tau_2}]^4 + {\rm Tr} \,( G \Sigma).
\end{equation}
A closer inspection reveals that a regulator above contributes only an
inessential $f$-independent constant\footnote{ Indeed, the variation of the
action $S_0[f] := - (N/2)\ln {\rm det} \,{\widetilde \Sigma}^0 - (N/2)
I[{\widetilde G}^0, {\widetilde \Sigma}^0]$ over $f$ vanishes, if $({\widetilde
G}^0, {\widetilde \Sigma}^0)$ is the reparametrized SYK saddle-point. Therefore,
one finds that $S_0[f] = - (N/2) \ln\det \Sigma^0 - (N/2) I[G^0,\Sigma^0]$ is
some constant. The last piece here, $I[{\widetilde  G}^0,{\widetilde  \Sigma}^0]
\equiv I[G^0,\Sigma^0]$, is explicitly reparametrization invariant. Hence the
determinant is also invariant just by itself. }, namely ${\rm tr}\ln \widetilde\Sigma^0 = {\rm tr}\ln \Sigma^0$. To keep
the notation slim, we will suppress its presence in much of our discussion below.

To prepare the  gradient expansion, we will first shuffle the time reparametrizations from
the 'large' self-energy to the 'small' symmetry-breaking terms, $\partial_\tau$ and $\epsilon_{\hat k}$, in the operator ${\cal D}$.
To do so, we introduce a generalized variant of the transformation~\eqref{eq:Mapping_M}, 
\begin{equation}
	\label{eq:Mapping_M_x}
	{\cal M}_\Delta:\, O_{\tau_1 \tau_2}^{x_1 x_2} \quad {\mapsto} \quad 
	\overline{O}_{t_1 t_2}^{x_1 x_2} =  {F_1'}^{\Delta/2} O_{\tau_1 \tau_2}^{x_1 x_2} {F_2'}^{\Delta/2}, \quad \tau_i = F(t_i, x_i),
\end{equation}
defined to be applicable to spatially  non-local operators $O_{\tau_1
\tau_2}^{x_1 x_2}$ with scaling dimension $\Delta$ and diffeomorphisms $F$
carrying space dimension (in Eq.~\eqref{eq:Mapping_M_x}, $F'_i \equiv F'(t_i,
x_i)$, as usual). We apply this transformation to each term in the
action~\eqref{eq:S_reg_f} individually, with account for the respective  scaling dimensions
$\Delta_\Sigma=3/2$ and $\Delta_G = 1/2$, to obtain
\begin{equation}
    \label{eq:S_reg_bar}
    S_{\rm reg}[f] = -\frac N 2 \ln {\rm det}\,\overline {\cal D} - \frac N 2 I[\overline G,\overline \Sigma].
\end{equation}
In particular,  the action $I$ remains invariant, as follows from the change
of differentials, $dt_i = f'_i d\tau_i$ in Eq.~\eqref{eq:I_G_Sigma}. Here, the
self-energy $\overline \Sigma$ is given by Eq.~\eqref{eq:Sigma_bar}, with the
field $b(t,x)$ being position dependent, and an analogous expression applies to
the Green's function $\overline G$. At the same time, the differential operator
${\cal D}$ transforms with a scaling dimension $\Delta=3/2$, and after the
mapping~\eqref{eq:Mapping_M_x} it becomes
\begin{equation}
	\label{eq:bar_D_def}
	{\cal M}_{3/2}:\, {\cal D} \,\, \,{\mapsto} \,\,\, \overline{\cal D} = \rho + j + \overline{\Sigma},
\end{equation}
where $\rho$ and $j$ denote the transformed time derivative operator
$\partial_\tau$ and the kinetic energy $\epsilon_{\hat k}$, respectively. It
will become clear shortly, that the former two afford an interpretation as energy
density and heat current, respectively.

Referring to Appendix~\ref{App:Vertex_Operators} for details,  the  operator $\rho$ is diagonal in space, i.e. $\rho^{x x'} =
\delta^{xx'}\rho_x$, where $\rho_x\equiv \rho$ is a first order differential
operator in time,
\begin{equation}
	\label{eq:bar_dtau}
	\rho = \frac 12 \left( b \overrightarrow{\partial_t} - \overleftarrow{\partial_t} b \right),
\end{equation}
with parametric space dependence from $b=b(x,t)$. Its action on smooth functions $g_t$ and $h_t$ is defined by the matrix elements
\begin{equation}
	\label{eq:OtMatrixElements}
	\bra{g}\rho \ket{h}= \int dt\, b ( g(\partial_t h)-(\partial_t g) h).
\end{equation}

The second operator, $j$, requires a bit more work. Unlike $\rho$, it has a non-trivial structure 
in both position and time. To define it, we introduce its Wigner symbol in position-momentum space as
		$j_k(x) = \int dy e^{ -i k y } j^{x + \frac y 2, x - \frac y 2}$,
cf.~Eq.~\eqref{eq:Wigner_function_E}.
In Appendix~\ref{App:Vertex_Operators} we show that the  lowest order Moyal expansion, of this Wigner symbol assumes the form
\begin{equation}
	\label{eq:j_k_x}
	j_k(x) = \epsilon_k/b + \frac i 2 \partial_k\epsilon_k \times \bigl(\ b \partial_x F \overrightarrow{\partial_t}  - \overleftarrow{\partial_t} b \partial_x F \bigr)
	\equiv j^{0}_k(x) + j^{1}_k(x),
\end{equation}
where $b=b(t,x)$ as above. We have also split $j_k(x)$ in two parts, containing
zero and one derivative operator, respectively. Similarly to Eq.~\eqref{eq:OtMatrixElements} above, the
matrix element of the latter is defined as
\begin{equation}
	\bra{g} j^{1}_k(x)\ket{h}= \frac i 2 \partial_k\epsilon_k \int dt\, \partial_x F  b ( g(\partial_t h)-(\partial_t g) h).
\end{equation}
Collecting terms, to lowest non-trivial order in gradients,  the  position-momentum Wigner
symbol of the operator~\eqref{eq:bar_D_def} is given by the
expansion:
\begin{equation} 
\label{eq:D_bar}
 \overline{\cal D}_k(x) = \rho + j^{1}_k(x) + h_k(x), \qquad 
 h_k(x) = \epsilon_k/b + \overline{\Sigma},	
\end{equation}
where $h_k(x)$ is the effective Hamiltonian introduced previously in Eq.~\eqref{eq:h_Wigner_Et}.

With the result~\eqref{eq:D_bar} at hand, we are now in a position to organize the gradient expansion of
the action~\eqref{eq:S_reg_bar} around the approximate saddle-point solution. For that purpose, we define
the propagator $\overline{\cal G}$ with a corresponding position-momentum Wigner symbol $\overline{\cal G}_k(x)$,
which obeys the Dyson equation $h_k(x) \star \overline{\cal G}_k(x) = -1$. This is a generalization
of our previous definition~\eqref{eq:1st_DE_bar} to the case of spatially inhomogeneous 
reparametrizations $f=f(\tau,x)$. One then rewrites the regularized action~\eqref{eq:S_reg_bar} in an equivalent form
by splitting it into the fluctuation contribution, 
\begin{equation}
\label{eq:S_fl_f}
    S_{\rm fl}[f] = - \frac N 2 {\rm tr}\,\ln\left(1 - (\rho + j^{1}) \overline{\cal G}\right) 
\end{equation}
and  one originating from the approximate saddle-point,
\begin{equation}
\label{eq:S_star_f}
    S_*[\overline G,\overline \Sigma] = -\frac N 2 {\rm tr}\,\ln(j^0+\overline\Sigma) -  \frac N 2 I[\overline G,\overline \Sigma],
\end{equation}
so that the sum of these two actions gives back $S_{\rm reg}[f]$. In 
Appendix~\ref{App:Rep_invariance_S_star} we verify that the second piece, $S_*[\overline
G,\overline \Sigma]$, does not  contribute to the AS
action~\eqref{eq:S_RI_b}. This  follows from the fact that $(\overline
G,\overline \Sigma)$  defines a  manifold of approximate
saddle-point solutions of the functional $S_*[G,\Sigma]$, 
parametrized by $f$. On the other hand,  the gradient expansion of the
first piece, Eq.~(\ref{eq:S_fl_f}), describes the effects of
fluctuations beyond the mean-field approximation.

The second-order expansion in $\rho$ and $j^1$ generates several terms, where those  contributing to the AS action~\eqref{eq:S_RI_b} are given by 
\begin{equation}
  \label{eq:S_fl_expansion}
    S_{\rm fl}[f] \to \frac{N}{2}{\rm tr}(\rho \overline{\cal G}) + \frac N 2 {\rm tr}(j^{1} \overline{\cal G} \rho \overline{\cal G}) + 
    \frac N 4 {\rm tr}(\rho \overline{\cal G} \rho \overline{\cal G}) \equiv S_{\rho}[f] + S_{j\rho}[f] + S_{\rho\rho}[f].
\end{equation}
Referring to Appendix~\ref{app:Gradient expansion} for details, we note that the
sum of the two terms, $S_{\rho}[f]+S_{j\rho}[f]$, defines  the imaginary part of
the AS action $S_\pm^{\rm AS}$, expressed as a functional of the pair of fields $(b, \partial_x
F)$. In particular, these two terms determine the central
charge~\eqref{eq:Cu_high_T}. The remaining term, $S_{\rho\rho}[f]$, provides the
real part, and it assumes  the form of the Schwarzian
action when expressed in terms of  $f(\tau,x)$.  
It sets a ratio $C/u$ and gives the scale of velocity~(\ref{eq:u_T}). 

\begin{figure}
\centering
	\includegraphics[width=0.7\linewidth]{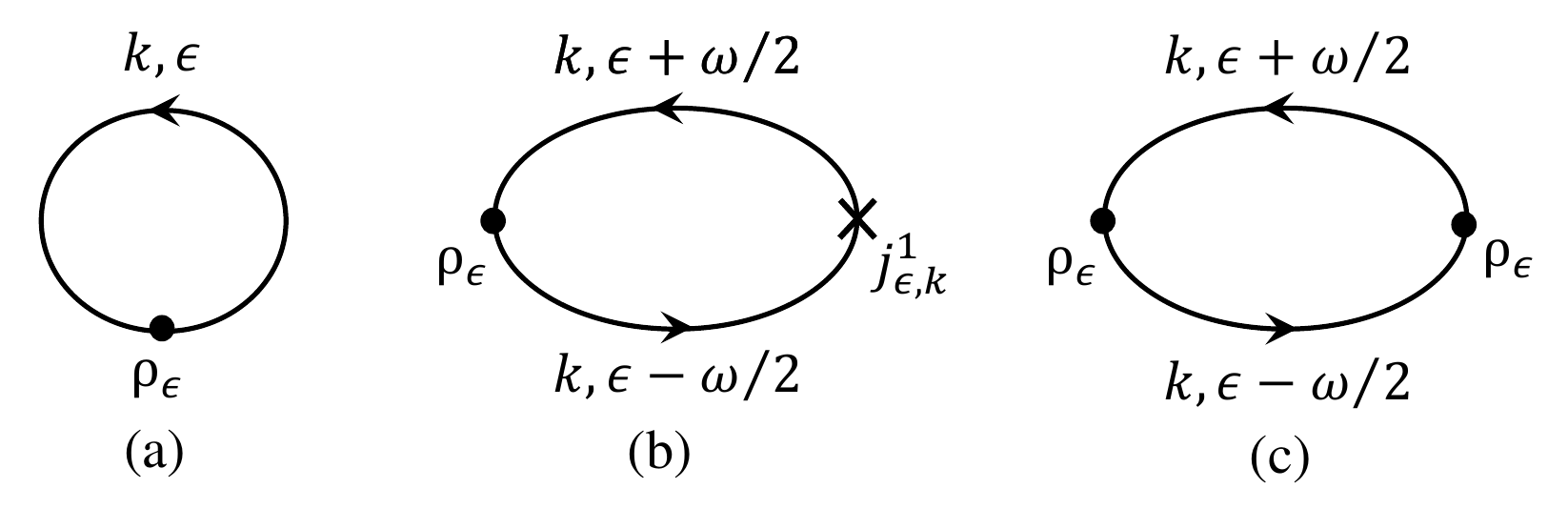} 
	\caption{Diagrams of the 2nd order gradient expansion contributing to the AS action.}
    \label{fig:Diagrams_1sr_2nd_order}
\end{figure}

The physical meaning of the terms contributing to the gradient expansion~\eqref{eq:S_fl_expansion} 
can be interpreted as follows. 
To leading order, the operators $\rho$ and  $j^1$ have the Wigner symbols 
\begin{equation}
\label{eq:Wigner_rho_j}
  \rho_{\epsilon,k} = - i b \epsilon + \dots,   \qquad
  j^1_{\epsilon,k} = (b \partial_x F) \cdot \epsilon  \partial_k \epsilon_k + \dots,
\end{equation}
identifying them as  energy density and heat current operators, the latter 
being proportional to the group velocity $v_k = \partial_k \epsilon_k$. Accordingly, the individual
 terms in the expansion~\eqref{eq:S_fl_expansion} describe correlation
functions of $\rho_{\epsilon,k}$ and $j^1_{\epsilon,k}$, whose Feynman diagrams are
shown in Fig.~\ref{fig:Diagrams_1sr_2nd_order}. A low frequency expansion of these
correlation functions in $\omega \ll \epsilon$ up to ${\cal O}(\omega^2)$ yields  
terms of the type $b'^2$ and $bb''$ in the action $S^{\rm AS}_\pm[b]$, see
Eq.~\eqref{eq:S_RI_b}. At the same time, the  integration region over the energy
$\epsilon$ is  wide, $T_\Lambda < |\epsilon| < J$, leading to the appearance of
$\log$-prefactors in the coupling constants $C$ and $u$. 
In this reading,  the
Schwarzian action is a density-density correlator, see
Fig.~\ref{fig:Diagrams_1sr_2nd_order}(c). Likewise, the central charge
and the kinematic part of the AS action are  defined by a
current-density correlator, Fig.~\ref{fig:Diagrams_1sr_2nd_order}(b).  

For completeness, we mention that the gradient expansion contains two extra
terms, not contributing to the AS action. These are a  first-order term,
$S_{j}[f]$, which reduces to a constant  when subjected to our regularization
scheme with a reparametrization dependent cut-off. There also is a  contribution
$S_{jj}[f]$ containing two spatial derivatives $\propto (\partial_x F)^2$. The
latter does not comply with ${\rm SL}(2,\mathds{R})$ invariance, and we
attribute it to the Jacobian of the transformation from initial to rotated
Majoranas under the mapping~\eqref{eq:Mapping_M_x}. Finally, for Model II 
the quantum fluctuations of $f(x,\tau)$ may compromise the SYK saddle point
at temperature scale $T_* \ll T_\Lambda$, which is defined by the competition
of the small parameter $k_1/k_0 \ll 1$ and the large ratio $J/T_\Lambda$,
see Eq.~\eqref{eq:T_star} for the explicit definition.
The latter limits a holographic duality of Model II to the range of temperatures $T_* < T < J$.
For further details we again refer to Appendix~\ref{app:Gradient expansion}.

To conclude this section, we note that the chiral SYK Hamiltonian~\eqref{eq:H_SYK4_D2}, which features right-moving Majorana fermions with $\partial_k \epsilon_k \geq 0$, gives rise to a single copy of the AS action, namely to $S_-[f]$. However, if the underlying topological insulator has either a Hall bar or Corbino geometry, the opposite boundaries will host Majorana edge states with opposite chiralities. In this setting, two independent copies of the AS action, $S_\pm[f]$, emerge, which is a scenario that takes place in a holographic dual of this model discussed in the next section.

\section{The AS action from gravity} 
\label{sec:grav}

In this section, we consider the theory describing fluctuations around the Euclidean BTZ black hole and provide the direct reduction of this theory to the AS action. This computation gets us in a position to match coupling constants as well as time reparametrizations emerging in the boundary and bulk framework, respectively. We begin with quick reviews of the phase space of  AdS$_3$ gravity in subsection \ref{sec:AdS3} and the Chern-Simons formulation of three-dimensional gravity in \ref{sec:GravityCS}. Expert readers can skip these subsections and directly turn to \ref{sec:CSAction}. There, we discuss the Euclidean BTZ black hole solution in the Chern-Simons formulation for a particular choice of analytic continuation. Finally, subsection \ref{sec:AS} derives the above-mentioned link to AS theory by quantizing gravity on top of the Euclidean BTZ background. 

\subsection{The phase space of asymptotically AdS$_3$ gravity}
\label{sec:AdS3}

To understand the role of the AS action in the context of asymptotically AdS$_3$ spacetimes, we start by reviewing some properties of three-dimensional pure gravity in the metric formalism. The most important difference to  gravity in higher dimensions is the topological nature of the three-dimensional theory, meaning that it has no local degrees of freedom \cite{Carlip:1998uc}. For example, mass sources spacetime curvature locally, but deformations of the metric away from the source are pure
gauge. This, however,  does  not mean that the theory without matter is trivial; its solutions can have non-trivial global properties like horizons or  mass and angular momentum, as we will briefly discuss.   

The Einstein-Hilbert action for three-dimensional pure gravity 
with negative cosmological constant $\Lambda<0$ is given by 
\begin{align}
	S=\frac{1}{16 \pi G} \int_\mathcal{M} \dd^3{x} \sqrt{-g}\qty(R-2\Lambda) \label{eq:SEH}. 
\end{align} 
It describes what we will refer to as ``AdS$_3$ gravity''. 
Here, $G$ is Newton's constant, and $g_{\mu \nu}$ is the metric on the three-manifold $\Mc$ with determinant $g$ and associated curvature scalar $R$.  
The variation of this action yields field equations that express constant negative curvature, $R_\mn = 2 \Lambda g_\mn$ or $R = -6/l^2$, where in the latter equation we introduced the AdS curvature radius $l$ by writing the cosmological constant of dimension mass squared as $\Lambda = -1/l^2$. 
As a consequence, the spacetime `looks the same everywhere', a high level of
symmetry manifesting itself in a $\textrm{SO}(2,2)$-isometry: all solutions are locally AdS$_3$. 
This fact in turn implies that all solutions with an
inequivalent causal structure must be obtainable from a `basic solution' via
identifications under the action of a discrete group~$\Gamma$ \cite{Banados1993,
	Carlip:1998uc}. This basic solution is given by global AdS$_3$ with the metric
\begin{align}
	\dd{s}^2=-\qty(1+\frac{r^2}{l^2})\dd{t}^2+\frac{1}{(1+\frac{r^2}{l^2})}\dd{r}^2+r^2 \dd{\phi}^2, \label{eq:Global}
\end{align}
where the coordinate ranges are $t\in \Rb$, $\phi \in (0,2\pi)$ and $r>0$.

Three-dimensional holography is generically considered subject to Dirichlet
boundary conditions. These are specified in terms of an induced metric on the asymptotic boundary $\Mc_{\infty}$ at spacelike infinity $r \to \infty$.
In this limit, the metric should reduce to
\begin{align}
	\dd{s}^2_\infty=\frac{l^2}{r^2} \dd{r}^2 + \frac{r^2}{l^2} \, dx^+ dx^-, 
    \label{eq:GlobalConformal}
\end{align}
where $dx^+ dx^-$ describes the flat two-dimensional metric, in light-cone coordinates $x^\pm = l \phi \pm t$, of the conformal boundary.

The problem famously allows for a black hole solution, which is called the BTZ black hole \cite{Banados1992}. It is labeled by two charges, namely its mass $M$ and angular momentum $J$ (with $\abs{J}<Ml$ to avoid naked singularities). The non-rotating BTZ solution is 
\begin{align}
	\dd{s}^2=-\frac{\xi(r)^2}{l^2}\dd{t}^2+\frac{l^2}{\xi(r)^2}\dd{r}^2+r^2 \dd{\phi}^2,\qquad \xi(r)\equiv \sqrt{r^2-r_+^2}, \label{eq:ds-btz}
\end{align}
with mass $M=r_+^2/(8Gl^2)$ a function of the location $r_+$ of the horizon, and coordinate ranges $t \in \mathbb R$, $\phi \in (0,2\pi)$ and $r>r_+$.   
This geometry will play an important role throughout. The surface area of the
horizon, $2 \pi r_+$, and the periodicity $\beta = 2\pi
l^2/r_+$ of Euclidean 
time $\tau = i t$ determine thermodynamic properties of the black hole: the 
Bekenstein-Hawking entropy $S_{BH} = 2 \pi r_+/(4G)$ and the Hawking temperature
$T = 1/\beta$.

The most general solution satisfying the Dirichlet boundary conditions is the Banados solution \cite{Banados1999} 
\ali{
     ds^2 = \frac{l^2}{r^2} dr^2 + \frac{r^2}{l^2} \left( dx^+ + \frac{l^2}{r^2} L(x^-) dx^- \right) \left( dx^- + \frac{l^2}{r^2} \bar L(x^+) dx^+ \right).
}
It is characterized by functions $L$ and $\bar L$, which quantify the gravitational energy of the solution
(e.g.~the BTZ black hole mass $M = (L + \bar L)/(4G)$ and angular momentum $J = (L-\bar L)/(4G)$ for constant horizon-dependent values of $L$ and $\bar L$ in that case). These geometries are referred to as `asymptotically AdS'. 

There is a class of diffeomorphisms $\xi$ which 
reach the boundary and change the near-boundary $L$ and $\bar L$, but only in such a way that they obey the prescribed `asymptotically AdS' fall-off behavior of the metric components $g_\mn(r)$ as $r \ra \infty$ 
(namely $L \ra L'$ and $\bar L \ra \bar L'$ but the solution remains of the Banados form, see e.g.~\cite{Roberts:2012aq}). 
Such diffeomorphisms become physical and obey the algebra of an asymptotic symmetry group. In this case, the Virasoro algebra with Brown-Henneaux central charge $C = 3l/2G$ \cite{Brown1986}. Indeed the Brown-Henneaux 
diffeomorphisms $\xi$ reduce on the boundary to conformal transformations of the light-cone coordinates $x^\pm \ra f_\pm(x^\pm)$. 
This is the famous Brown-Henneaux result that identifies the role of two-dimensional conformal symmetry in (classical) AdS$_3$ gravity, interpreted today as a precursor of AdS/CFT. 

Once a global structure of  spacetime (e.g.~global AdS or BTZ)   
has been fixed, the perturbations around it are asymptotically realized degrees of freedom referred to as `boundary gravitons'.

The situation parallels that realized in topological field theory: While
topological field theory on a space without boundary is gauge invariant, this
feature is lost in the presence of a boundary $\Mc_\infty$. Gauge
transformations which do not reduce to the identity at the boundary are governed
by an effective boundary action  on this surface, which upon quantization
defines an infinite-dimensional Hilbert space. In the gravitational context, the
role of the latter is taken by the AS action \cite{Alekseev1989, Alekseev1990}
\cite{Cotler2018,Cotler:2020ugk,Henneaux2020,Mertens2022}, whose derivation we proceed to discuss in the following sections.  
For this, we make use of the CS gauge theory description of AdS$_3$ gravity,
which makes the above analogy quite explicit. 

\subsection{Three-dimensional gravity as Chern-Simons theory}
\label{sec:GravityCS}

The Einstein-Hilbert action has long been known to have an equivalent Chern-Simons description \cite{Witten1988,ATownsend}. 
We present a short review of this equivalence for \eqref{eq:SEH} 
in order to set notation, and refer the interested reader to the standard review \cite{Donnay2016} for more details. 

First, one rewrites the Einstein-Hilbert action \eqref{eq:SEH} in a first order form, meaning in terms of degrees of freedom that appear in the action with first order derivatives rather than second order ones (as for the metric 
in the metric formulation). 
These degrees of freedom are the frame field or vielbein, in this case dreibein, $e^a$ and spin connection $\omega^a_{\phantom{a}b}$, with Latin letters for the frame index or `internal index' $a=0,1,2$. 
The vielbein, sometimes referred to as the `square root of the metric', is the object that expresses the existence of a local inertial frame $\eta_{ab}$ in each point of the manifold, $ds^2 = \eta_{ab} e^a e^b$. 
It is a one-form $e^a = e^a_\mu dx^\mu$, whose coefficients $e^a_\mu$ allow transformation between spacetime indices $\mu$ and frame indices $a$, 
\ali{
	g_{\mu \nu}=e_{\mu}^a e_{\nu}^b \eta_{ab}, \qquad \eta_{ab}=g_{\mu \nu} e_{a}^\mu e_{b}^\nu=\operatorname{diag}(-1,1,1).  \label{eq:metric} 
} 
The spin connection $\omega^a_{\phantom{a}b}$ is a one-form $\omega^a_{\phantom{a}b} = \omega^a_{\phantom{a}b \mu} dx^\mu$, 
with $\omega^a_{\phantom{a}b \mu}$ taking over the role of the Christoffel symbols in the metric formulation, 
such as in the definition of a covariant derivative. 
In terms of these variables, the Einstein-Hilbert action \eqref{eq:SEH}  becomes 
\ali{
	S=-\frac{1}{16 \pi G} \int \epsilon_{abc} e^a \wedge\left(d \omega^{bc}+ \omega^b_{\phantom{a}d} \wedge \omega^{dc}+\frac{1}{3 l^2}  e^b \wedge e^c\right).  
	\label{eq:first-order}
}

Variation with respect to $e$ and $\omega$ impose respectively constant curvature and vanishing torsion, as an equivalent formulation of Einstein's field equations. 

The first order action \eqref{eq:first-order} can be further rewritten as a CS action for the group $\mathrm{SO}(2,2)$, by further combining $e^a$ and $\omega^a_{\phantom{a}b}$ into a CS gauge field valued in $\mathrm{SO}(2,2)$ (and identifying $l/4G$ with the CS level $k$). Its invariance, up to boundary terms, under $\mathrm{SO}(2,2)$ 
gauge transformations 
reflects the local isometry of the locally AdS$_3$ solutions of pure AdS$_3$ gravity discussed in the last section. 
The isomorphism $\mathrm{SO}(2,2) \simeq (\SL_L \times \SL_R) / \mathbb{Z}_2$
finally allows to introduce instead 
two $\mathfrak{sl}(2,\mathbb{R})$ valued CS gauge fields $A=A^a j_a$ and $\bar{A}=\bar{A}^a \bar{j}_a$, with the coefficients $A^a$ and $\bar{A}^a$ related to the gravitational degrees of freedom by 
\ali{
     A^a=\frac{1}{2}\epsilon^{abc}\omega_{bc}+\frac{e^a}{l}, \qquad  \bar{A}^a=\frac{1}{2}\epsilon^{abc}\omega_{bc}-\frac{e^a}{l} .  \label{eq:Aofeomega} 
}
They are real and independent. 
The $\mathfrak{sl}(2,\mathbb{R})$ generators $j_a$ and $\bar j_a$ satisfy  
\ali{
	 \operatorname{tr}(j_a j_b)=\frac{1}{2}\eta_{ab}, \qquad   \quad [j_a,j_b]=\epsilon_{abc} \eta^{cd} j_d \label{eq:j-commutation} 
} 
with $\epsilon_{012}=1$, and similarly for the barred sector. 
In terms of these variables, the action \eqref{eq:first-order} takes the form of 
the difference of Chern-Simons actions for $A$ and $\bar{A}$: 
\ali{
	S = S_{CS}[A]-S_{CS}[\bar{A}], \qquad S_{CS}[A]=-\frac{k}{4\pi}\int_\mathcal{M} \tr(A \wedge \dd{A} + \frac{2}{3} A \wedge A \wedge A ) \label{eq:SCS-grav} 
}
with the level given by $k = l/(4G)$. 
The equations of motion now read $F=\dd{A}+A \wedge A=0$ (and similarly for $\bar{F}$), which together impose constant curvature and vanishing torsion. 
Essential to the gauge theory interpretation of three-dimensional gravity is the identification of the frame indices with CS Lie algebra indices in \eqref{eq:Aofeomega}. 
The CS gauge fields as such are one-forms $A^a = A^a_\mu dx^\mu$ and $\bar{A}^a = \bar{A}^a_\mu dx^\mu$ (with $\mu = t, \phi, r$). 

At the quantum level, the considered CS theory is $Z = \int
\mathcal D (A, \bar{A}) e^{i \qty(S_{CS}[A]-S_{CS}[\bar{A}])}$ (where the path integration is over the appropriate quotiented CS gauge group). 
Of course, the above argument at the level of the action provides us with an equivalence 
of the classical theories, and not of the full quantum theories. 
For the purposes in this paper, it will be sufficient that the equivalence holds for the path integral in the neighborhood of classical saddles. 

\subsection{The BTZ black hole saddle of Chern-Simons theory} \label{sec:CSAction}

The reviewed CS description \eqref{eq:SCS-grav} of three-dimensional gravity will allow us to establish the connection to the AS action. Since the boundary theory is developed in a finite temperature framework, we first analytically continue to Euclidean signature. In this section we discuss the analytic continuation, and the Euclidean BTZ solution for the resulting action. 

We analytically continue to Euclidean time $\tau$ by setting $t = -i \tau$ as well as $A_t = i A_\tau$ and $\bar{A}_t = i \bar{A}_\tau$. Since 
\ali{ 
     A_t dt = A_\tau d\tau,  \label{eq:cont}
} 
the action $S$ in form notation \eqref{eq:SCS-grav} remains the same under this procedure. That is with the understanding that $A^a = A^a_\mu dx^\mu$ and $\bar{A}^a = \bar{A}^a_\mu dx^\mu$ are now one-forms on a Euclidean manifold $\mathcal M$, with the spacetime index taking values $\mu = \tau, \phi, r$.

From the gravity perspective, 
the fact that $S$ does not pick up a factor of $i$ means that under this continuation we continue to integrate over Lorentzian `target space' metrics $g_\mn = \eta_{ab} e^a_\mu e^b_\nu$ (with an $iS$ in the exponent) rather than over 
Euclidean `target space' metrics $g_\mn = \delta_{ab} e^a_\mu e^b_\nu$ (with typically $(-S_E)$ in the exponent). As such, the above analytic continuation is different from the one discussed more traditionally in the context of Euclidean quantum gravity \cite{Witten:1989ip}. What is crucial here is that our CS gauge field $A_\mu^a$ has two types of indices: a `worldsheet' spacetime index $\mu$ and a `target space' index $a$ valued in the Lie algebra. 
In \cite{Witten:1989ip}, and as detailed nicely in the appendix of \cite{Bunster2014}, one not only analytically continues $\mu$ from $t$ to $\tau$ in the same way as discussed above, but also continues the index $a$ 
from being $\mathfrak{sl}(2,\mathbb{R})$ to $\mathfrak{sl}(2,\mathbb{C})$ valued. The latter symmetry reflects the local isometry of locally AdS$_3$ solutions of pure AdS$_3$ gravity in Euclidean signature. 
However, from a condensed matter theory perspective, it is very unnatural to change the group upon analytically continuing the theory. So we proceed with the continuation as outlined in the previous paragraph, 
with $j_a$ and $\bar j_a$ being still $\mathfrak{sl}(2,\mathbb{R})$ generators. 
Note that this is the same analytic continuation as the one used in \cite{Cotler:2020ugk}  based on similarity to JT/Schwarzian discussions.  
What we require is that the Euclidean BTZ metric can be obtained as a saddle of our continued CS theory. As we will shortly see this is indeed the case, by allowing a complex $e^a$ and thus complex $(A^a,\bar{A}^a)$ solution.

The Euclidean CS theory we employ is thus \eqref{eq:SCS-grav} with $\mathcal M$ a Euclidean manifold spanned by coordinates $(\tau,\phi,r)$. 
We take the three-manifold to be of the form $\Sigma \times S^1$ with $\Sigma$ a disk, 
\ali{
	\mathcal M = D \times S^1.  
}
This is in analogy to the seminal discussion of CS theory on $\Sigma \times \mathbb{R}^1$ (with different choices of $\Sigma$) 
in \cite{Elitzur89}, where the $\mathbb{R}^1$ direction naturally provides a 
time for a canonical formulation of the theory. In our case, $S^1$ similarly provides a Euclidean timelike direction for a canonical perspective. We take this direction to be labeled by $\phi$ (not $\tau$, crucially different from \cite{Cotler2018}). The remaining coordinates $(r,\tau)$ span the disk $D$. 
That is, the manifold on which the CS gauge field lives is a solid torus with $\tau \sim \tau + \beta$ labeling the contractible cycle with period $\beta$, and $\phi \sim \phi + 2 \pi$ the uncontractible cycle with period $2 \pi$. This worldsheet torus is then characterized by the modular parameter $\mop=\frac{i \beta}{2\pi l}$\footnote{Throughout, we use boldface notation $\mop$ for the modular parameter and leave $\te$ for imaginary time.}.  

As in the seminal work \cite{Elitzur89}, the CS action can be brought in a form where the gauge field coordinate $A_\phi$ in the direction away from $\Sigma$ acts as a Lagrange multiplier. To achieve this, we proceed to split the gauge field 
\ali{ 
    A=A_\phi \dd{\phi}+ \tilde{A}_i \dd{x^i}  \label{eq:angqu}
    } 
with $x^i=\{r, \tau\}$ and define $\Tilde{F}={F}_{ r \te} \dd{r} \wedge
\dd{\te}$\footnote{We define $\epsilon^{\tau \phi r}=1$.}. The action $S_{CS}[A]$ 
becomes  
\begin{align}
	I_{CS}[A]=-\frac{k}{2\pi}\int_\mathcal{M} \dd{\phi} \wedge \tr(-\frac{1}{2}\tilde{A}\wedge \partial_\phi \tilde{A}+ A_\phi \tilde{F}) \label{eq:SCS-ham} . 
\end{align}
Because a boundary term has been dropped to obtain this expression, we employ a new letter $I_{CS}$ for this action. It is equal to $S_{CS}$ up to a boundary term of the form $i A_\phi A_\tau$. 
The new form will allow a rewriting of the theory as effective WZW theory at the boundary. 

\begin{figure}[t]
	\centering
	\includegraphics[width=0.7\linewidth]{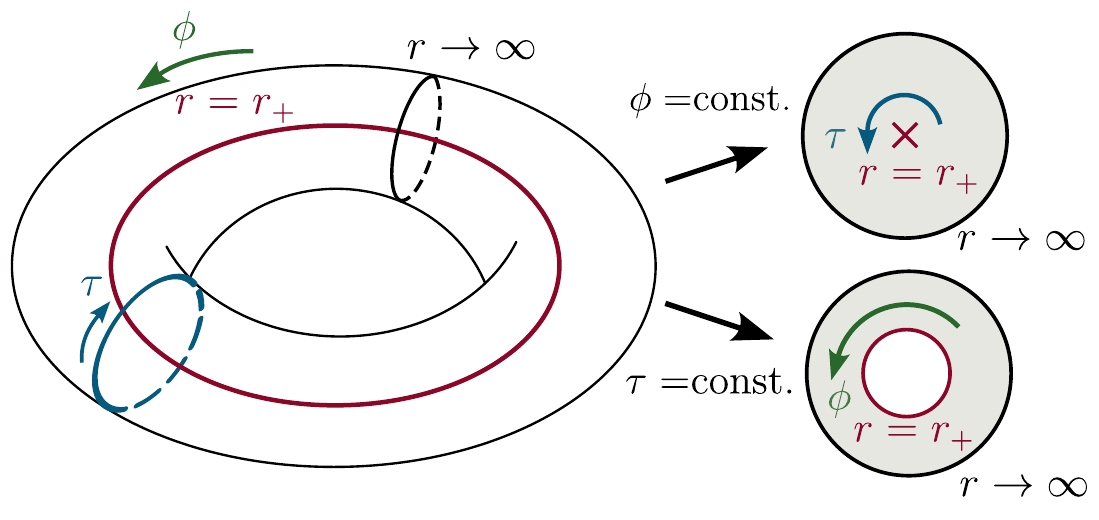}
	\caption{The topology of the Euclidean BTZ black hole \eqref{eq:ds-btz-Eucl} with horizon at $r = r_+$ is that of a solid torus with contractible $\tau$-cycle and uncontractible $\phi$-cycle. 
	}
	\label{fig:BTZ-torus}
\end{figure} 
The action $I_{CS}[A] - I_{CS}[\bar{A}]$ gives rise to a well-defined variational problem subject to chiral boundary conditions when it is supplemented by the boundary action 
\ali{
	I_{\mathcal{M}_\infty}[A,\bar{A}]= \frac{i k}{4\pi} \int_{\mathcal{M}_\infty} \dd^2{x} \qty( \operatorname{tr}({A}_{\te}^2)+\operatorname{tr}(\bar{A}_{\te}^2)). \label{eq:S-bdy}
	}
Indeed, the variation of the total action 
\begin{align}
	I=I_{CS}[A]-I_{CS}[\bar A]+I_{\mathcal{M}_\infty}[A,\bar A]   \label{eq:I-full}  
\end{align}
is given by 
\begin{align}
	\delta I=(\text{E.O.M})+ \frac{k}{2\pi}  \int_{\mathcal{M}_\infty}\dd^2{x} \qty(\delta A_{\te}(-A_\phi+i A_{\te})+\delta \bar{A}_{\te}(\bar{A}_\phi+i \bar{A}_{\te})),  
\end{align}
the boundary contribution of which vanishes for the boundary conditions $A_\phi=i A_{\te}$ and $\bar{A}_\phi=-i \bar{A}_{\te}$ at $\Mc_{\infty}$.  
These are the Euclidean versions of the chiral boundary conditions $A_\phi=A_t$ and $\bar{A}_\phi=-\bar{A}_t$ typically used in the CS treatment of gravity \cite{Donnay2016, Banados1998,Banados:2016zim}.

A particular solution of the variational problem with chiral boundary conditions is 
\begin{align} 
	A_{\,*}=\begin{pmatrix}
		\frac{\dd{r}}{2\xi(r)} &\frac{\dd{\wb}}{2l^2}\qty(r -\xi(r)) \\ 
		\frac{\dd{\wb}}{2l^2}\qty(r +\xi(r)) & -\frac{\dd{r}}{2 \xi(r)}
	\end{pmatrix}
	, \quad
	\bar{A}_{\,*}=\begin{pmatrix}
		-\frac{\dd{r}}{2\xi(r)} &-\frac{\dd{w}}{2l^2}\qty(r+ \xi(r)) \\ 
		-\frac{\dd{w}}{2l^2}\qty(r- \xi(r)) &\frac{\dd{r}}{2 \xi(r)}
	\end{pmatrix}, \label{eq:CSBTZ}	
\end{align}
where $\xi(r)$ is defined in Eq.~\eqref{eq:ds-btz} and we have defined light-cone coordinates and their Wick rotations
\begin{align}
	x^+=t+l\phi \mapsto -i \tau + l \phi= \wb \qquad x^-=-t+l\phi \mapsto i \tau + l \phi= w. \label{eq:hol-coord}
\end{align}
By \eqref{eq:Aofeomega}, the solution \eqref{eq:CSBTZ} can be straightforwardly checked to describe the (non-rotating) Euclidean BTZ metric 
\ali{
	\dd{s}^2=\frac{\xi(r)^2}{l^2}\dd{\tau}^2+\frac{l^2}{\xi(r)^2}\dd{r}^2+r^2 \dd{\phi}^2,\qquad \xi(r)\equiv \sqrt{r^2-r_+^2} \label{eq:ds-btz-Eucl} 
} 
with $\tau \in (0,\beta)$, $\phi \in (0,2\pi)$ and $r>0$. 
It is characterized by a contractible $\tau$-circle, as visualized in Fig.~\ref{fig:BTZ-torus}. 
The other way around, to construct the CS gauge field form of this metric, one reads off the dreibeins from the metric, then constructs the spin connections using the zero torsion condition, and finally combines those into chiral and anti-chiral gauge fields through \eqref{eq:Aofeomega}. 

Solutions to the equations of motion $F=0$
are pure gauge fields $A=g^{-1}dg$ with $g \in
\SL$. For the Euclidean BTZ solution, we have
\begin{align}
	g_{\,*}=\begin{pmatrix}
		\rho_{\,*} \cosh\frac{\pi\wb}{\beta} &\rho_{\,*}^{-1} \sinh\frac{\pi \wb}{\beta} \\ 
		\rho_{\,*} \sinh\frac{\pi \wb}{\beta} &\rho_{\,*}^{-1} \cosh\frac{\pi \wb}{\beta}
	\end{pmatrix}, \qquad
	\bar{g}_{\,*}=\begin{pmatrix}
		\rho_{\,*}^{-1} \cosh\frac{\pi w}{\beta} &\rho_{\,*} \sinh\frac{\pi w}{\beta} \\ 
		\rho_{\,*}^{-1} \sinh\frac{\pi \wb}{\beta} &\rho_{\,*} \cosh\frac{\pi w}{\beta}
	\end{pmatrix}, 
	\label{eq:g-BTZ}
\end{align} 
with $\rho_{\,*}=(r/r_+ + \xi(r)/r_+)^{1/2}$. Here we note that for $\tau \to \tau+\beta$, $g_{\,*}$ and $\gb_{\,*}$ are multi-valued as elements of $\SL \times \SL$ with $(g_{\,*},\gb_{\,*}) \to (-g_{\,*},-\gb_{\,*})$, but this is cured via the diagonal $\mathbb{Z}_2$ quotient appearing in the gauge group discussed above~\eqref{eq:Aofeomega}. The single-valuedness is equivalent to a trivial holonomy around the $\tau$-cycle and we will be careful to preserve it in our fluctuation analysis. For $\phi \to \phi + 2\pi$,  $g_{\,*}$ and $\gb_{\,*}$ are multi-valued, indicating a non-trivial $\phi$-holonomy which encodes mass and angular momentum of the black hole \cite{Banados1998}.

As can be read off from \eqref{eq:CSBTZ}, the Euclidean BTZ saddle of our CS theory has complex components, in particular $A_\tau = A_\tau^a j_a$ and $\bar{A}_\tau = \bar{A}_\tau^a \bar j_a$ are pure imaginary (with $j_a$ and $\bar j_a$ as discussed being real $\mathfrak{sl}(2,\mathbb{R})$ generators).   
Later we will consider real fluctuations around this complex saddle. 
As a final consistency check,  we calculate the on-shell action of the BTZ
solution~\eqref{eq:CSBTZ}, to which only the boundary term of \eqref{eq:I-full} contributes. Substitution of Eq.~\eqref{eq:CSBTZ} leads to
\begin{align}
	-i I_{\mathcal{M}_\infty}[A,\bar A]=\frac{k}{4\pi} \int_{\mathcal{M}_\infty}\dd^2{x} \qty( \operatorname{tr}(A_{\te}^2)+\operatorname{tr}(\bar{A}_{\te}^2))=\beta M - S_{BH},
\end{align} 
reproducing, as it should, the result of the Euclidean, on-shell Einstein-Hilbert action for the BTZ black hole. This evaluates to the free energy with the familiar Bekenstein-Hawking entropy $S_{BH}=2 \pi r_+/(4G)$. 

Interpreting the BTZ solution \eqref{eq:ds-btz-Eucl} as being asymptotically indistinguishable from the 
global AdS solution suggests a set of fall-off conditions on the general metric that gives a precise meaning to the notion of an ``asymptotically AdS spacetime''. These are the Brown-Henneaux boundary conditions \cite{Brown1986}. In gauge field language they are given by the condition that a solution displays the following behavior at $r \rightarrow \infty$: 
\begin{align}
	A=\begin{pmatrix}
		\frac{\dd{r}}{2r}+O(r^{-2}) & O(r^{-1}) \\ 
		\frac{r\dd{\wb}}{l^2}+O(r^{-1}) & -\frac{\dd{r}}{2r}+O(r^{-2})
	\end{pmatrix}
	, \qquad
	\bar A =\begin{pmatrix}
		-\frac{\dd{r}}{2r}+O(r^{-2)} & -\frac{r \dd{w}}{l^2}+ O(r^{-1}) \\ 
		O(r^{-1}) &\frac{\dd{r}}{2r}+O(r^{-2})
	\end{pmatrix} .  \label{eq:CSbcs}
\end{align} 
With these boundary conditions, the asymptotic symmetry group (of residual
diffeomorphisms in metric formulation, or gauge transformations in the CS
formulation) 
famously is that of the Virasoro symmetry of 2D CFT, with the combination of
gravitational parameters $3l/(2G)$ taking the role of a central charge $C$
\cite{Brown1986}. It is these Brown-Henneaux conditions that we will impose on
fields describing real fluctuations around the BTZ saddle.

\subsection{AS action for fluctuations around the BTZ black hole} \label{sec:AS}

With the  Chern-Simons action \eqref{eq:I-full} at hand, we now proceed to the
AS representation of the theory.
We employ the same strategy as that of
\cite{Cotler2018}, but applied to the Euclidean BTZ rather than global AdS$_3$.
Compared to the AS action for fluctuations around global AdS$_3$ derived in
\cite{Cotler2018}, ours will have interchanged roles of $\tau$ and $\phi$. This
is important for the match to the AS action~\eqref{eq:Actions_Equivalence} we
obtain from the boundary perspective. The AS
action~\eqref{eq:Actions_Equivalence} or \eqref{eq:S_graviton} appeared before
in gravitational discussions, e.g.~in
\cite{Mertens2022,Mertens2018,Henneaux2020}, without an explicit derivation
employing the techniques of \cite{Cotler2018}. It is this derivation we provide
now. 

The gauge field we discuss lives on a manifold with the topology of a solid
torus $D \times S^1$ with $D$ the disk ($\{r,\tau\} \times \{\phi\}$). The AS
construction amounts to the application of a standard protocol~\cite{Elitzur89}
reducing the theory to one defined on the torus boundary with coordinates
${(\phi,\tau)}$. In a first step, the integration over $A_\phi$ imposes the
constraint $\tilde{F}=0$, i.e. $\tilde{A}\equiv g^{-1} \tilde{\dd} g$ where
$g\in \SL$ is again single-valued in $\tau$, and $ \tilde{\dd}$ is the exterior
derivative with respect to $\{r,\tau\}$. We note that this representation
implies  a local redundancy $g \to \upsilon(\phi) g$ with $\upsilon \in \PSL$,
which we will comment on at the end of the section.

The substitution of this representation into the action~\eqref{eq:I-full} leads to the sum of two chiral
Wess-Zumino-Witten models \cite{ Coussaert1995} (see, e.g., Ref.~\cite{Donnay2016} for 
details)
\begin{align}
I=I_{+}[g]+I_{-}[\gb]. \label{eq:Swzw}
\end{align}
Here, 
\begin{align}
I_{ \pm}[g]=\frac{k}{2 \pi}\left(-i \int_{\mathcal{M}_\infty} d^2 x \operatorname{tr}\left(\left(g^{-1}\right)^{\prime} \partial_{ \pm} g\right) \pm \frac{1}{6} \int_{\mathcal{M}} \operatorname{tr}\left(g^{-1} d g \wedge g^{-1} d g \wedge g^{-1} d g\right)\right),  \label{eq:Swzwbis}
\end{align}
where $\partial_\pm = \frac 12 (\partial_\te \mp  i \partial_x)$ with $x=l
\phi$, we use the shorthand notation $f'=\partial_\tau f$
and the path integral measure is now $\mathcal{D}g$, the Haar-measure on $\PSL$. 
The first term $I_\pm[g]$ is defined on the boundary,
and the second is locally exact, meaning that it, too, affords a boundary
representation. In this way the theory becomes one of gauge field fluctuations
supported by the system boundary. 

To make this reduction concrete, we  employ the Gauss decomposition
\begin{align}
g=\begin{pmatrix}
1 & 0 \\ 
h & 1
\end{pmatrix} \begin{pmatrix}
\lambda & 0 \\ 
0 &\lambda^{-1}
\end{pmatrix}
\begin{pmatrix}
1 & \Psi \\ 
0 & 1
\end{pmatrix}
, \qquad \gb=\begin{pmatrix}
1 & \overline{h} \\ 
0 & 1
\end{pmatrix} \begin{pmatrix}
1/\overline{\lambda} & 0 \\ 
0 & \overline{\lambda} 
\end{pmatrix}
\begin{pmatrix}
1 & 0 \\ 
\overline{\Psi} & 1
\end{pmatrix}, \label{eq:GaussP}
\end{align}
with $h,\Psi \in \Rb$ and $\lambda>0$, and likewise for the barred sector.
Compatibility with the Brown-Henneaux boundary conditions~\eqref{eq:CSbcs} is
established via the relations
\begin{align}
\lambda=\sqrt{\frac{r}{l^2 h'}}, \quad \Psi=-\frac{l^2 h''}{2rh'}, \qquad \bar{\lambda}=\sqrt{\frac{r}{l^2\bar{h}'}}, \quad \bar{\Psi}=-\frac{l^2\bar{h}''}{2r\bar{h}'} \label{eq:g-constraints}
\end{align}
where $h\equiv \eval{h}_{\Mc_\infty}$ is independent of $r$, and $h'>0$ for $g$ to be real.  In this way, the theory has collapsed to a single real degree of freedom, $h=h(\tau,x)$. 
In this representation, the  
remaining bulk term in \eqref{eq:Swzwbis} becomes a boundary term
\begin{align*}
 	\frac{1}{3}\int_{\Mc} \operatorname{tr}\big((g^{-1}dg )^3\big) &=\int_{\Mc} d\lambda^2 \wedge d\Psi \wedge dh = \int_{\Mc_{\infty}} \lambda^2 d\Psi \wedge dh\, 
 \end{align*}
and with the constraints~\eqref{eq:g-constraints} in place the  WZW-action assumes the form 
\begin{align}
\label{eq:WZW_action_Ih}
    I_{ \pm}[h]=\frac{i C}{12 \pi}\int_{\Mc_{\infty}}\!\!\!\!d^2 x \qty(\frac{3}{2}\frac{h'' \partial_\pm h'}{h'^2}- \frac{\partial_\pm h''}{h'}) =
    \frac{i C}{12 \pi}\int_{\Mc_{\infty}}\!\!\!\!d^2 x\qty(\frac{1}{2}\frac{h'' \partial_\pm h'}{h'^2} -  \left(\frac{\partial_\pm h'}{h'}\right)')
\end{align}
Specifically, in the Gauss-parametrization of the BTZ stationary solution, up to a factor of $i$, 
we have $ h_{*}=\tan( \pi (\tau+ix)/\beta)$. Generalizing to functions which are $i)$ continuously 
connected to the saddle and $ii)$ keep $g$ single-valued as a function of $\tau$, fluctuations are described by 
\begin{align}
    h(\tau,x)=\tan(\frac{\pi f(\tau,x)}{\beta}) \label{eq:F-definition}
\end{align}
in terms of a reparametrization of the imaginary 
time cycle,  $f(\te,x) \sim f(\te,x) + \beta$ and $f'(\te,x)>0$. The fluctuation measure is obtained by evaluating the Gauss measure (which originally reads as $\mathcal{D}g=\mathcal{D} H \mathcal{D} \lambda \mathcal{D} \Psi$) at the boundary subject to the constraints  \eqref{eq:g-constraints} leading to $\mathcal{D}f  \qty(\prod_{\tau,x}f')^{-1}$. 

Substituting the trajectory~\eqref{eq:F-definition} into the boundary WZW action~\eqref{eq:WZW_action_Ih}, one
recovers the AS action $S_\pm^\beta[f]$ by virtue of relation~\eqref{eq:Actions_Equivalence}. 
We have thus reached our goal, a representation of the partition sum as 
\begin{align}
	Z(\mop)=&\int_{\substack{f(\tau,x=0)=f(\tau,x=L) \\ \bar{f}(\tau,x=0)=\bar{f}(\tau,x=L)}}
	\frac{ \mathcal{D} f(\tau,x)  \mathcal{D}\bar{f}(\tau,x)}{\prod_{x,\tau}f'(\tau,x) \bar{f}'(\tau,x)} e^{-S^\beta_+ [f]- S^\beta_- [\bar{f}]}, \label{eq:PIFull}
\end{align}
with
 \begin{align}
S^\beta_\pm [f] =  \frac{C}{24\pi} \int\limits_0^\beta d{\te} \int\limits_0^{2 \pi l } 
\dd{x} \left(  \frac{f'' \partial_{\pm} f'}{ f'^2} - \frac{ 4\pi^2 }{\beta^2} f' \partial_{\pm}  f\right), \qquad f' 
\equiv \partial_\te f \label{eq:SASE},
 \end{align}
and the coupling constant $C=\frac{3l}{2G} \gg 1$ given by the Brown-Henneaux
central charge. This path integral is identical to that discussed in \cite{Mertens2022}, but our interpretation and derivation is different: In our discussion, it is the fluctuation theory of the BTZ saddle-point. We note again that it describes real fluctuations around a complex saddle. 

Before concluding this section, a few comments are in place. First, in the 
representation Eq.~\eqref{eq:PIFull}, the $\PSL$ redundancy $g(x,\te) \to \upsilon(x)
g(x,\te)$ that is hardwired into the theory acts as a fractional linear
transformation
\begin{align} \label{eq:f-quotient}
\tan(\frac{\pi}{\beta}f) \to \frac{a(x) \tan(\frac{\pi}{\beta}f) +b(x)}{c(x) \tan(\frac{\pi}{\beta}f) +d(x)}, \qquad ab-cd=1.
\end{align}
From here on, we will exclusively work with  $\PSL$ quotients and ignore this
transformation throughout. The action Eq.~\eqref{eq:SASE} is identical to the
effective action~\eqref{eq:S_graviton} for the chiral SYK model, except that
presently we are working with a velocity scale $u$ set to unity and with the
identification $L \equiv 2\pi l$. 
Secondly, in
Appendix~\ref{App:Liouville-one-loop} we show that Gaussian integration around
the stationary solutions leads to the partition sum
 \begin{align}
 Z(\mop)=\abs{\chi_{0}(-1/\mop)}^2, \qquad \chi_{0}(\mop)=q^{-\frac{c}{24}} \prod_{n=2}^{\infty}\frac{1}{1-q^n}, \qquad q=e^{2\pi i \mop}, \label{eq:Z-vacuum}
 \end{align}
 which is the vacuum character of a CFT with central charge $c=C+13$ in the dual channel.
 Finally, a  Laplace transform of the partition function~\eqref{eq:Z-vacuum} to an energy representation recovers the expected Cardy growth of the density of states $\rho(E) \sim e^{2\pi
\sqrt{\frac{c}{6}E}}$ at high energies $E \gg 1$, reproducing the
microcanonical Bekenstein-Hawking entropy. 

Let us finally comment on the matching of coupling constants between the AS actions derived from the bulk and the boundary. The bulk theory contains three different length scales: $G, l$ and $\beta$, where time is measured in units of length. Consequently, the partition function depends only on two dimensionless ratios: 
$C={3l}/{2G}$ and $\mop={i \beta}/{2\pi l}$, which must be equivalently identified in the boundary theory.
There, the central charge $C$ is given by Eqs.~\eqref{eq:Cu_high_T}, while the modular parameter can be extracted from the underlying geometry as $\mop={i \beta u}/{L}$. Here, the system size is related to the AdS$_3$ radius $l$ in the holographic setup by the simple relation $L=2\pi l$, while the velocity $u$ can be determined from 
the same Eqs.~\eqref{eq:Cu_high_T} and~\eqref{eq:u_T}. Also, given $C$, $L$ and $\beta u$ as boundary data, 
the effective Newton’s constant $G$ in the bulk can then be reconstructed as $G = 3l/2C$.

\section{Correlation functions from Liouville theory}\label{Sec:Liouville} The
goal of this section is to evaluate the effect of the reparametrization averaged
with the action~\eqref{eq:S_graviton} on various correlation functions of the
model. In Subsection~\ref{Sec:Liouville-map}, we begin by introducing our method
of choice for calculating observables, namely the Liouville map. This approach
is then applied in Subsection~\ref{sec:light} to the two-point function of
Majorana operators at coinciding spatial points, and in
Subsection~\ref{sec:heavy} to `clusters' of Majorana operators, which allow for
more interesting effects in the correlation function as compared to single
operators. Finally, in Subsection~\ref{sec:OTOC}, we calculate the OTOC between
a cluster of Majorana operators with a single operator, as a means to diagnose chaos.

The  building blocks for all subsequent calculations are generalizations
of the reparameterized two-point function~\eqref{eq:G_rescale}, which we define as
\begin{align} \label{eq:bilocal}
    \mathcal{O}_{\tau_1 \tau_2}^{2l}(x)= \qty( \frac{ \partial_{\tau_1} f_1 \partial_{\tau_2} f_2}{ \sin^2 \qty(\pi (f_1-f_2)/\beta)})^l, \qquad f_i=f(\tau_i,x)
\end{align}
and from which we recover $\widetilde{G}_{\tau_1\tau_2}$ for $l=1/4$.
These bilocal operators are interesting from the holographic point
of view, since conformal correlation functions as
e.g.~\eqref{eq:GS0_SYK_t} are known to describe
particle propagation between two points anchored on the boundary  (at equal
spatial position) along a bulk geodesic. Dressing these objects with
reparametrizations governed by the AS action is  equivalent 
in the bulk to dressing the particles with gravitons 
\cite{Fitzpatrick2014, Fitzpatrick2015, Fitzpatrick2016, Chua2023}.  

The calculations of observables simplify in the limit of large $N$, leading to a
large coupling constant $C$ in front of the AS action via Eq.~\eqref{eq:Cu_high_T}.
The largeness of this coupling enables a classical approach, wherein correlation
functions are computed in terms of  on-shell solutions $f=f_0$. We also compare
this approach to results obtained by more elaborate methods of conformal field theory, differing from ours only
in anomalous corrections to the scaling dimensions of the operators appearing in
the correlation functions. In executing this program,  we will need to
distinguish between two types of correlations functions: those containing
operator insertions $\mathcal{O}$ which are `light' in the sense that they do
not affect the stationary phase solutions $f_0$ (up to corrections vanishing in
the $C\to \infty$ limit), and others with `heavy' operators which do modify the
solutions. (In the holographic context, the difference between  heavy and
light particles is whether they  back-react on the geometry or not
\cite{Hulik2016}.) We will show that the classical treatment of heavy operator
insertions is a highly economical approach, alternative to previous strategies
based on perturbation theory in the reparametrization mode formalism
\cite{Cotler2018, Haehl2018} or  the Liouville bootstrap \cite{Mertens2017,
Mertens2018}.

Within our approach, the complexity of the problem is shifted to the computation
of equations of motion, in the presence of reparameterized two-point functions
${\cal O}^{2l}_{\tau_1 \tau_2}(x)$. To simplify this step,  we take preparative
measures: We first map the AS theory to Liouville theory with boundaries, for which
the equations of motion simplify considerably \cite{Menotti2004, Hulik2016}.
Having found the solutions on the Liouville side, we  map them back to obtain
the saddle-point reparametrization $f_0(\tau,x)$.  In the following, we explain
this strategy in  detail. 

\subsection{The Liouville map} \label{Sec:Liouville-map} The equivalence between 
AS theory and Liouville theory with boundaries we use builds on the work of
\cite{Mertens2017, Mertens2018}. It posits the identity of path integrals
\footnote{The relation holds for $S_{\pm}$ independently. When following
\cite{Mertens2018} and expressing $f(\tau,x)$ in terms of two functions
$A(\tau,x)$ and $B(\tau,x)$ defined for $0<\tau<\beta/2$ and $-\beta/2<\tau<0$
respectively, the roles of $A$ and $B$ are exchanged in the actions $S_{+}$ and
$S_{-}$ and one hence has to also exchange their roles in the field
transformation.} 

\begin{align} \label{eq:PIEquiv}
\int\frac{ \mathcal{D} f }{\prod_{x,\tau}f'(\tau,x)} (\cdots) e^{-S_{\pm}^{\beta}[f]} = \int \mathcal{D} \phi  (\cdots) e^{-S_L[\phi]},
\end{align}
where the ellipses symbolically represent matching operator insertions. 
This
connection is established via a mapping of field variables,
$\phi(\tau,x)=\phi(f(\tau,x))$, defined for positive $\tau\in [0,\beta/2]$ as
\begin{align}
	\phi(\tau,x)&=\ln(\frac{- \partial_\tau f_+ \partial_\tau f_-}{\sin^2({\pi(f_+-f_-)}/{\beta})}), \qquad f_\pm=f(\pm \tau,x).   \label{eq:Liouville-field}
\end{align} 
The $\phi$-representation of the fluctuation action is given by \footnote{To avoid confusion with the quantities on the disk which are defined with a prime, we denote temporal derivatives as $\partial_\tau$ in this section.}
\begin{align} \label{eq:SL}
    S[\phi]=\frac{C}{24 \pi u} \int_0^{L} \int_0^{\beta/2} \dd{x} \dd{\tau}\left( \frac{( \partial_\tau \phi)^2}{4}+\frac{(u\partial_x \phi)^2}{4}+ \qty(\frac{2\pi}{\beta})^2 e^{\phi}- \partial_\tau^2 \phi \right)\,, 
\end{align}
which is a variant of the action of Liouville field theory in the semiclassical
regime $C\gg 1$~\cite{Zamolodchikov1995}, and is therefore denoted $S_L[\phi]$ in \eqref{eq:PIEquiv}. Importantly, the constants appearing
in Eq.~\eqref{eq:SL} are direct imports from the microscopic model discussed in
section~\ref{sec:Chiral-SYK}, where we have reintroduced the velocity scale
$u(1/\beta)$ given by Eq.~\eqref{eq:u_T}. In
Appendix~\ref{App:Liouville-one-loop} we corroborate the equivalence to the
$f$-representation by demonstrating that without operator insertions,
integration over $\phi$ recovers the partition function~\eqref{eq:Z-vacuum}.

\begin{figure}
        \centering
	\includegraphics[width=0.9\linewidth]{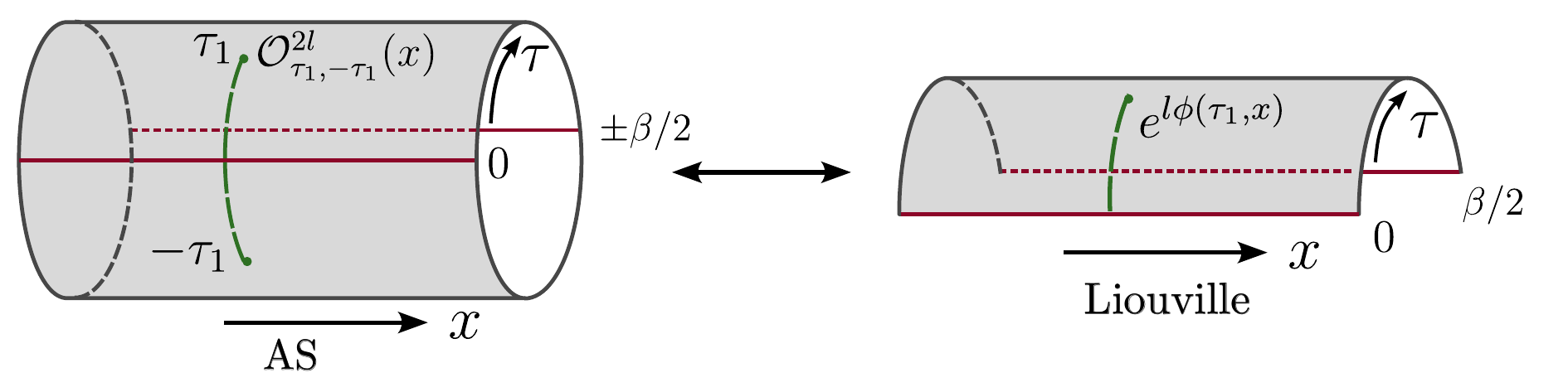} 
	\caption{ Left: Imaginary time $\tau$ runs in the interval $ \tau \in
		(-\beta/2,\beta/2)$ along the cylinder. Bilocal operators are placed
		symmetrical around $\tau=0$ with legs at the two points
		$(\tau_1,-\tau_1)$ with $\tau_1>0$. Right: After mapping the system to
		Liouville theory, the range of the imaginary time cycle is halved to
		$\tau \in (0,\beta/2)$. The bilocal operator becomes a local vertex
		operator with behavior for $\tau_1 \to 0, \beta/2$ prescribed by the
		ZZ-boundary conditions.
		 }
		\label{fig:Liouville-Map} 
\end{figure}

Before proceeding, we need to take a closer look at the space-time interval
underlying the $\phi$-theory, $[0,\beta/2]\times [0,L]$, see
Fig.~\ref{fig:Liouville-Map}. Eq.~\eqref{eq:Liouville-field} implies a
divergence $\phi(\tau,x)\to \infty$ at $\tau=0,\beta/2$ corresponding to
Dirichlet, or ZZ-boundary conditions $e^{-\phi(0,x)}=e^{-\phi(\beta/2,x)}
\rightarrow 0$ in the parlance of Liouville field theory. The rationale behind
restricting $\phi$ to the positive imaginary time domain is likewise evident
from Eq.~\eqref{eq:Liouville-field}: the $\phi$-field at $\tau>0$ encodes
information on reparametrization configurations both at $\tau$ and the mirror
symmetric time $-\tau$. The same symmetry implies a constraint on  temporally bilocal operator insertions ${\cal
O}^{2l}_{\tau_1\tau_2}(x)$ as in~\eqref{eq:bilocal} in 
the theory:  only
configurations symmetric around the origin ${\cal O}^{2l}_{\tau_1, -\tau_1}(x)$
afford a consistent $\phi$-representation. Finally, the
$f$-representation of our  operators  diverges at coinciding
temporal arguments due to the presence of denominators $\sim 1/\sin(\pi
(f_\tau-f_{-\tau})/\beta)^\alpha$. 
In the $\phi$-language, these divergences are reflected in the above-mentioned boundary conditions. 

Within the set of temporally symmetric
observables, it is convenient to switch to a holomorphic
representation~\eqref{eq:hol-coord}, i.e. 
\begin{align}
  {\cal O}^{2l}_{\tau_1\tau_2}(x)=e^{l \phi(\tau,x)} \equiv \V{l}(w,\wb) \qquad  w=i \tau+ x/u, \quad \qquad \wb=-i \tau + x/u.  \label{eq:bilocal-liouville}
\end{align} 
This transformation conveniently eliminates the
 $\SL$ `target space' redundancy~\eqref{eq:f-quotient} of the reparametrization
 mode\cite{Mertens2018}, i.e. the Liouville representation describes
 the $f$-theory with $\mathrm{SL}(2,\Bbb{R})$ appropriately modded out.
 
 On any domain related to the strip by a
 holomorphic coordinate  transformation defined by $(w, \wb)  \mapsto
 (\xi(w),\overline{\xi(w)}) \equiv (\xi,\xib)$ with $\xi, \xib$
 \textit{dimensionless}, the field $\phi$  transforms as \cite{Seiberg1990,
 Ginsparg1993}
 \begin{align}
	 \qty(\frac{2 \pi}{\beta})^2 e^{\phi(w, \overline{w})}=e^{\phi'(\xi, \overline{\xi})}\abs{\pdv{\xi(w)}{w}}^2 \label{eq:Liouville-Trafo}.
 \end{align}
 Here the bar denotes complex conjugation, and we multiplied by a factor of
 $(2\pi/\beta)^2$ to keep the field $\phi'$ dimensionless.
 Eq.~\eqref{eq:Liouville-Trafo}  reveals the  interpretation of $\phi$ as a Weyl factor
 of a metric with $e^\phi \dd{w} \dd{\wb}$ invariant under conformal
 transformations. It also implies $\Delta_l=2l$ as the  engineering
  dimension of the operators Eq.~\eqref{eq:bilocal-liouville}.

Turning to operator insertions, depending on the value of $l$, one distinguishes between  `light' operators with
$l= \mathcal{O}(1)$ and `heavy' ones, $l=\mathcal{O}(C) $\cite{Seiberg1990}. To
understand their difference, note that the insertion of $n$ vertex operators at
coordinates $w_i$ with degree $l_i$ is described by a
modified action
\begin{align}
	\label{eq:ActionGeneralizedToSources}
	S[\phi] \mapsto S[\phi]- \sum_i^n l_i \phi(w_i).
\end{align} 
Accordingly, the equations of motion $\delta_\phi S=0$  generalize to
\begin{align} \label{eq:LiouvSemicl}
    \frac 12 \partial_a \partial^a \phi= \qty(\frac{2\pi}{\beta})^2 e^{\phi}- \frac{24 \pi}{C} \sum_i^n l_i \delta^{(2)}(w-w_i)\,.
\end{align}
Light operators affect the equations of motion only negligibly in terms of a
$1/C$ correction, while the heavy ones need to be accounted for in the solution.
As a consistency check, we note that the action in the absence of sources,
$S_0= - (\pi C/12 u) ( L/\beta) $, obtained  by evaluating the
AS action~\eqref{eq:SASE} on the trivial solution $f_0(\tau,x)=\tau$, follows
equivalently by  substituting the  solution~\eqref{eq:phi-0-strip} of the
sourceless Eq.~\eqref{eq:LiouvSemicl}  into~\eqref{eq:SL}.

The most general correlation function we will work with is between a light
operator of conformal weight $l_1$ and a heavy one of conformal weight $l_2$.
Classically, it reads 
\begin{align} \label{eq:Corr-general} 
\expval{\V{l_1}(w_1) \V{l_2}(w_2)}=e^{l_1 \phi(w_1;l_2,w_2) -S(l_2,w_2)+S_0}\,.
\end{align}
Here,  
$\phi(w;l_2,w_2)$ is the solution to the Liouville
equation~\eqref{eq:LiouvSemicl} in the presence of a single heavy source of
weight $l_2$ at $w_2$,  $\exp(S_0)$ normalizes the correlation function, and we  suppressed the $\wb$-dependence for 
readability. The evaluation of its
action, 
\begin{align}
    S(l_2,w_2)\equiv S[\phi(\circ; l_2,w_2)]-l_2 \lim_{w  \to w_2} \phi(w;l_2,w_2)
\end{align}
requires a regularization procedure because of singularities in the limit $w\to
w_2$\footnote{In order for $S(l_2,w_2)$ not to diverge close to the source, one
needs to supplement the action with a counter term \cite{Zamolodchikov1995,
Menotti2004, Menotti2006}, which also modifies the behavior of $S(l_2,w_2)$
under conformal transformations. Details on the regularization are given in
Appendix~\ref{App:Liouville-one-source}.}.  Finally, we will describe the
high-temperature regime $(\beta u)/L \ll 1$ of interest to us  by taking the
limit of a spatially infinite system, $L \to \infty$.

\subsection{Single Majorana operator} \label{sec:light}

We begin with an analysis of the correlation function~\eqref{eq:G20} between two
(`light')  single Majorana operator insertions at the same point in space,
$x=x'$, and times $\tau_1$ and $\tau_2$. In this case, the need to modify the
action does not present itself, and the computation of the correlation function
reduces to taking the  average of \eqref{eq:G_rescale} over the
reparametrization $f$. Focusing on the leading $(0+1)d$ SYK contribution $G^0$,
we have
\begin{equation}
	G^{xx}_{\tau_1 \tau_2} \equiv G(\tau_{12}) = \expval{ \widetilde{G}_{\tau_1\tau_2}}_\text{AS}, \label{eq:G2}
\end{equation}
 where we neglected the spatial argument due to translational invariance in $x$,
 and $\tau_{1,2}=\pm \tau_0/2$. Turning to the  Liouville representation, we include the factor $(\pi/\beta)^{1/2}$ of the finite
 temperature theory and obtain
\begin{align} \label{eq:G2-single} 
    G(\tau_0)=n_0 \qty(\frac{2\pi}{\beta J})^{1/2} \expval{\V{1/4}(w_0)} , 
    \qquad n_0= -    \frac{ {\rm sgn}(\tau_{12}) k_0}{\sqrt{4 \pi}},
\end{align}
where $w_0=i \tau_0/2+x/u$ and the exponent $1/4$ determines the engineering
scaling dimension of the Green's function as $\Delta_G = 2\times
1/4 =1/2$. 

We next analyze this functional average within the stationary phase approach. 
Inspection of Eq.~\eqref{eq:Corr-general} shows that in the absence
of a heavy source, $\mathcal{V}_{l_2}$, the action contributions $-S(0,0)+S_0=0$
cancel, and we are left with the exponential $e^{\phi_0(w)/4}$, where $\phi_0$
is the solution to the Liouville equations of motion~\eqref{eq:LiouvSemicl}
without source insertion. The concrete computation of such solutions is best
done in a disk-geometry, related to the strip by a conformal
transformation~\cite{Menotti2006, Hulik2016}. In the following, we review this
map, and discuss how it enables the concrete evaluation of the mean field
theory.

\begin{figure}
    \centering
    \includegraphics[width=1\linewidth]{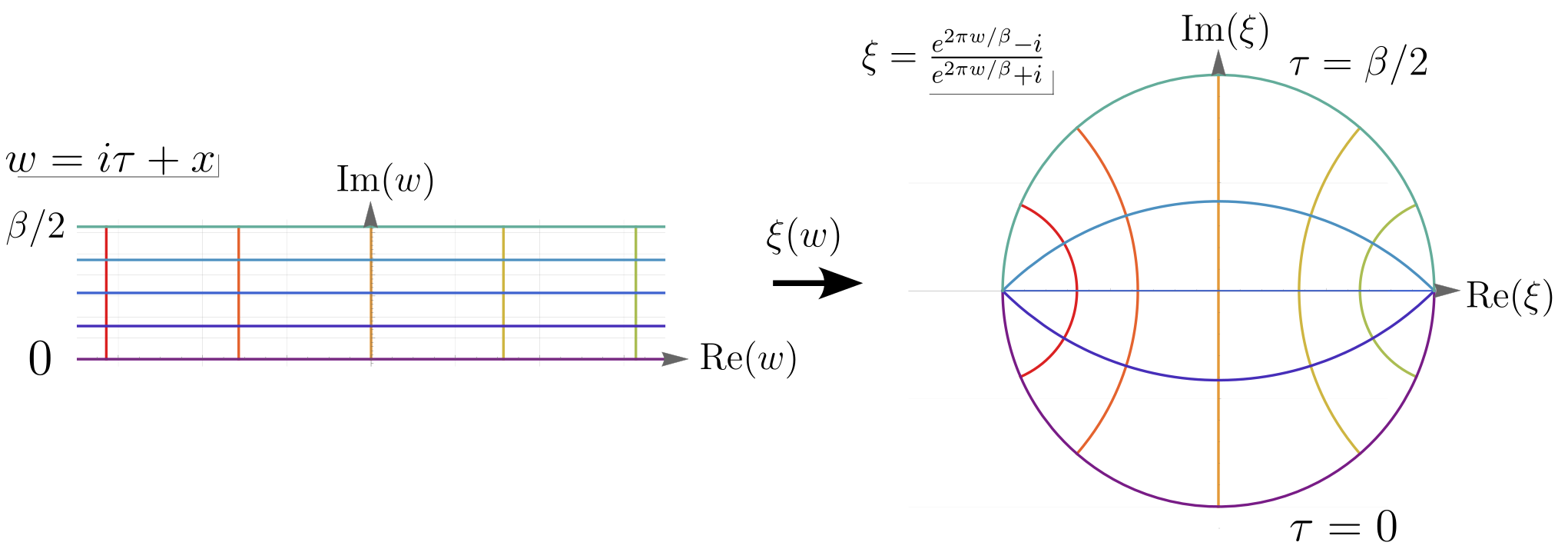}
    \caption{Behavior of constant $\tau$- and $x$-slices on the half-strip under the mapping~\eqref{eq:Cayley-map}.}
    \label{fig:mappings}
\end{figure}

Our map between strip and disk  is given by (cf. Figure~\ref{fig:mappings})
\begin{align}
\xi(w)=\frac{e^{\frac{2\pi}{\beta}w}-i}{e^{\frac{2\pi}{\beta}w}+i}\,.  \label{eq:Cayley-map}
\end{align}
Then the mapping $(w,\wb) \mapsto
(\xi(w),\overline{\xi(w)})=(\xi,\xib)$ maps the full strip to
the interior of the unit disk. Labeling functions
$\psi'(\xi)$ defined on the disk by a prime,  the translation back to the strip
is defined as  $\psi(w)=(\xi^{-1}\circ  \psi' \circ \xi)(w)$, or implicitly
through $\xi(\psi(w))=\psi'(\xi(w))$.

One can show~\cite{Hulik2016} that in disk coordinates the general solution $
\phi'(\xi)$ of the Liouville equation coupled to sources, Eq.~\eqref{eq:LiouvSemicl}, is given by
\begin{align}
   e^{\phi'(\xi, \overline\xi)}=\frac{ 4\abs{\partial_\xi f'(\xi)}^2}{(1-\abs{f'(\xi)}^2)^2}\, , 
   \label{eq:Liouville-General-solution}
\end{align}
where $f'(\xi)$ is a meromorphic function, whose specific profile is determined by the
inhomogeneity. In addition, the asymptotic
condition $|f'(\xi)|^2\to 1$ for $|\xi|\to 1$  implements the divergence
$e^{\phi'(\xi)}\to \infty$ required by the ZZ boundary conditions.

From this result, the solution  on the strip is found by inverse transform
 under the $\xi$-map, via Eq.~\eqref{eq:Liouville-Trafo}. Defining a function
 $f(w)$ through $\xi(f(w))\equiv f'(\xi(w))$, the condition $\overline{f(w)}=f(\overline{w})$ holds. Using this feature, a straightforward calculation summarized in Appendix~\ref{sec:L_field_strip} leads to 
\begin{align}
 \phi(w, \overline{w})=\ln(\frac{\abs{\partial f(w)}^2}{ \sin^2 \qty({i \pi (f(\wb)-f(w))}/{\beta})})\,. 
 \label{eq:Strip-general-solution}
\end{align}
In this representation, ${f}(w)$ has the status of a  conformal transformation
of the strip. Comparing it with the definition of the Liouville field
Eq.~\eqref{eq:Liouville-field}, we are led to the identification 
\begin{align}
    f(\tau,x) = - i f(w), \label{eq:rep-identification}
\end{align}
between the solutions of the Liouville equation and reparametrization
transformations in the AS theory.

We next  apply these structures to the computation of the light operator
correlation function Eqs.~\eqref{eq:G2} and \eqref{eq:G2-single}. In this case,
we just need the stationary solution in the absence of sources. On the disk,
this  is the identity map, $f'(\xi)=\xi$, implying that $f'(w)=w$ is the
identity, too. (With Eq.~\eqref{eq:rep-identification} we find that, up to irrelevant ${\rm SL}(2, \mathds{R})$
transformations, this result
translates to the trivial reparametrization,
$f(\tau,x)=\tau$.). Finally, Eq.~\eqref{eq:Strip-general-solution} implies
\begin{align} \label{eq:phi-0-strip}
\phi (w, \overline{w})= - \ln({\sin^2\qty({i\pi (\wb-w)}/{\beta})}) \,,
\end{align}
leading to
\begin{align}
    G(\tau_0)=n_0\, \qty(\frac{2\pi}{\beta J})^{1/2} \abs{ \sin{\frac{\pi \tau_0}{\beta} }}^{-1/2}.
\end{align}
We observe the absence of corrections to the scaling exponent within the 
variational framework. This finding, expected to become exact in the $C\to \infty$ limit, is consistent with boundary CFT results\cite{Teschner2000, Teschner2001}
\begin{align} \label{eq:Delta-light}
    \Delta_{l_1}= 2l_1+\frac{6l_1}{C}(1-2l_1),
\end{align} 
i.e. the mean field value modified by a correction vanishing in the semiclassical limit. 

\subsection{Majorana clusters} \label{sec:heavy} 

\begin{figure}
    \centering
    \includegraphics[width=0.5\linewidth]{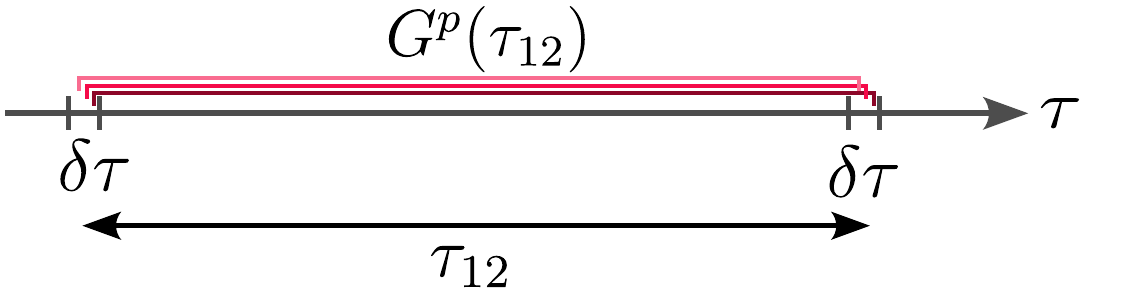}
    \caption{Construction of a cluster of Majorana fermions with $\delta \tau/\tau_{12} \ll 1$.}
    \label{fig:clusters}
\end{figure}

We next  generalize  to the insertion of a large number, $p$, of Majorana operators. Following Ref.~\cite{Bagrets2016}, we consider  the $p$ fields grouped in time into clusters of almost
coinciding times $- \tau_0/2+ \eta_j$ and $+\tau_0/2+ \eta'_j$, $j=1,\dots, p$,
where $|\eta_j|,|\eta'_p|<\delta \tau$ are small, see Fig.~\ref{fig:clusters}.
Taking the limit of vanishingly small separations, we obtain
\begin{align}
     G^{[p]}(\tau_{0})\overset{\text{def.}}{=}&N^{-p}\sum_{i_1 \cdots i_p} \langle \gamma^{i_1}(\tau_0/2+\eta_1) \gamma^{i_1}(-\tau_0/2+\eta'_1,x) \cdots \gamma^{i_p}(\tau_0/2+\eta_p) \gamma^{i_p}(-\tau_0/2+\eta'_p,x)\rangle \nonumber \\
     = &
	\expval{ G^p_{\tau_0/2,-\tau_0/2}[f]}_\text{AS}=n_0^p \, \qty(\frac{2\pi}{\beta J})^{p/2} \V{p/4}(w_0,\wb_0) , \label{eq:Cluster-full}
\end{align}
In this way, we define a cluster operator of engineering dimension
$l_2=p/4 $, which in the case $p\sim C$ becomes heavy. 
Using again~\eqref{eq:Corr-general}, but now  at $l_1=0$, the correlation function becomes
\begin{align}
    G^{[p]}(\tau_0) = n_0^p\,  \qty(\frac{2\pi}{\beta J})^{p/2} e^{ -S(l_2,w_0)+S_0} \label{eq:Cluster-approx}.
\end{align}
In order to calculate this object, we need to evaluate the Liouville action in presence of a source at an arbitrary position. 

To start, consider an
operator insertion of weight $l_2/C= \text{const} \lesssim 1$\footnote{Operator insertions with $pl > C/12$ create holes in the
underlying manifold (in our case the unit disk) \cite{Seiberg1990, Hulik2016}.
Such system deformations are not included  microscopic model, and we excluded it
by imposing the bound.} at strip coordinate $w_0=i\beta/4$, or $\xi=0$ in the disk picture. As shown  in Appendix~\ref{App:One-source}, the solution is
given by
\begin{align} \label{eq:LivOneSourceZero}
 f'(\xi)=\xi^{1-2\delta}, \qquad \quad e^{\phi'(\xi;0,l_2)}\overset{\text{Eq.}~\eqref{eq:Liouville-General-solution}}{=}
 \frac{4(1-2\delta)^2}{\qty(\abs{\xi}^{2\delta}-\abs{\xi}^{2-2\delta})^2},
\end{align}
where $\delta=6l_2/C$. We next shift the
source from the origin to the  disk coordinate $\xi_0=\xi(w_0)$ corresponding to our strip coordinate $w_0$. This is achieved by
application of the $\textrm{SU}(1,1)$-isometry on the Poincaré disk 
\begin{align}
\label{eq:Poincare_disk_map}
	h(\xi;\xi_0) =\frac{\xi_0-\xi}{1-\xib_0 \xi},
\end{align}
leading to \cite{Menotti2004, Menotti2006}
\begin{align}
\phi'(\xi;\xi_0,l_2)=\phi'(h(\xi;\xi_0);0,l_2)+\ln\abs{\dv{h(\xi;\xi_0)}{\xi}}^2.\label{eq:h-def}
\end{align}
With the general disk-solution at hand and using the conformal mapping~\eqref{eq:Cayley-map},
one can relate the disk action $S'(l_2, \xi_0)$, evaluated on the configuration~\eqref{eq:h-def},
to the desired action $S(l_2, w_0)$ on the strip. Referring to Appendix~\ref{App:Liouville-one-source} 
for somewhat intricate details of this procedure, we simply state the final result:
\begin{align} \label{eq:SL-one-source-explicit-main}
    \!\!\!\! S(l_2, w_0)&=S_0 - \ln U(l_2)  - \frac{ C \delta  (1-\delta) }{6} \ln( \frac{\beta^2}{4\pi^2}
     \abs{h_w'(\xi(w),\xi_0)}^2_{w=w_0})
    +\frac{C \delta^ 2}{6}\ln \abs{\frac{2\pi }{\beta J}}^2,
\end{align}
Here the prefactor $U(l_2)$ with $U(0)=1$ 
stems from the $\xi$-independent part
of the on-shell action and is given by Eq.~\eqref{eq:one-point-coeff}.
With the help of definitions for mappings $\xi(w)$ and $h(\xi)$,
see Eqs.~\eqref{eq:Cayley-map} and~\eqref{eq:Poincare_disk_map},  
a straightforward evaluation using a chain rules for derivatives yields
\begin{equation}
    \frac{\beta^2}{4\pi^2} \abs{h_w'(\xi(w),\xi_0)}^2_{w=w_0} = \frac 14 \times
    \frac{1}{{ \sin^2(\pi \tau_0/\beta)}},
\end{equation}
which is independent of the spatial position $x$ as expected.
Plugging the result for the action $S(l_2, w_0)$ into Eq.~\eqref{eq:Cluster-approx}, we obtain 
for the correlator
\begin{align} \label{eq:Heavy-correlator}
	G^{[p]}(\tau_0) = 
    n_0^p\, U(l_2) \,
    \qty(\frac{\pi}{\beta J})^{2 l_2 \left(1-\frac{6 l_2}{C}\right)} \abs{ \sin{\frac{\pi \tau_0}{\beta} }}^{-2 l_2 \left(1-\frac{6 l_2}{C}\right)},
\end{align}
predicting a significant change of the scaling dimension $\Delta_{l_2}\to
\Delta_{\textrm{s.cl.}}\equiv 2l_2 \left(1-6l_2/C  \right)$, of order $\mathcal{O}(C)$ in case of heavy operators. 
Comparing to the CFT framework, there the prefactor $U(l_2)$ is
called the ZZ one-point coefficient\cite{Schomerus2005},
and the scaling dimension is given
by\cite{Teschner2000, Teschner2001}
\begin{align}
\label{eq:Delta_Q}
    \Delta^Q_{l_2}= \Delta_{\textrm{s.cl.}}+\frac{6 l_2}{C} \equiv 2l_2 + \frac{6 l_2}{C}(1-2 l_2).
\end{align}
Again, we observe consistency with the semiclassical result up to subleading terms. 
Within the path integral approach employed here, one can also demonstrate~\cite{Menotti2006} that
a quantum correction ${6 l_2}/{C}$ to the semiclassical scaling dimension $\Delta_{\textrm{s.cl.}}$
arises solely from the Gaussian fluctuations around the saddle-point trajectory~\eqref{eq:h-def}.
In other words, the result~\eqref{eq:Delta_Q} is one-loop exact.

\begin{figure}[t]
    \centering
    \includegraphics[width=0.5\linewidth]{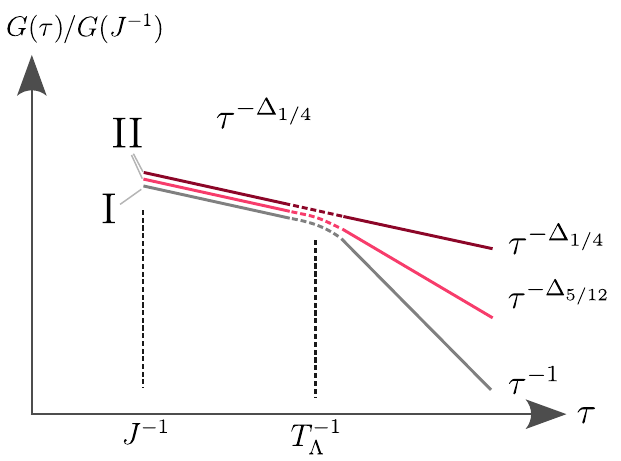}
    \caption{Log-log plot of the Majorana two-point functions for Model I (grey) and Model II (dark and light red) with $\gamma=3/2$. While both models show the same behavior for times shorter than $T_\Lambda^{-1}$, the relevance of the kinetic term in Model I leads to a crossover to the Fermi-liquid regime at larger time scales. 
    For Model~II, beyond this time scale the Green's functions of localized and mobile Majoranas 
    show algebraic decays with exponents $\Delta_{1/4}=\frac 12  (1+3/2C)$ and $\Delta_{5/12}=\frac 5 6 (1+2/C)$, respectively.
    }
    \label{fig:two-point}
\end{figure}

\subsection{Majorana two-point functions}
Having identified the relevant scaling dimensions,  
we are now in a position to discuss the two-point correlation
function~\eqref{eq:G20} of Majoranas operators. We begin with the linearly
dispersive  Model I whose correlation function at  short times $J^{-1} \ll \tau
\ll T_\Lambda^{-1}$ decays with the mean-field SYK exponent $\Delta_G = 1/2$ (cf. section ~\ref{subsec:Models}).
Reparametrization fluctuations change this exponent 
to $\Delta^Q_{1/4}$, Eq.~\eqref{eq:Delta-light}, leading to
$\mathcal{O}(C^{-1})$ corrections relative to the mean-field result. For longer
times, $\tau \gtrsim T_\Lambda^{-1}$, reparameterization fluctuations are gapped out, and Model I exhibits Fermi liquid behavior,
with an exponent $\Delta_{\rm FL} = 1$ (cf. the  gray line in Fig.~\ref{fig:two-point}).

For Model II, we can analyze two correlators: the full two-point function,
defined by Eq.~\eqref{eq:G20}, and the one associated only with mobile
Majoranas, given by Eq.~\eqref{eq:lambda_Majoranas_def}. The full correlator is
again described by the mean-field SYK exponent $\Delta_G = 1/2$ for the entire
time range, which is shifted to $\Delta^Q_{1/4}$ when quantum fluctuations are
included (see the dark red line in Fig.~\ref{fig:two-point}, plotted on a
log-log scale). By contrast, the second correlator, $g^{xx}_\tau$, which refers
to delocalized Majoranas, decays with a mean-field exponent $\Delta = 1/2 +
1/2\gamma$ at long times $\tau \gtrsim T_\Lambda^{-1}$. For our case of
interest, where $\gamma = 3/2$, reparametrization fluctuations change this
value to $\Delta^Q_{5/12}$. As a result, the qualitative behavior of
$g^{xx}_\tau$ (see the light red line in Fig.~\ref{fig:two-point}) closely
resembles that of the Majorana two-point function of Model I.

It is now worth discussing the relation between the above results for the Majorana correlation function and the well-studied problem of electron tunneling in mesoscopic physics. As we have seen in sections~\ref{sec:light} and \ref{sec:heavy}, the net effect of reparametrization fluctuations is to change the mean-field scaling exponent of the two-point Majorana function, $\Delta_l=2l$, to its quantum counterpart, as given in Eq.~\eqref{eq:Delta_Q}, where the induced correction scales as $1/C$ (and thus vanishes in the limit $N \to \infty$).
The analogous phenomenon in condensed matter physics is broadly known as electron tunneling in the presence of an 'electromagnetic environment'~\cite{Nazarov_Blanter:2009}. This encompasses a large class of problems studying how electron-electron interactions suppress the single particle Green’s function (or tunneling rate) in low-dimensional correlated electron systems. Examples include two-dimensional 
disordered films~\cite{Nazarov:1989, Kamenev:1999}, quasi-one-dimensional disordered wires~\cite{Mora:2007}, one-dimensional Luttinger liquids out of equilibrium~\cite{Ngo_Dinh:2012}, and compressible quantum Hall (QH) edge states~\cite{Levitov:2001}. In all these cases, the solution strategy closely follows the approach discussed above: one determines the optimal fluctuation of the electron density induced by a delta-source representing the injection of an extra electron charge into the system, and then evaluates the action cost of such an optimal fluctuation. This action typically scales as $S(\tau_0) \propto \ln^d (\tau_0)$, where $d=1,2$ is the system dimension and $\tau_0$ is the tunneling time. A similar scaling behavior appears in Eq.~\eqref{eq:SL-one-source-explicit-main} in the limit $\tau_0 \ll \beta$. The precise nature of the 'electromagnetic environment' depends on the specific problem and may, for instance, include statistical Chern-Simons gauge fields when describing composite fermions~\cite{Levitov:2001}. In this context, reparametrization fluctuations in the chiral SYK model, which correspond to boundary gravitons in the dual holographic setting, can be viewed as a direct analog of such 'electromagnetic environment' in the Majorana system, with its strength controlled by the inverse central charge $1/C$ or, in other words, by the gravitational Newton's constant $G$ in the bulk.

\subsection{OTOC in the Heavy-Heavy-Light-Light limit} \label{sec:OTOC} 
We finally probe early-time chaos of our model by calculating the contribution of reparameterization fluctuations to the OTOC. This observable has been extensively studied in SYK chains \cite{Gu2017, JianYao17, Sachdev17, Cai2018, Lian2019}. In particular, systems exhibiting an emergent reparameterization symmetry generically show growth of OTOCs with the maximal Lyapunov exponent $\lambda=\frac{2\pi}{\beta}$ \cite{Gu2017, JianYao17, Sachdev17, Cai2018}. While in the referenced works, the propagator of the soft modes in the Gaussian approximation has a diffusive pole, the one corresponding to the AS action is of ballistic nature \cite{Cotler2018, Haehl2018}. In both cases, one expects the same maximally chaotic behavior \cite{Blake2018, Blake20182}, as we explicitly confirm now. To this end, we consider the OTOC between a cluster of Majorana operators and a single Majorana operator. In the low-energy theory,
this quantity assumes the form of a two-point function of a heavy and a light vertex
operator, cf. Eq.~\eqref{eq:Corr-general}, referred to as a Heavy-Heavy-Light-Light
(HHLL) correlation function in the following. While this quantity can be computed within the reparametrization mode formalism~\cite{Cotler2018}, we use the Liouville approach for our analysis. Though these calculations
operate within the general framework of Liouville theory as developed before,
their detailed execution requires a sequence of multiple analytic continuations.
Casual readers are invited to skip this construction and directly turn to our
final result Eq.~\eqref{eq:OTOCHL-approx}.

For a general field theory, the OTOC is defined  as \cite{Maldacena2015}
\begin{align}
    \tilde{F}(t,x)=\Tr[y V y W(t,x) y V y W(t,x) ] \label{eq:OTOC-definition}\,, \quad y^4=\frac{1}{Z}e^{-\beta H},
\end{align}
where  $t=-i \tau$ is real time and $V\equiv V(0,0)$ and $W(\tau,x)$
are  local operators. This function can be rewritten as
\cite{Bagrets2017}
\begin{align}
     \tilde{F}(t,x)=\eval{\Tr[y^4 T_{\tau} V(\tau_1,0) V(\tau_2,0) W(\tau_3,x) W(\tau_4,x) ]}_{\tau\to i t} \,, \label{eq:OTOC-2nd-form}
\end{align}
where we analytically continued to imaginary times
\begin{align}
    \tau_1=\frac{3 \beta}{4}, \quad \tau_3=\frac{\beta}{2}+\tau, \quad \tau_2=\frac{\beta}{4}, \quad \tau_4=\tau\,, \qquad \tau_1>\tau_3>\tau_2> \tau_4\,, \label{eq:sigma-otoc}
\end{align}
and  exchanged the order of the second and the third operator in the previous line. For our calculation, we take
\begin{align}
    V(\tau,0)=\gamma_\mathbf{i}^p(\tau,0) \equiv\prod_{k=1}^p \gamma_{i_k}(\tau,0), \qquad \mathbf{i}=\{i_1,.., i_p \}, \qquad W(\tau,x)= \gamma_j(\tau,x),
\end{align}
with $p=\mathcal{O}(C)$, where we sum over indices appearing twice in the correlation function and neglected the time increments in the definition of $V$ as compared to~\eqref{eq:Cluster-full}. Further, to simplify the calculation, one regularizes this object as
\begin{align}
     F(t,x)=\eval{\frac{\sum_{ \mathbf{i},j} \Tr[y^4 T_{\tau} \gamma_ \mathbf{i}^p(\tau_1,0) \gamma_ \mathbf{i}^p(\tau_2,0) \gamma_j(\tau_3,x) \gamma_j(\tau_4,x) ]}{\sum_{ \mathbf{i}} \Tr[y^4 T_{\tau} \gamma_ \mathbf{i}^p(\tau_1,0) \gamma_ \mathbf{i}^p(\tau_2,0)] \sum_{j} \Tr[y^4 T_{\tau} \gamma_j(\tau_3,x) \gamma_j(\tau_4,x)]}}_{\tau\to i t} \,. \label{eq:OTOC-Majoranas}
\end{align}
\begin{figure}[t]
    \centering
    \includegraphics[width=0.65\linewidth]{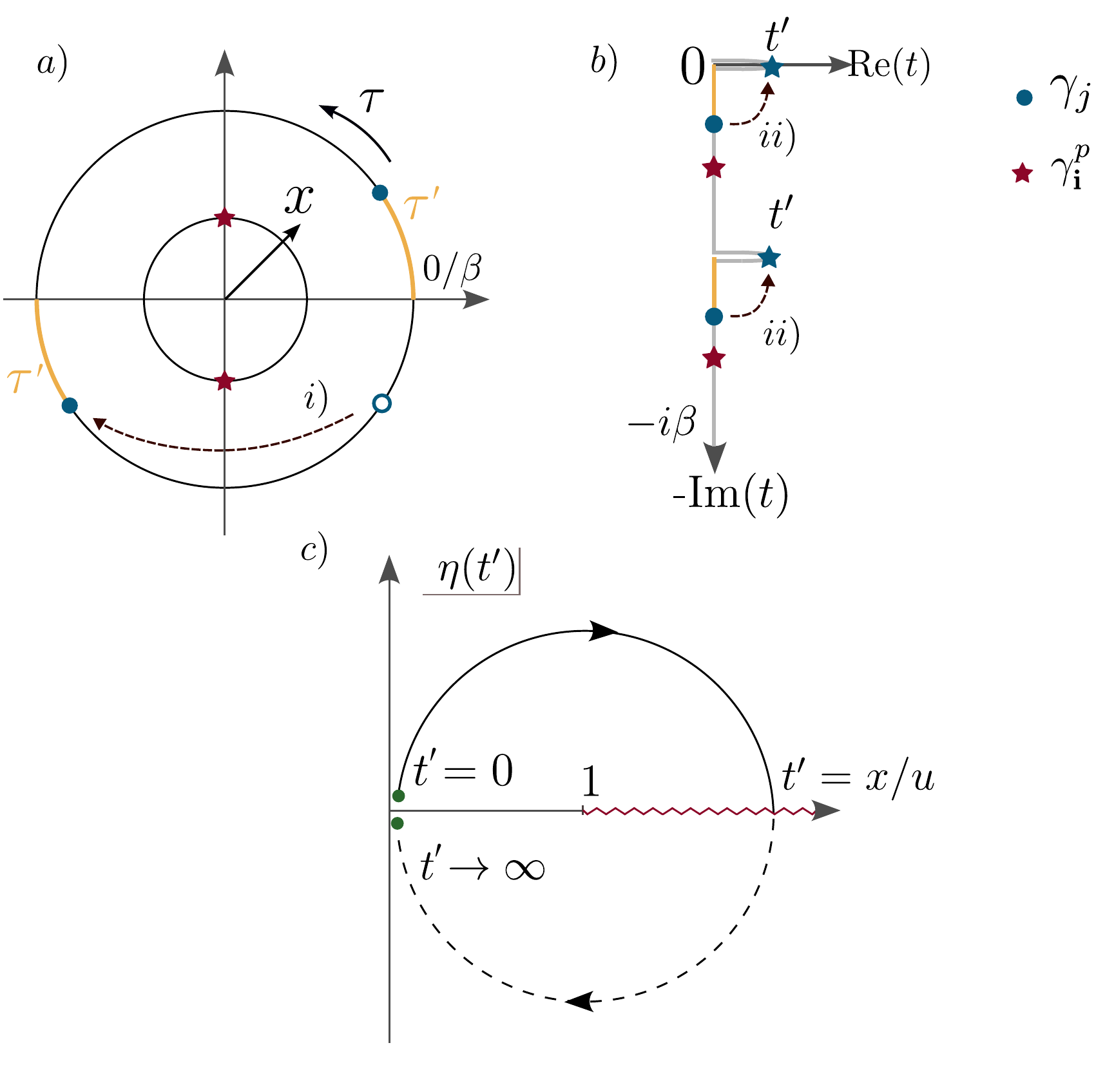}
    \caption{Configuration of operators in the OTOC calculation (time argument of the OTOC is denoted by $\tau' /t'$): In $a)$ the original mirrored configuration is broken via the move $i)$ to bring the operators into the configuration~\eqref{eq:sigma-otoc}. In $b)$ the complex time-contour of the correlation function after the analytic continuation $\tau' \to it'$ by the move $ii)$ is shown. In $c)$, the path of the cross-ratio $\eta(t')$ is shown, where the segments on the first and on the second Riemann sheet are indicated by full and dashed lines respectively.}
    \label{fig:OTOC-times}
\end{figure}
The question now is how to transfer this object to the Liouville framework.
While the times $\tau_1$ and $\tau_2$ in the configuration~\eqref{eq:sigma-otoc}
are mirrored on the upper half-plane and $V$ hence realizes a heavy operator as
defined in Eq.~\eqref{eq:Cluster-full}, this is not true for the times
$\tau_3$ and $\tau_4$ corresponding to $W$. Fortunately, as described in
\cite{Lam2018}, this does not pose a problem since the non-mirrored
configuration can always be obtained from the mirrored one by an analytical
continuation. As illustrated in Figure~\ref{fig:OTOC-times} $a)$, on the plane
with coordinates $z=e^{2\pi w \beta^{-1} }$ with $w=i \tau+x$, one starts from
$w_3'=\wb_4=-i \tau+x$ and then continues $w_3' \mapsto w_3=i (\tau+\beta/2)+x$.
In a next step, one further continues to the mixed time OTOC configuration $\tau
\mapsto it$ via the move $ii)$ shown in Figure~\ref{fig:OTOC-times} $b)$.
Within the saddle-point approximation employed here, the possibility of such an analytical 
continuation is ensured by the analytic dependence of the Liouville field~\eqref{eq:Strip-general-solution} 
on the complex coordinates $(w, \overline{w})$, as discussed in more details in Appendix~\ref{sec:L_field_strip}.

To implement these two steps in our context, we start from the  two-point function on the strip
\begin{align} \label{eq:normalized4pt}
  g_4(w_2,\wb_2,w_4,\wb_4) \equiv \frac{\expval{\V{l_1}(w_4,\wb_4) \V{l_2}(w_2,\wb_2)}}{\expval{\V{l_1}(w_4,\wb_4)}\expval{ \V{l_2}(w_2,\wb_2)}},
\end{align}
where the conformal weights now scale like $l_1 = \mathcal{O}(1)$ and $l_2 =
\mathcal{O}(C)$ as in \eqref{eq:Corr-general} and we made the $\wb$-dependence
explicit. Also note that the normalization of the four-point function cancels against the
conformal transformation factors~\eqref{eq:Liouville-Trafo} when moving between
coordinate systems, i.e. we can switch freely between $\xi$- and
$w$-coordinates. By~\eqref{eq:Corr-general}, the logarithm of the full
correlation function is then given by
\begin{align}
 \ln g_4=l_1(\phi'_1(\xi_4;l_2,\xi_2)-\phi_0'(\xi_4)),
\end{align}
where we understand all disk-coordinates appearing as functions of the
strip-coordinates $\xi=\xi(w)$. Using  Eqs.
$\eqref{eq:LivOneSourceZero}$ and $\eqref{eq:h-def}$ for the one-source solution
and also taking the limit $\delta \to 0$ for $\phi_0'$, we obtain
\begin{align}
   l_1(\phi'_1(\xi_4;l_2,\xi_2)-\phi_0'(\xi_4))&=l_1 \ln(\frac{(1-2\delta)^2(1-\abs{h}^2)^2\abs{h}^{(-4\delta)}}{\qty(1-\abs{h}^{2-4\delta})^2}) \nonumber\\
    &=2 l_1 \ln(\frac{(1-2\delta) \eta (1-\eta)^{-\delta}}{1-(1-\eta)^{1-2\delta}})\,, \label{eq:HHLL-exponent}
\end{align}
where we used again the definition $\delta=6l_2/C$, $h=h(\xi_4; \xi_2)$ and introduced the conformal cross-ratio $\eta=1-\abs{h(\xi_4;\xi_2)}^2 $. 
This object is now exactly the logarithm of the HHLL vacuum conformal block of a CFT, encoding interactions due to pure gravity \cite{Fitzpatrick2015, Fitzpatrick2016}. We now perform the move $i)$ $\xib_4 \mapsto \xi_3$, which amounts to inserting the times~\eqref{eq:sigma-otoc} and the corresponding positions into the conformal cross-ratio. From here, we perform move $ii)$ $\tau \mapsto i t$ and pick $x>0$ to observe $\expval{ \V{l_1}(w_4,\wb_4)}$ crossing the light-cone of $\expval{ \V{l_2}(w_2,\wb_2)}$ at $t=x/u$. The cross-ratio can then be simplified to
\begin{align}
    \eta(t)= \frac{2i}{i-\sinh{\frac{2 \pi}{\beta}(t-x/u)}}\,. \label{eq:crossratiofull}
\end{align}
As a function of $t$, it traces out a circle in the complex plane centered around $\eta=1$, as illustrated in Fig.~\ref{fig:OTOC-times} $c)$: It starts at $\eta \approx 0$, moves in the first quadrant to $\eta=2$ at $t=x/u$ and then traces out the lower half in the fourth quadrant on the way back to $\eta \approx 0$ for $t \gg x/u$. When hitting the light-cone at $t=x/u$, the cross-ratio crosses the branch cut of the root in the denominator of~\eqref{eq:HHLL-exponent} and hence is defined on the second Riemann sheet of the function. There, the behavior of the correlation function changes qualitatively compared to that of a time-ordered correlation function. The OTOC is for all times well-approximated by the function \cite{Roberts2014, Fitzpatrick2016}
\begin{align}
F(t,x)= \eval{g_4(w_2,\wb_2,w_4,\wb_4)}_{\text{OTOC}} \simeq \frac{1}{\qty(1+\frac{6 \pi l_2 }{C}e^{\frac{2\pi}{\beta}(t-x/u)})}\,. \label{eq:OTOCHL-approx}
\end{align}
From here we can read off the maximal
Lyapunov exponent $\lambda={2\pi}/{\beta}$ and the butterfly velocity $v_B=u$.  
While the first result agrees with previous calculations of the OTOC in holographic CFTs in 
the same limit of parameters \cite{Fitzpatrick2016, Cotler2017} and reflects the universality
mentioned before~\cite{Blake2018, Blake20182}, the second one is more non-trivial. 
The identification $v_B=u$, see Eqs.~\eqref{eq:Cu_high_T} and ~\eqref{eq:u_T},  
implies that $v_B$ is approximately independent of $x$ and $t$, which needs to be 
contrasted to the more complex form of $v_B(x,t)$ found in the chiral variant of the SYK model proposed in Ref.~\cite{Lian2019}. Also, 
in our model $v_B$ shows a weak $J$-dependence, as compared, e.g., to the random hopping models of \cite{JianYao17} and \cite{Cai2018}, where $\lim_{J/V \to \infty}v_B(J/V)=0$ holds with $V$ being the average hopping, i.e. strong interactions suppress information scrambling\footnote{Both models posses two independent hopping constants $V$ and $V'$, which we assume to be equal in this discussion.}.
In essence, the functional form of the OTOC says that in a strongly
chaotic setting with large $\lambda \propto T$ operators separated by a distance
$x>0$ remain uncorrelated, i.e. $C(x,t)\equiv1-F(x,t)\approx 0$, for times  $t<x/u$. At larger times, the
OTOC quickly reaches its saturation value $C(t,x)\approx 1$, which in our
normalization means that the two operators in question become correlated. A
new insight gained from the present analysis is that we explicitly constructed
operators of the microscopic theory, namely Majorana clusters, which map to vertex
operators in the universal Liouville variables. These operators serve as
probes into the chaotic behavior of the underlying chiral SYK model and can also be used as building blocks for more general correlation functions.

\section{Summary and discussion} \label{sec:discussion} 

The quest for a boundary theory of pure AdS$_3$ gravity is a longstanding
problem. A crucial step towards a better understanding was the identification of the 
AS action as a proposed theory of fluctuations around gravitational saddle
points~\cite{Cotler2018}. The structural similarity between the AS action and
the Schwarzian action in one dimension lower provided  strong evidence for the
existence of a JT/SYK-like duality between gravity and a dimensionally extended
variant of a model in the SYK universality class. In this work, we have
constructed this theory and its holographic correspondence to the bulk at a
level of explicitness previously realized in one dimension lower. Adapting the
rationale of the two-dimensional holographic correspondence, our construction
identified the AS action as the common low energy theory of a chain of SYK
islands coupled to realize a particular type of chiral dispersion, and the BTZ
black hole saddle of AdS$_3$ gravity. The construction of this theory from 
`microscopically defined' parent theories
allowed us to match coupling constants
and to compute the concrete form of correlation functions, 
notably out-of-time-order correlation functions as witnesses of early 
time chaotic dynamics. 

On the boundary, our starting point was a chain of  SYK dots coupled in such a
way that their Hamiltonian included a one body term with uniformly 
`chiral'
dispersion $\partial_k \epsilon(k)\ge 0$. Microscopically, such conditions
cannot be realized in stand-alone quasi one-dimensional lattices,  discrete
symmetries excluding a uniformly increasing dispersion. However, systems
exhibiting all signatures required by our construction --- strong randomness,
interactions, and chirality --- are realizable at topological insulator
boundaries, e.g., the edges of quantum Hall droplets. While our discussion has
not been focusing on aspects of concrete experimental feasibility, a point to
emphasize is that the microscopic one-dimensional boundary theory underlying our
construction is realizable in principle within the framework of condensed matter
physics. Similarly to the situation in one dimension lower, it exhibits an
infinite dimensional symmetry under local reparameterizations of time,
spontaneously broken at the mean level to a residual $\textrm{SL}(2,\Bbb{R})$,
and explicitly broken by time derivatives and inter-site coupling. A lowest
order gradient expansion in these symmetry breaking operators then produced the
AS action, where the condition of exactly retained local
$\textrm{SL}(2,\Bbb{R})$ symmetry implied stringent conditions on the allowed
contents of that action.   

Turning to the gravity side, our goal was to derive a matching AS action, now as the boundary fluctuation theory of BTZ black hole solutions. 
Indeed, earlier constructions in the literature \cite{Cotler2018, Mertens2022,Henneaux2020} 
had identified the AS action as a fluctuation theory common to a variety of gravitational settings, using different methods. 
For our set-up, we 
opted to follow the well-established principal strategy of \cite{Cotler2018}, but applied beyond the global AdS$_3$ case. 
We started from the Chern-Simons representation of three-dimensional gravity, considered it for field configurations representing fluctuations around a BTZ black hole, and from there projected to the boundary in Euclidean signature. 
The resulting AS action could then be compared to that following from the SYK construction, to obtain an identification matching microscopic system parameters (basically, $C = 3l/2G$). 
The net result of this construction is a bulk-boundary correspondence equaling that in one
dimension lower in its level of explicitness (In fact, a straightforward 
dimensional reduction reduces to the familiar \cite{Maldacena2016,Mertens162} 
SYK/JT duality in terms of the
Schwarzian action \cite{Mertens2018}.)

As in the two-dimensional case, the present theory describes a holographic
duality at time scales comparable to $N$, the number of local constituents of 
the SYK dots, or `singly perturbative' time scales in the parlance of gravity.
The two-dimensional holographic correspondence is also established via a second
bridge, building on chaotic fluctuations on much later time scales, $\sim e^{N}$,
comparable to the Hilbert space dimension of the SYK dots. This link has been established in terms of matrix theory
series representations (`singly non-perturbative'), and elements of topological
string theory (`doubly non-perturbative'). While elements of random matrix
theory have been identified for three-dimensional gravity in Ref. \cite{Cotler2021}, there remains the challenge to establish a corresponding link to a microscopic boundary theory. 

\section*{Acknowledgments}
We thank Thomas Mertens for valuable discussions. KW also thanks Suting Zhao for useful conversations. This work has been supported by the Deutsche Forschungsgemeinschaft (DFG, German Research Foundation) within CRC network TR 183 (Project Grant No. 277101999) as part of subproject B04.

\begin{appendix}
\numberwithin{equation}{section}

\section{Shklovskii's model of IQHE egde modes}
\label{app:Soft_IQHE_edge}

We summarize below the main results of Shklovskii's approach~\cite{Chklovskii:1992, Chklovskii:1993} 
to the formation of edge states at the boundary of 2DEG in the IQHE regime
in the situation of the so-called 'smooth' edge, i.e. when an electron density at the boundary gradually approaches to zero
on a spatial scale which is much larger than the magnetic length $l_B$. Under such condition 
and when effects of Coulomb interaction are included on a level of Hartree-Fock approximation the spatial profile of Landau levels (LL) close
the edge happens to be of the form shown in Fig.~\ref{fig:Flat_band_model}, which is characterized by a wide (almost) flat region 
and the power-law exponent $\gamma=3/2$. 
\begin{figure}[t]
	\begin{center}
		\includegraphics[width=0.65\columnwidth]{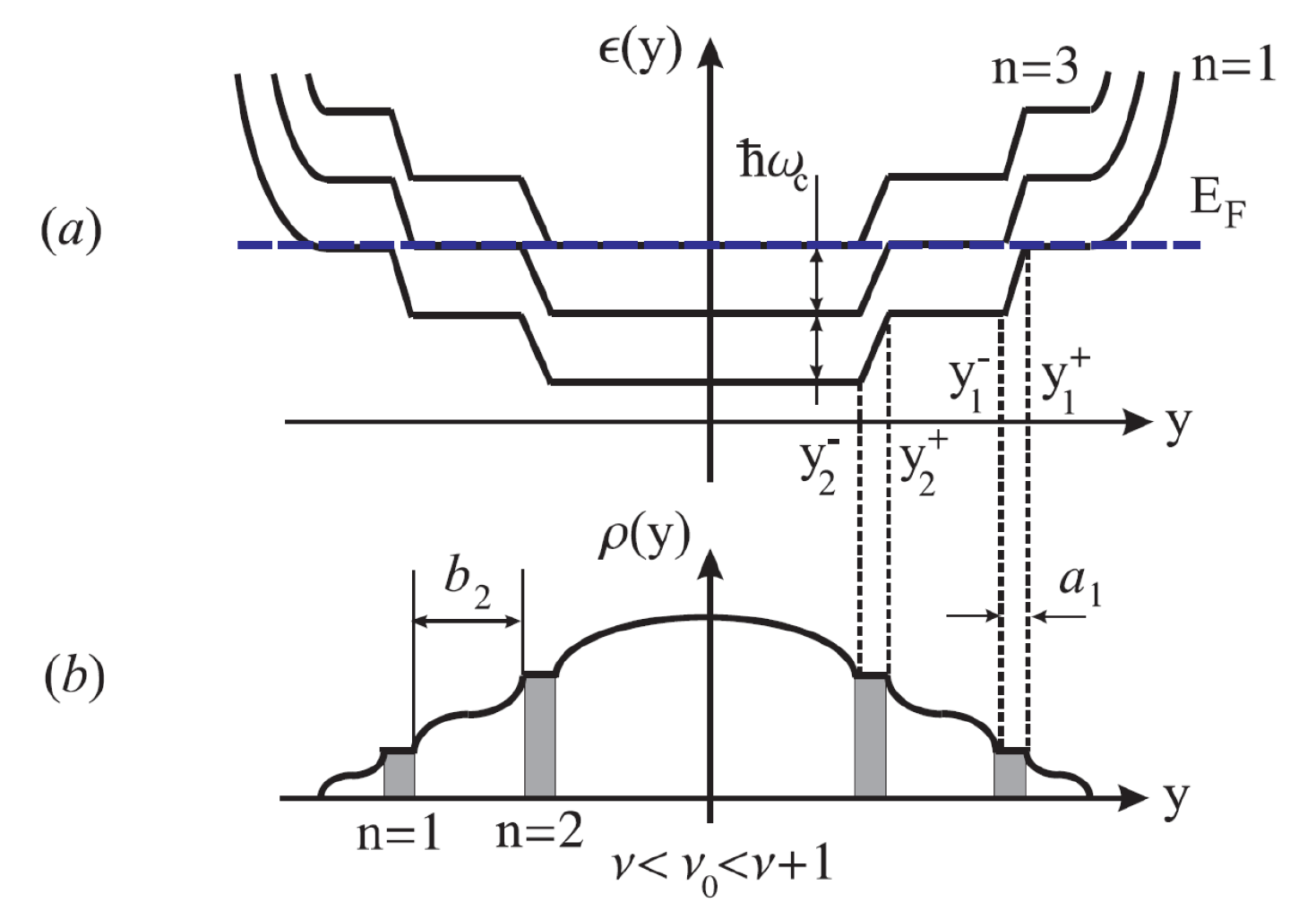}
		\caption{The structure of edge states in IQHE (taken from Ref.~\cite{Ngo_Dinh:2012}).
             A cross-section through the center of a narrow 2DEG strip with a smooth gate confined electron density:
			(a) Landau level energies as the function of a transverse coordinate $y$ and 
			(b) electron density $\rho(y)$.
			In addition to $\nu$ completely filled LLs, the central region can sustain a partially filled Landau level. 
			We define $a_k = y_k^+ - y_k^-$ as the width of the $k$-th incompressible strip (shown in grey), 
			which has the integer filling factor $k$. 
			The compressible regions (shown in white) correspond to areas with non-integer filling fraction and have the width $b_k = y_{k-1}^- - y_{k}^+$. The illustration above corresponds to the case $\nu=2$.  
		}
		\label{fig:IQHE_edge}
	\end{center}
\end{figure}
In more concrete terms, the 2DEG in the QHE regime is divided into compressible and incompressible strips. The filling
factor in the $n$-th incompressible strip is integer. These strips are separated by much wider regions of
compressible Hall liquid with a non-integer filling factor (compressible strips). In Fig.~\ref{fig:IQHE_edge} we show
a corresponding sketch of a typical electron density profile $\rho(y)$ corresponding to the narrow in $y$-direction and infinite along $x$-direction
strip of a gate confined 2DEG so that both left and right propagating edge modes emerge on respective boundaries.
Let us denote by $y_k^{\pm}$ the boundaries between compressible and incompressible regions. 
Then $a_k = y_k^+ - y_k^-$ is the width of the $k$-th incompressible strip, while $b_k = y_{k-1}^- - y_k^+$
is the width of the $k$-th compressible one. In the situation of gate-induced
confinement of 2DEG in the QHE regime the widths $b_k \gg a_B$, with $a_B = \epsilon/(m_{\rm eff} e^2)$ being the effective
Bohr radius. At the same time $a_k$ scales as  $a_k \sim \left(b_k a_B\right)^{1/2}$, so that
in general the condition $b_k \gg a_k \gg a_B$ is satisfied. 
In this picture compressible regions play the role of edge channels --- the self-consistent electrostatic potential
is constant through the compressible strips and can be controlled by connecting them to external leads.
Exact positions $y_k^\pm$ and widths of those strips were found in Refs.~\cite{Chklovskii:1992, Chklovskii:1993}.

It is also worth mentioning here another important spatial scale in QHE --- the magnetic length $l_B = (c/eB)^{1/2}$.
In the IQHE regime we have $l_B \gg a_B$. This condition can be also written as
\begin{equation}
	\hbar \omega_c = \frac{e B}{m c} \ll \frac{e^2}{\epsilon l_B},
\end{equation}
i.e. the distance between the Landau levels is smaller that a typical Coulomb energy on the magnetic length scale. 
Thus the Bohr radius $a_B$ plays a role of the short distance cut-off. 

One of the main results of the Shklovskii's theory are the following singularities in the spatial profiles of electron density $\rho(y)$ and
electron energies close to the strip boundaries $y_k^\pm$. Namely, a density profile shows square-root-like behavior,
\begin{equation}
	\rho(y) \sim \mp |y-y_k^\pm|^{1/2} + k n_L, \qquad  y \to y_k^\pm \pm 0, \qquad n_L = 1/(2\pi l_B^2),
\end{equation}
(here $n_L$ is an electron density on the 1st LL), while the corresponding energy profile of the k-th and (k+1)-th Landau levels close the Fermi energy reads as
\begin{equation}
	\label{eq:e_y_32}
	\epsilon(y) \sim E_F  \pm |y-y_k^\pm|^{3/2}, \qquad  y \to y_k^\pm \mp 0,
\end{equation}  
and we have omitted dimensionful prefactors in front of singular terms for brevity (also, the sketch of LLs in Fig.~\ref{fig:IQHE_edge} doesn't reflect the exponent $3/2$ properly).

Now, it is well-known that for two-dimensional Landau problem one can relate the momentum and position dependencies of eigenstates. Namely, if we fix the
gauge to be $A_x = - B_z y$ and $A_y=A_z=0$, then in the absence of disorder a momentum $k_x$ along $x$-axis is a good quantum number, thus
the wave function $\psi_{n k_x}(x,y)$ is characterized by the continuous momentum $k_x$ and a discrete Landau level index $n$.
In this gauge and for large $k_x \gg 1/l_B$ the wave function forms a narrow thread extended along $x$-axis but localized on a scale $l_B$ in $y$-direction and centered around
\begin{equation}
	y_{k_x} = k_x l_B^2.
\end{equation} 
When the self-consistent electrostatic potential changes adiabatically on a scale $l_B$, such that even an incompressible strip is sufficiently wide, $a_k \gg l_B$, 
the spectrum of Landau problem reads
\begin{equation}
	\epsilon_{n k_x} = \epsilon_n(y_{k_x}) \equiv \epsilon_n( k_x l_B^2),
\end{equation}
where energies $\epsilon_n(y)$ are those shown in Fig.~\ref{fig:IQHE_edge}. On taking into account the relation~(\ref{eq:e_y_32}) one
reproduces the model of the flat band in Fig.~\ref{fig:Flat_band_model} with the exponent $\gamma=3/2$. In particular, for a given compressible strip of width $b_k$ the corresponding momentum $k_0$ becomes $k_0 = b_k/(2l_B^2)$.

\section{Solution of mean-field equations in the IR limit} 
\label{app:mean-field}
This Appendix provides technical details for the derivation of equations~(\ref{eq:S_E}) and (\ref{eq:GS_t}).
They follow from the solution of the mean-field equation for the 'flat-band' model in the strong dispersive
with the use of a perturbation theory w.r.t. a small parameter $k_1/k_0 \ll 1$.

\subsection{Perturbative scheme}

We solve the system of mean-field equations~(\ref{eq:MF}) by neglecting the $i\epsilon$- term in the 1st equation. This approximation will be justified a posteriori. A simple evaluation of the integral (\ref{eq:MF_GF}) yields
\begin{eqnarray}
	\label{eq:G_xx}
	G_\epsilon &\simeq& -  \frac{1}{2\pi}\int\limits_{-\pi/a}^{\pi/a} \frac{dk}{\epsilon_k+ \Sigma_\epsilon}
	\simeq -\frac{k_0}{\pi\Sigma_\epsilon}  +  \frac{1}{\pi} \int_0^{+\infty} \frac{\Sigma_\epsilon \,dk}{\Lambda^2 (k/k_1)^{2\gamma} + [i\Sigma_\epsilon]^2}  \nonumber\\ 
	&=& - \frac{k_0}{\pi\Sigma_\epsilon}\left( 1 + b_1(\gamma)
	\frac{k_1}{k_0}\left(\frac{i\pi\Sigma_\epsilon}{\Lambda}\right)^{1/\gamma} \right),
\end{eqnarray}
where we've introduced a constant $b_1(\gamma)$ defined by an integral
\begin{equation} \label{eq:b-def}
	b_1(\gamma)=  \frac{1}{\pi^{\gamma}} \int_0^{+\infty} \frac{dx}{x^{2\gamma}+1} = \frac{\pi^{1-1/\gamma}}{2\gamma \sin \frac{\pi}{2\gamma}}.
\end{equation}
It was used above that the integral in (\ref{eq:G_xx}) 
is dominated by momenta satisfying $k \ll k_1$ as long as 
$|\epsilon|\ll T_\Lambda $, thus the upper limit of integration can be extended to infinity. 
Because of $\Sigma_\epsilon$ is yet unknown function of $k_1/k_0$, the result (\ref{eq:G_xx}) needs to be understood as the starting point for the self-consistent calculation of both $G_\epsilon$ and $\Sigma_\epsilon$.

Taking the leading term in (\ref{eq:G_xx}) we get $G_\epsilon \Sigma_\epsilon= - k_0/\pi$. 
Together with the relation $\Sigma_{\tau} = (J^2/ k_0^{3})\, G_\tau^3$ 
it yields the zeroth order SYK-type solution~(\ref{eq:GS0_SYK_E}).
To find the 1st order correction to the SYK solution one may use the ansatz~\eqref{eq:S_E} 
for the energy dependence of $\Sigma_\epsilon$ with yet undefined coefficient $\sigma_1(\gamma)$. Its self-consistence will be verified later. Expanding~(\ref{eq:G_xx}) up to linear order in $k_1/k_0$,
the Green's function becomes
\begin{equation}
	\label{eq:G_E1}
	G_\epsilon = G^{0}_\epsilon\left[1 + \Bigl(b_1(\gamma) - \sigma_1(\gamma)\Bigr) \left(\frac{k_1}{k_0}\right) \left|\frac{\epsilon J}{\Lambda^2}\right|^{1/2\gamma}+ \dots \right]
\end{equation}
On defining the function $C(\Delta)$ as in~(\ref{eq:C_Delta}), one can further translate this expression into the time domain using the Fourier transform~\eqref{eq:FT},
\begin{equation}
	\label{eq:G_t1}
	G_\tau= G^{0}_\tau \left[ 1 + \Bigl(b_1(\gamma) - \sigma_1(\gamma)\Bigr) b_2(\gamma) \left(\frac{k_1}{k_0}\right) \left|\frac{J}{\Lambda^2\tau}\right|^{1/2\gamma}  +
	\dots\right]\,,
\end{equation}
where $b_2(\gamma) = {C(1-\Delta_1)}/{\sqrt{2\pi}}$ and $\Delta_1 =1/2 + 1/(2\gamma)$. 
At this stage we can use the self-consistent equation, $\Sigma_\tau = (J^2/ k_0^{3})\, G^3_\tau$, 
which yields the self-energy
\begin{equation}
	\label{eq:S_t1}
	\Sigma_\tau = \Sigma^{0}_\tau \left[ 1 + 3 \Bigl(b_1(\gamma) - \sigma_1(\gamma)\Bigr) b_2(\gamma) \left(\frac{k_1}{k_0}\right) \left|\frac{J}{\Lambda^2\tau}\right|^{1/2\gamma}  +
	\dots\right].
\end{equation}
The latter can be converted back to the energy domain using once again the relation ~(\ref{eq:FT}).
The final result reads
\begin{equation}
	\Sigma_\epsilon = \Sigma^{0}_\epsilon \left[ 1 + \Bigl(b_1(\gamma) - \sigma_1(\gamma)\Bigr) b_2(\gamma) b_3(\gamma) \left(\frac{k_1}{k_0}\right) 
	\left|\frac{\epsilon J}{\Lambda^2}\right|^{1/2\gamma}  +
	\dots\right],
\end{equation}
where $b_3(\gamma) = {3 C(1+\Delta_1)}/{(2\sqrt{2\pi})}$. 
The latter expression indeed matches the initial ansatz for $\Sigma_\epsilon$ in the 
form of equation~(\ref{eq:S_E}) and yields the linear equation for the unknown $\sigma_1(\gamma)$:
\begin{equation}
	\sigma_1(\gamma) = \Bigl(b_1(\gamma) - \sigma_1(\gamma)\Bigr) b_2(\gamma) b_3(\gamma),
\end{equation}
and its solution is given by~(\ref{eq:gs2}). 
After that the final expressions for coefficients $g_1(\gamma)$, $\tilde g_1(\gamma)$ and
$\tilde\sigma_1(\gamma)$ follow from equations~(\ref{eq:G_E1}), (\ref{eq:G_t1}) and (\ref{eq:S_t1}), respectively.
For example,
\begin{equation}
	g_1(\gamma) = b_1(\gamma) - \sigma_1(\gamma) = \frac{b_1(\gamma)}{1 + b_{2}(\gamma) b_{3}(\gamma)},
\end{equation}
in agreement with Eq.~(\ref{eq:gs1}). One also observes that dropping $i\epsilon$--term from the Dyson equation
was a legitimate assumption, since the found correction to the self-energy scales as $\sim |\epsilon|^{\Delta_1}$ with $\Delta_1 <1$. 

\subsection{Table of integrals and coefficients} 
\label{eq:Table}
The coefficients in a perturbative solution of the Dyson's equation can be expressed via 
the function
\begin{equation}
	\label{eq:C_Delta}
	C(\Delta) = 2 \cos\left(\frac{\pi\Delta}{2}\right)\Gamma(1-\Delta), \qquad C(\Delta)C(1-\Delta) = 2\pi,
\end{equation}
which specify the following direct Fourier transform,
\begin{equation}
	\label{eq:FT}
	\int_{-\infty}^{+\infty} \frac{{\rm sgn}(\tau)}{|\tau|^\Delta} e^{i\epsilon\tau}\, d\tau = i C(\Delta) {\rm sgn} (\epsilon)|\epsilon|^{\Delta-1}, 
\end{equation}	
and its inverse,
\begin{equation}
	\int_{-\infty}^{+\infty} i{\rm sgn} (\epsilon)|\epsilon|^{\Delta-1} e^{- i\epsilon\tau}
	\frac{d\epsilon}{2\pi} = \frac{C(1-\Delta)}{2\pi} \frac{{\rm sgn}(\tau)}{|\tau|^\Delta}.
\end{equation}
Let us further define  
$\Delta_1 = {1}/{2} + {1}/{(2\gamma)}$
to be the (twice) scaling dimension of the mobile Majoranas and introduce auxiliary functions
\begin{align}
	b_1(\gamma) &= \frac{\pi^{1-1/\gamma}}{2\gamma \sin \frac{\pi}{2\gamma}}, & b_2(\gamma) &= \frac{C(1-\Delta_1)}{\sqrt{2\pi}}, \nonumber \\
	b_3(\gamma) &= \frac{3 C(1+\Delta_1)}{2\sqrt{2\pi}}, &
	b_{23}(\gamma) &= b_2(\gamma) b_3(\gamma) \equiv \frac{3}{2\Delta_1}\tan\frac{\pi \Delta_1}{2}.  
\end{align}
Then $g$-functions, defining the asymptotic behavior of correlators $G_\epsilon$ and $G_\tau$ , read
\begin{equation}
	\label{eq:gs1}
	g_1(\gamma) = \frac{b_1(\gamma)}{1 + b_{23}(\gamma)}, \qquad \tilde g_1(\gamma) = \frac{b_1(\gamma) b_2(\gamma) }{1 + b_{23}(\gamma)}, 
\end{equation}
while $\sigma$-functions, quantifying subleading corrections to $\Sigma_\epsilon$ and $\Sigma_\tau$, evaluate to
\begin{equation}
	\label{eq:gs2}
	\sigma_1(\gamma) = \frac{b_1(\gamma) b_{23}(\gamma) }{1 + b_{23}(\gamma)}, \qquad 
	\tilde \sigma_1(\gamma) = 3 \tilde g_1(\gamma). 
\end{equation}
Here the monotonous functions $g_1(\gamma)$ and $\tilde g_1(\gamma)$ are equal to zero at $\gamma=1$ and saturate to $1/4$ at $\gamma \to \infty$.
The function $\sigma_1(\gamma)$ has limiting values $\sigma_1(1) = 1/2$ and $\sigma_1(+\infty) = 3/4$ and it displays a weak minimum around $\gamma =5/4$. 

\section{Reparametrizations as similarity transformations}
\label{App:Rep_S}
It is instructive to formulate the reparametrization of bilocal operators and, in particular, the
transformation~(\ref{eq:Mapping_M}) in a language of matrix algebra. Consider the set of time dependent
functions $g:S^1\to S^1, \tau\to g(\tau)\equiv g_\tau$ as a linear space with
scalar product $\bra{g}\ket{h}=\int d\tau\,g_\tau h_{\tau'}$, and orthonormal
basis $\{\ket{t}\}$, $\bra{\tau}\ket{\tau}=\delta_{\tau-\tau'}$. We may
interpret the reparametrization, $\tau\mapsto t = f(\tau)$, and its inverse, $t \mapsto \tau = F(t)$, 
as the introduction of a new basis $\{\ket{t}\}$, which is obtained from the old one through a
transformation matrix 
\begin{align}
	\bra{t}\ket{\tau}&\equiv S_{t,\tau}=\delta_{t-f(\tau)}\,{f'_\tau}^{1/2},\cr 
	\bra{\tau}\ket{t}&\equiv S^{-1}_{\tau,t}=\delta_{\tau-F(t)}\,{F'_t}^{1/2},
\end{align}
where the scaling factors multiplying the $\delta$-function make the transformation unitary. 
Indeed, one can check it by evaluating
\begin{equation}
   [S^{\dagger}]_{\tau,t}  =  S^*_{t,\tau} = \delta_{t-f(\tau)}\,{f'_\tau}^{1/2} = \delta_{\tau-F(t)}\,{F'_t}^{1/2} 
   = S^{-1}_{\tau,t},
\end{equation}
as required, where it was used that $f'_\tau F'_{f(\tau)}=1$. With this definition, the orthonormality of the new basis is established as 
\begin{align*}
	\bra{t}\ket{t'}=\int d\tau\, \bra{t}\ket{\tau}\bra{\tau}\ket{t'}=\int d\tau S_{t,\tau}S^{-1}_{\tau,t'} 
	=\int d\tau \delta_{t-f(\tau)}{f'_\tau}^{1/2} \delta_{\tau-F(t')}{F'_{t'}}^{1/2}=\delta_{t-t'}.
\end{align*}
Also note that the $f$-scaling factors commute with the transformation in the
sense that $F'_t S_{t,\tau}=S_{t,\tau}f'_\tau$. With these structures in place,
the time reparametrization can be understood as a change of basis. Specifically, one
can represent the mapping ${\cal M}_\Delta$ as a superposition of the unitary $S$ and 
a diagonal (in time) congruent transformations. To see it explicitly, one may evaluate
\begin{equation}
	\bra{\tau_1} \overline O \ket{\tau_2} = \left( S^{-1} \overline O S \right)_{\tau_1 \tau_2} = 
	{f'_{\tau_1}}^{\!\!\!1/2}\, \overline O_{t_1 t_2}\, {f'_{\tau_2}}^{\!\!\!1/2}, \qquad t_i = f(\tau_i),
\end{equation}
where the `matrix products' are defined as $(S^{-1} \overline O  S)_{\tau \tau'}=\int dt
dt' S^{-1}_{\tau,t}\overline O _{tt'}S_{t',\tau'}$. Then from definition of the mapping ${\cal M}_\Delta$
it follows that
\begin{equation}
   O_{\tau_1 \tau_2} = {f'_{\tau_1}}^{\!\!\!\frac{\Delta-1}{2}}\, (S^{-1} \overline O  S)_{\tau_1 \tau_2}\,
   {f'_{\tau_2}}^{\!\!\!\frac{\Delta-1}{2}} \equiv
   \bigl( {f'}^{\frac{\Delta-1}{2}} S^{-1} \overline O  S {f'}^{\frac{\Delta-1}{2}} \bigr)_{\tau_1 \tau_2}.
\end{equation}
Consequently, the inverse relation to above reads
\begin{equation}
	\overline O_{t_1 t_2} = 
	\bigl( {F'}^{\frac{1-\Delta}{2}} S O  S^{-1} {F'}^{\frac{1-\Delta}{2}} \bigr)_{t_1 t_2},
\end{equation}
which shows that in the language of matrix algebra 
${\cal M}_\Delta = {F'}^{\frac{1-\Delta}{2}} S$ so that one can write $\overline O = {\cal M}_\Delta  O {\cal M}_\Delta^T$, 
and thereby our proposition is proved.

\section{Gradient expansion of the $G\Sigma$--action}
\label{app:Gradient expansion}
In this Appendix we collect the technical details of our derivation of the AS action from the $G\Sigma$--functional, which was outlined in Sec.~\ref{sec:gradients_GSigma}.

\subsection{Vertex operators}
\label{App:Vertex_Operators}
We begin by deriving the vertex operator $\rho$, which is the image of the time derivative operator with matrix elements
$[\partial_\tau]_{\tau_1 \tau_2}=\delta'(\tau_1 - \tau_2)$. From the definition~(\ref{eq:Mapping_M}), it follows that one can express $\rho_{t_1 t_2} = {F'_1}^{3/4} \delta'(F_1-F_2) {F'_2}^{3/4}$,
which requires understanding of the kernel $ \delta'(F_1-F_2)$. 
By acting with this kernel on an arbitrary function of time $f(t)$, one can verify that 
\begin{equation}
	\label{eq:delta_prime}
	\delta'(F_1-F_2) = \frac{1}{F_1'}  \delta'(t_1-t_2) \frac{1}{F_2'}
\end{equation} 
holds. Hence, the latter gives us the matrix elements $\rho_{t_1 t_2}  =  {F_1'}^{-1/4} \delta'(t_1-t_2) {F_2'}^{-1/4}$.
One can further associate
$\delta'(t_1-t_2)$ with the kernel of a symmetrized differential operator $\tfrac 12 (\overrightarrow{\partial_t} - \overleftarrow{\partial_t})$,
whose action is defined similarly as in~(\ref{eq:bar_dtau}), but with $b=1$. By setting  $ a \equiv {F'}^{-1/4} = \sqrt{b}$, we finally find
\begin{equation}
	\rho = \frac 12 a_t ( \overrightarrow{\partial_t} a_t) -  \frac 12 (a_t \overleftarrow{\partial_t}) a_t = 
	\frac 12 ( b_t \overrightarrow{\partial_t} -  \overleftarrow{\partial_t} b_t),
\end{equation}
as was stated in Eq.~(\ref{eq:bar_dtau}).

To derive the second vertex operator, the heat current density $j$, one needs to analyze how the time-local kinetic energy operator ${\varepsilon}_{\tau_1 \tau_2}^{x_1 x_2} = \varepsilon_{x_1 x_2}\,\delta_{\tau_1 \tau_2}$ changes under a substitution $\tau_i \to F(t_i, x)$. Here
\begin{equation}
	 \varepsilon_{xx'} = \int_k e^{ i k (x-x')} \epsilon_k 
\end{equation}
is the position representation of the chiral dispersion relation. Further calculations below now become somewhat 
more involved. To start, we define the kinetic energy in the new time frame,
\begin{equation}
	{\cal E}_{t_1 t_2}^{x_1 x_2}  = \varepsilon_{x_1 x_2}\,\delta (F_{t_1,x_1}- F_{t_2,x_2}),
\end{equation}
and evaluate its Wigner symbol with respect to spatial coordinates in a first-order gradient expansion.
Using the integral representation for the $\delta$-function,
\begin{equation}
	{\cal E}_{t_1 t_2}^{x_1 x_2} = \frac{1}{2\pi}\int e^{i \alpha F_1} \varepsilon_{x_1 x_2} e^{- i \alpha F_2} \,d\alpha, \qquad  F_i \equiv F(t_i, x_i),
\end{equation} 
which properly orders the product of three operators, one can further apply the Moyal expansion to find,
\begin{eqnarray}
	{\cal E}_{t_1 t_2}(x,k) &=& \epsilon_k\, \delta(F_1- F_2) -  
	\frac{1}{4\pi}\int ( \partial_x F_1 + \partial_x F_2 ) {\partial_k \epsilon_k}\, e^{i \alpha (F_1  - F_2)} \alpha d\alpha + {\cal O}(\hbar^2) \nonumber \\
	&=& \frac{\epsilon_k}{F_1'} \delta_{t_1 t_2} +   \frac i 2 {\partial_k \epsilon_k}\,( \partial_x F_1 + \partial_x F_2 ) \delta'(F_1-F_2) +  {\cal O}(\hbar^2),
\end{eqnarray}  
where we have now redefined $F_i$ to be $F_i = F(t_i, x)$ with $x$ being a coordinate variable of the Wigner transform. At this stage one can simplify $\delta'(F_1-F_2)$ according to (\ref{eq:delta_prime}). Further, using the definition~(\ref{eq:Mapping_M_x}) with $\Delta = 3/2$, one finds for the heat current operator
\begin{eqnarray}
	\label{eq:bar_V}
	{j}_{t_1 t_2}(x,k) &=& {F_1'}^{3/4}{\cal E}_{t_1 t_2}(x,k)  {F_2'}^{3/4} 
	 = \epsilon_k {F_1'}^{1/2} \delta_{t_1 t_2} \nonumber \\
	 &+&\frac i 2 {\partial_k \epsilon_k} \times 
	{F_1'}^{-1/4} \bigl( \partial_x F_1 \,\delta'(t_1-t_2) + \delta'(t_1-t_2)\, \partial_x F_2 \bigr) {F_2'}^{-1/4}.
\end{eqnarray}
This expression can be further simplified. For that, we consider the kernel of an auxiliary operator,
\begin{equation}
	J_{t_1 t_2} = g_{t_1} h_{t_1} \delta'(t_1-t_2) g_{t_2} + g_{t_1}  \delta'(t_1-t_2)  h_{t_2} g_{t_2}, 
\end{equation}
where, as before, $g_t$ and $h_t$ are two arbitrary functions with periodic boundary conditions in time.
On substituting $\delta'(t_1-t_2) \to \tfrac 12 (\overrightarrow{\partial_t} - \overleftarrow{\partial_t})$, one finds (after some simplification) that $J_{t_1 t_2}$ is, in fact, the kernel of a simpler differential operator
\begin{equation}
	J = \frac 12 (g^2 h)  \overrightarrow{\partial_t} - \frac 12 \overleftarrow{\partial_t} (g^2 h).   
\end{equation}
In order to decode (\ref{eq:bar_V}), one further sets $g \to {F'_t}^{-1/4} = \sqrt{b_t}$ and $h \to \partial_x F$, which leads to
\begin{equation}
	j(x,k) = {\epsilon_k}/{b} + 
	\frac i 2 {\partial_k \epsilon_k} \times \bigl( b \partial_x F \, \overrightarrow{\partial_t} - 
	\overleftarrow{\partial_t} \, b \partial_x F  \bigr) \equiv j^0_k(x) + j^1_k(x).
\end{equation} 
In this way, we reproduce the vertex operator $j_k(x)$, defined by Eq.~(\ref{eq:j_k_x}) in the main text.

\subsection{Gradient correction to the Wigner symbol of the Green's function} 
\label{App:Gradient_correctiom_Moyal}

In this subsection, we identify temporal gradient corrections to the Wigner symbol $\overline{\cal G}_{k,\epsilon}$ 
of the exact propagator $\overline{\cal G}$, which is defined by the relation 
$\overline{\cal G} = -  (j^0 + \overline{\Sigma})^{-1}$, where $j^0_{k,\epsilon}(s,x) = \epsilon_k/b_{s,x}$ 
denotes the corresponding Wigner symbol of the current operator and
$\overline \Sigma_\epsilon(s) = \Sigma(\epsilon; \Lambda/b_s)$ refers to the one of the self-energy.

We first note that 
if $A_\epsilon(s)$ and $B_\epsilon(s)$ are the Wigner
symbols of two operators $\hat A_{t_1 t_2}$ and $\hat B_{t_1 t_2}$, then the
Weyl product $A_\epsilon(s)\star B_\epsilon(s)$ corresponds to the Wigner symbol of $(A
B)_{t_1 t_2}$. The same definition applies to the spatial domain.
Using this framework, the exact propagator can be defined via the relation
$\overline{\cal G}_{\epsilon,k}(s) \star h_{\epsilon,k}(s) = -1$, where $h_{\epsilon,k}(s)$ is 
the effective Hamiltonian~\eqref{eq:h_Wigner_Et}. To resolve this equation, note that the Weyl product 
admits the following formal representation,
\begin{eqnarray}
 (A \star B)(\epsilon,s) &=& A_\epsilon(s) e^{-\tfrac{i\hbar}{2}(\overleftarrow{\partial_s}
 \overrightarrow{\partial_\epsilon} - \overleftarrow{\partial_\epsilon}
 \overrightarrow{\partial_s})} B_\epsilon(s), 
\end{eqnarray}
which can be employed to construct its semiclassical expansion in powers of $\hbar$.
Our immediate goal is to determine gradient corrections in time, as the inclusion of spatial gradients will be addressed separately 
(see Appendix~\ref{app:1st-order_gradient_terms}). Note that in the leading approximation, used consistently throughout the derivation of the AS action, the self-energy $\overline \Sigma_\epsilon(s)$ can be approximated by its SYK limit~\eqref{eq:GS0_SYK_E} and is thus independent of the 'slow' time $s$. Accordingly, it is instructive to evaluate the following restricted Weyl product:
\begin{eqnarray}
    A_\epsilon \star \phi(s) \star B_\epsilon &=&   A_\epsilon  e^{\tfrac{i\hbar}{2}\overleftarrow{\partial_\epsilon}\,
 \overrightarrow{\partial_s}} \phi(s) e^{-\tfrac{i\hbar}{2}\overleftarrow{\partial_s}\,
 \overrightarrow{\partial_\epsilon}} B_\epsilon \\
 &=& A_\epsilon \phi(s) B_\epsilon - i \phi'(s)\, A_\epsilon \overleftrightarrow{\partial_\epsilon}  B_\epsilon  -
 \frac 12 \phi''(s)\,   A_\epsilon \overleftrightarrow{\Box}_\epsilon B_\epsilon + {\cal O}(\hbar^3), \nonumber
\end{eqnarray}
where the first order derivative is properly anti-symmetrized, $\overleftrightarrow{\partial_\epsilon} = \frac{1}{2}(\overrightarrow{\partial_\epsilon} - \overleftarrow {\partial_\epsilon})$, 
while the second order differential operator is defined by the relation
\begin{equation}
\label{eq:D_Alembert_operator_omega}
A_\epsilon \overleftrightarrow{\Box}_\epsilon\, B_\epsilon : = \frac{d^2 }{d\omega^2} \Bigl( A_{\epsilon + \omega/2} B_{\epsilon - \omega/2}\Bigr)\Bigl|_{\omega=0} = 
\frac 14 A''_{\epsilon} B_{\epsilon} + \frac 14 A_{\epsilon}  B''_{\epsilon} - \frac 12 A'_\epsilon B'_\epsilon. 
\end{equation}
The above considerations lead to the following simple rule: one may replace the Wigner symbol
$j^0_{k,\epsilon}(s,x)$ by the corresponding differential operator,
\begin{equation}
    j^0_{k,\epsilon}(s,x) \to   \frac{\epsilon_{k}}{b} + \delta \hat j^0_{k}, \qquad  \delta \hat j^0_{k} =
    - i \epsilon_{k} \left(\frac 1 b\right)' \overleftrightarrow{\partial_\epsilon}  - 
     \frac{\epsilon_{k}}{2}\left(\frac 1 b\right)'' \overleftrightarrow{\Box}_\epsilon + {\cal O}(\hbar^3),
\end{equation}
and approximately evaluate the Wigner symbol $\overline{\cal G}_{\epsilon,k}$ using a perturbation series
over the gradient correction $\hat \delta j^0_{k}$ as follows,
\begin{equation}
	\overline{\cal G}_{\epsilon,k} = \overline G_{\epsilon,k} + \overline G_{\epsilon,k}  \delta \hat j^0_{k}\overline G_{\epsilon,k}
	+ \overline G_{\epsilon,k} \delta \hat j^0_{k}\overline G_{\epsilon,k}  \delta \hat j^0_{k}\overline G_{\epsilon,k} + \dots,
\end{equation}
where $\overline G_{\epsilon,k} \simeq - (\epsilon_k/b + \Sigma_\epsilon^0)^{-1}$ is the zeroth order Green's function.
After straightforward simplifications, one finds:
\begin{eqnarray}
\label{eq:delta_G}
    \overline{\cal G}_{\epsilon,k}  &=& \overline G_{\epsilon,k}  + \delta  \overline{\cal G}_{\epsilon,k}  +   {\cal O}(\hbar^3), 
    \nonumber \\
    \delta  \overline{\cal G}_{\epsilon,k} &=&  - \frac{\epsilon_{k}}{2}\left(\frac 1 b\right)''  \overline G_{\epsilon,k}
    \overleftrightarrow{\Box}_\epsilon \overline G_{\epsilon,k} + \frac {\epsilon^2_{k}}{2} {\left(\frac 1 b\right)'}^2
    \left(\partial_\epsilon \overline G{}_{\epsilon,k} \overleftrightarrow{\partial_\epsilon} \,\overline G_{\epsilon,k}^2 \right).
\end{eqnarray}
The correction of order ${\cal O}(\hbar)$ is absent here, as previously noticed in Eq.~\eqref{eq:bar_G_h}.

\subsection{Reparametrization invariant part of the $G\Sigma$-action}
\label{App:Rep_invariance_S_star}

In this Appendix, we demonstrate that within the temperature range $T_\Lambda < T < J$, the action $S_0[\overline{G}, \overline{\Sigma}]$, defined by Eq.~\eqref{eq:S_star_f}, is reparametrization invariant up to gradient terms of order ${\cal O}(b'^2, bb'')$ and therefore does not contribute to the low-energy AS action. On the other hand, the explicit symmetry breaking introduced by the kinetic energy term  ($\epsilon_k$) in the microscopic Hamiltonian leads to the emergence of a non-local-in-time and relevant perturbation to the AS action in the infrared limit, i.e. at $T < T_\Lambda$. This observation is significant only in the case of {Model II}.
We then proceed to determine the conditions under which such a relevant perturbation can be considered small, ensuring that the AS action remains a valid description of the chiral SYK model over a broad range of temperatures $T_* < T< T_\Lambda$, 
with the infrared temperature cutoff $T_*$ identified below.

For future reference, we represent Eq.~\eqref{eq:S_star_f}
in the equivalent form:
\begin{equation}
\label{eq:S_star}
    S_*[\overline{G}, \overline{\Sigma}] =  \frac N 2 {\rm tr}\, \ln \overline{\cal G}
    - \frac N 2 I[\overline{G}, \overline{\Sigma}],
\end{equation}
where $\overline{\cal G} = - (j_0 + \overline{\Sigma})^{-1}$ is the exact Green's function of the effective Hamiltonian~\eqref{eq:D_bar}. For the convenience of the reader, we repeat the transformed Green's function and 
self-energy expressed in terms of their mean-field solutions~\eqref{eq:Eqs_MF}, which read
\begin{equation}
    \overline{G}^{xx}_{t_1 t_2} = G(t_1-t_2; \Lambda^x_{t_1 t_2}), \qquad
    \overline\Sigma^{xx'}_{t_1 t_2} = \Sigma(t_1-t_2; \Lambda^x_{t_1 t_2})\delta_{xx'},
\end{equation}
where the rescaled kinetic energy scale reads $\Lambda^x_{t_1 t_2} = \Lambda (b_1 b_2)^{-1/2}$,
and we abbreviated $b_i = b(t_i, x)$. We also introduce a momentum-resolved Green's function of the mean-field solution,
\begin{equation}
\label{eq:G_mf_Ek}
        G_{\epsilon, k}(\Lambda) = - \left[\epsilon_k + \Sigma(\epsilon; \Lambda)\right]^{-1}.
\end{equation}
It can be used to construct the approximate Wigner symbol of the propagator $\overline{\cal G}$
through the relation
\begin{equation}
\label{eq:cal_G_Moyal}
    \overline{\cal G}_{\epsilon, k}(\Lambda) =  G_{\epsilon, k}(\Lambda/b) + {\cal O}(\hbar^2), \qquad b=b(x,s)
\end{equation}
which is valid up to second-order gradient corrections. The latter were
identified in the previous section and are explicitly given by Eq.~\eqref{eq:delta_G}.

\subsubsection{Local approximation}

With the above definitions at hand, one can now check that the action~\eqref{eq:S_star} is $b$-independent
within a {\it local approximation} (possible corrections due to temporal gradients will be accounted for later). 
To this end, we set $\Lambda^x_{t_1 t_2} \to \Lambda/b_{s,x}$, where $s=(t_1+t_2)/2$ is the center of mass time
and approximate $\ln \, {\rm det}\,\overline{\cal G}$ using the Moyal symbol~\eqref{eq:cal_G_Moyal}. These two steps lead to the action,
\begin{equation}
\label{eq:S_star_Lambda}
    {S}_*(\Lambda) =  \frac{N}{2}  \int_{s,x} \int_{k,\epsilon} \ln\, {G}_{\epsilon, k}(\Lambda/b)  
    - \frac{ N J^2}{8 k_0^3} \int_{x, t_{1,2}} \!\!\! [\overline{G}^{xx}_{t_1 t_2}]^4 
    - \frac{N}{2}\int_{x,t_{1,2}} \!\!\! \overline{G}^{xx}_{t_1 t_2}  \overline{\Sigma}^{xx}_{t_2 t_1},
\end{equation}
which should be regarded as a function of the kinetic energy scale $\Lambda$ as indicated above.
For $\Lambda \to 0$, the action equals the $f$-independent constant, 
\begin{equation}
     S_*(0) = - (N/2)\, {\rm tr}\,\ln \Sigma^0 -(N/2) I[\Sigma^0, G^0],
\end{equation}
with $(\Sigma^0, G^0)$ being the SYK saddle-point~\eqref{eq:MeanFieldSingleGrain}. One can then find the
flow equation when $\Lambda$ changes:
\begin{equation}
\label{eq:dS_dLambda}
    \Lambda \partial_\Lambda S_*(\Lambda) = \frac{N}{2}\!\!\!\int\limits_{|\epsilon|<J/b^2}\!\!\!
    \frac{d\epsilon}{2\pi}\int_{s,x,k} (\epsilon_k/b) 
    {G}_{\epsilon, k}(\Lambda/b).
\end{equation}
where it was used that $\epsilon_k \propto \Lambda k^\gamma$.
During the derivation of the above relation, one can disregard partial derivatives $\partial_\Lambda  \overline{G}$ and $\partial_\Lambda  \overline{\Sigma}$, as within the local approximation, the pair $(\overline{G}, \overline{\Sigma})$ becomes the exact saddle-point of the action~\eqref{eq:S_star_Lambda}. We have also restricted the energy integral using a reparametrization dependent UV cut-off. This regularization scheme was previously used to derive the Schwarzian action from the SYK model~\cite{Bagrets2016}, and we will provide a few comments on it below. To proceed, note that the mean-field self-energy exhibits the following scaling property in the energy representation,
\begin{equation}
\Sigma(\epsilon;\Lambda/b) = b^{-1}\Sigma(\epsilon b^2; \Lambda),
\end{equation} 
which follows from Eq.~\eqref{eq:GS_b}. Hence, for the Green's function one can write the relation
\begin{equation}
\label{eq:Green_rescale}
G_{\epsilon, k}(\Lambda/b) = - b \left[ \epsilon_k + \Sigma(\epsilon b^2; \Lambda)\right]^{-1}
= b G_{\epsilon b^2, k}(\Lambda).
\end{equation} 
On further introducing a new energy integration variable, $\widetilde \epsilon = \epsilon b^2$, one arrives at 
the flow Eq.~\eqref{eq:dS_dLambda} in the $b$-independent form,
\begin{equation}
\label{eq:dS_dLambda_1}
    \Lambda \partial_\Lambda S_*(\Lambda) = \frac N 2 \int {d\tau dx} 
     \!\!\!\int\limits_{|\widetilde\epsilon\,|<J} 
    \frac{d\widetilde\epsilon dk}{(2\pi)^2} {\epsilon_k}{G_{\widetilde \epsilon, k}(\Lambda)},
\end{equation}
where it was used that $d\tau = F' ds = ds/b^2$. Hence, one concludes that $S_*(\Lambda)$ is reparametrization invariant as well. 

Let us now comment on the origin of the reparametrization-dependent cut-off. To this end, consider a typical operator bilocal in time 
with scaling dimension $\Delta$:
\begin{equation}
{O}^\Delta_{\tau_1 \tau_2}(x) = \left[\frac{f'_1 f_2'}{(f_1-f_2)^2} \right]^{\Delta/2} \to\quad
{\rm Re}\, \left[\frac{f'_1 f_2'}{(f_1-f_2 + i \delta_{12})^2} \right]^{\Delta/2}.
\end{equation}
The singular kernel here requires a regularization at short times, which can be achieved by shifting the time argument 
into the complex plane, as it is reflected in the expression above, with some cut-off scale $\delta_{12} \sim J^{-1}$.
Further on, one should require that in the UV limit, when $t_1\to
t_2$, the $f$-dependence drops out from ${O}_{\tau_1 \tau_2}(x)$ in the 
leading order (the next-to-leading term produces the Schwarzian), since reparametrizations, by definition,
are an IR effect. The latter suggests to define $\delta_{12} = (f'_1 f'_2)^{1/2}/J$.
In the local approximation, it becomes $\delta=b^2/J$, which also translates into the UV cut-off $J'=J/b^2$ in the energy
domain.

\subsubsection{Inclusion of gradient corrections}

We now address the fate of gradient corrections, appearing when one goes beyond the local approximation used above. These corrections can be categorized into two types.
The first type is related to the substitution $\Lambda^x_{t_1 t_2} \to \Lambda/b_{s,x}$, employed in the previous subsection. 
With $t_{1,2} = s \pm t/2$, this substitution introduces an error:
\begin{equation}
    \delta \Lambda = \Lambda^x_{t_1 t_2} - \Lambda/b_{s,x} = (\Lambda t^2)\times \frac{{b'}^2 - b b''}{ 8 b^3} + {\cal O}(t^4),
    \qquad b'=\partial_s b,
\end{equation}
which in turn results in the following variation to the action~(\ref{eq:S_star}), 
\begin{equation}
\label{eq:delta_S_star}
    \delta S_*[\overline{G}, {\overline \Sigma}] = {\rm Tr }\left( \frac{\delta  S_*}{\delta \Sigma}\Bigl|_{X_*} \partial_\Lambda \overline \Sigma + 
    \frac{\delta  S_*}{\delta  G}\Bigl|_{X_*} \partial_\Lambda \overline G \right) \delta \Lambda \to 0,
\end{equation}
where $X_*=(\overline G, \overline \Sigma)$ is the saddle point solution within the local approximation, namely
$\overline G^x_{t_1t_2} = G(t,\Lambda/b_{s,x})$, and the same applies to $\overline \Sigma^x_{t_1 t_2}$.
However, this variation vanishes because the mean-field solution $X_*$ is an exact saddle point of the
action~\eqref{eq:S_star} for an arbitrary reparametrization $b$; see also the discussion in Sec.~\ref{sec:Rep_Inv_Soft_Modes}.
It is worth noting that this observation parallels the remark made during the derivation of Eq.~\eqref{eq:dS_dLambda}. 
Consequently, up to terms of order ${\cal O}({b'}^2, bb'')$, there is no contribution to the low-energy action in this case.

Another type of gradient corrections, not yet included into our analysis,
arises when the action $S_L[b] = (N/2)\,{\rm tr}\ln \overline{\cal G}$ in Eq.~\eqref{eq:S_star}
is expanded in terms of the variation $\delta \overline{\cal G}$, which defines the exact Wigner symbol~\eqref{eq:delta_G}
of the Green's function. A straightforward implementation of this program encounters infrared divergences 
at temperature scales well below $T_\Lambda$, as the direct gradient expansion in terms of slow and 
fast degrees of freedom becomes unjustified in this limit. Since our analysis of {Model I} ---
with a linear dispersion $\epsilon_k \sim k$ --- is restricted to the high temperature range $T_\Lambda < T <J$, 
the discussion that follows is only relevant for {Model II}, which features flat bands.

When properly regularized, the logarithmic action $S_L[b]$ expanded over the gradient correction $\delta \overline{\cal G}[b]$
generates at small temperatures $T\ll T_\Lambda$ an infinite series of ${\rm SL}(2,\mathds{R})$-invariant and 
nonlocal-in-time operators. The simplest among them takes the following form:
\begin{equation}
\label{eq:bilocal_action}
  {\cal S}[f] \sim  - N k_1 T_\Lambda^{-1/2\gamma}  \int_{x,\tau_{1,2}} \!\!\!\!  
  \beta^\Delta {\cal O}^\Delta_{\tau_1\tau_2}(x), \qquad 
  \Delta = 2 + 1/2\gamma,
\end{equation}
with the time-bilocal operator defined by Eq.~\eqref{eq:bilocal}. The tilde symbol ($\sim$) indicates
that here and in what follows we omit all numerical constants of order unity. 
Note that the above action stems entirely from itinerant chiral Majorana modes, which explains its prefactor
that includes a small momentum $k_1$, see Fig.~\ref{fig:Flat_band_model}.
The overall coupling constant here has a positive scaling dimension, 
${\cal D} = 1 - 1/2\gamma \geq 1/2$ (we set $[x]=[\tau]=-1$), indicating the relevance of this perturbation to the AS action 
in the infra-red limit. We relegate a derivation of the result~\eqref{eq:bilocal_action} 
and its holographic interpretation to our subsequent communication and proceed below with formulating the conditions 
under which such a relevant perturbation can be considered small compared to the AS action.

Treating the bilocal operator in Eq.~\eqref{eq:bilocal_action}  as a perturbation, we can average it over quantum fluctuations governed by the AS action, $\langle {\cal O}^\Delta_{\tau_1\tau_2}(x) \rangle_{\rm AS}$, to obtain a small correction to the free energy. The technical procedure for calculating such averages was outlined in Sec.~\ref{sec:heavy}. We know that the engineering scaling dimension $\Delta$ undergoes transmutation to the quantum dimension $\Delta_Q = \Delta + 3 \Delta(1-\Delta)/C$.
As a result, the free energy per length acquires the non-analytic dependence on temperature,
\begin{equation}
\label{eq:free_energy}
   \frac{{\cal F}(T)}{L} = - \frac{\pi C}{12} \times  \left[\, \frac{T^2}{u_0} \ln \frac{J}{T}  + \frac{T^2_\Lambda}{u_1}
   \left( \frac{T}{T_\Lambda}\right)^\delta\,\right], \qquad \delta = \Delta_Q-1.
\end{equation}

Here, the first term arises from the Schwarzian action with a temperature dependent velocity~\eqref{eq:u_T},
and we have introduced a second velocity scale,
\begin{equation}
    u_1 \sim \frac{C T_\Lambda}{N k_1} \sim \frac{\Lambda^3}{J^2 k_1} \ln\frac{J}{T_\Lambda},
\end{equation}
to characterize the magnitude of the perturbative correction. Initially, the latter can be considered small,
provided the condition $u_0/u_1 \sim (J/T_\Lambda) (k_1/k_0) \ll 1$ is satisfied.
This ensures that at the crossover temperature scale $T_\Lambda$, the Schwarzian contribution to the free energy dominates.
Furthermore, based on physical grounds, the exponent $\delta$ must be positive, which imposes a lower bound on the 
acceptable central charge,
\begin{equation}
    C > C_* = 6 + 3/(2\gamma).
\end{equation}
On the other hand, the upper bound for the exponent, $\delta < 3/2$, holds by construction. 
As one can see from the derived relation~\eqref{eq:free_energy}, the perturbation theory breaks down
around the infra-red temperature scale $T_* \ll T_\Lambda$, which can be estimated from the
following implicit relation,
\begin{equation}
\label{eq:T_star}
    \frac{T_*}{T_\Lambda} = \left(\frac{u_0}{u_1} \ln^{-1} \frac{J}{T_*} \right)^{\frac{1}{2-\delta}}.
\end{equation}
which is found by approximately equalizing two different contributions to ${\cal F}(T)$.
At smaller temperatures, $T<T_*$, the AS action deformed by the relevant operator~\eqref{eq:bilocal_action}
calls for a non-perturbative treatment, possibly within the RG scheme similar to the 
one developed earlier for the model of granular SYK arrays~\cite{Altland:2019}. 

To summarize this subsection, we have verified that if a velocity ratio satisfies the condition
$u_0/u_1 \ll 1$, then a holographic duality between the chiral SYK model comprising flat bands and the AS
action may hold in the parametrically wide range of temperatures $T_* \ll T \ll J$.

\subsection{First-order gradient terms}
\label{app:1st-order_gradient_terms}
In this subsection, we are aiming to analyze linear in gradient terms,
\begin{equation}
    S_\rho[f] + S_j[f] = \frac N 2 {\rm tr} (\rho \overline{\cal G}) +  \frac N 2 {\rm tr} (j^1 \overline{\cal G}),
\end{equation}
which are obtained from the first-order expansion of the action~\eqref{eq:S_fl_f}. We start our calculations
from the action $S_\rho[f]$, which involves the energy density operator~\eqref{eq:bar_dtau}. Following
Pruisken, we double the power of the Green function in this action. To this end, one introduces the propagator
$\overline{\cal G}(\mu)$, 
which depends on an auxiliary energy $\mu$, so that one can write
\begin{equation}
    \overline{\cal G} = - \int_{-\infty}^0 d\mu\, \overline{\cal G}^2(\mu),
    \qquad \overline{\cal G}(\mu) = (\mu - j_0 - \overline{\Sigma})^{-1}.
\end{equation}
Based on the above identity, the action $S_\rho[f]$ can be equivalently rewritten as 
a sum of two (right and left) contributions:
\begin{equation}
    S_\rho[f] = {S}^R_\rho[f] + S^L_\rho[f] = 
    - \frac N 4 \int_{-\infty}^0 d\mu\, {\rm tr} \left[ \overline{\cal G}(\mu) 
    (b \overrightarrow{\partial_t} - \overleftarrow{\partial_t} b) \overline{\cal G}(\mu)\right],
\end{equation}
where the trace operation implies an integration over temporal- and spatial indices. 
From here we switch to the Wigner representation by writing
\begin{eqnarray}
S^R_\rho[f] &=&  -  \frac N 4 \int_{-\infty}^0 d\mu  
\int_{s,x} \int_{k,\epsilon} \, \overline{\cal G}_{\epsilon,k}(\mu) \star b_s \star
(-i\epsilon + \tfrac 1 2 \overrightarrow{\partial_s}) \overline{\cal G}_{\epsilon,k}(\mu), \\
S^L_\rho[f] &=&  - \frac N 4 \int_{-\infty}^0 d\mu  
\int_{s,x} \int_{k,\epsilon}  
\overline{\cal G}_{\epsilon,k}(\mu) (-i\epsilon - \tfrac 1 2 \overleftarrow{\partial_s}) \star b_s \star
\overline{\cal G}_{\epsilon,k}(\mu),
\end{eqnarray}
where it was used that time derivative operators, when acting in the Wigner representation, become
\begin{equation}
{\partial_{t_1}} =  -i\epsilon + \tfrac 1 2 {\partial_s}, \qquad
{\partial_{t_2}} = i\epsilon + \tfrac 1 2 {\partial_s}.
\end{equation}
In what follows we are aiming to employ the 1st order Moyal expansion to the
convolutions $\overline{\cal G}(\mu) \star b_s  \star  \partial_s\overline{\cal G} (\mu)$ and 
$\partial_s \overline{\cal G} \star b_s \star \overline{\cal G}(\mu)$ stemming, respectively, from the right- and left actions.
When doing so, one may substitute the exact Wigner symbol $\overline{\cal G}_{\epsilon,k}(\mu)$ by its local approximation, 
\begin{equation}
\label{eq:G_ek_rescaling}
    \overline G_{\epsilon,k}(\mu; \Lambda) \equiv  G_{\epsilon,k}(\mu; \Lambda/b) = \left[\mu - \epsilon_k/b - \Sigma(\epsilon; \Lambda/b)\right]^{-1},
\end{equation}
cf. Eqs.~\eqref{eq:G_mf_Ek} and \eqref{eq:cal_G_Moyal}. Similarly to Eq.~\eqref{eq:G_mf_Ek},
we keep the spatio-temporal dependence of $G_{\epsilon,k}(\mu; \Lambda)$ implicit. To proceed, we introduce the time derivative
\begin{equation}
	F(\mu) = \partial_s \overline{G}(\mu) = \overline{G}^2(\mu) \epsilon_k (1/b_s)',
\end{equation}
where further on all dependencies on $(k,\epsilon)$ and $(x,s)$ are omitted for brevity.
Then, the first-order Moyal expansion takes the form
\begin{equation}
\overline{\cal G}(\mu) \star b_s \star \partial_s \overline{\cal G}(\mu) =  \overline{\cal G}(\mu) b_s \partial_s \overline{\cal G}(\mu) + 
\frac i2 \left[\overline{G}(\mu) \partial_k  F(\mu) - F(\mu) \partial_k \overline{G}(\mu) \right]\partial_x b_s + {\cal O}(\hbar^2),
\end{equation}
with a similar expression being valid for the other convolution, where $\overline{\cal G}$ and its partial derivative are swapped. Taking into account that
the momentum derivative simplifies to
\begin{equation}
\partial_k\overline{G}(\mu) = \overline{G}^2(\mu) \partial_k \epsilon_k/b,
\end{equation}
we find the following gradient expansion for the action $S_\rho[f]$: 
\begin{equation}
\label{eq:S_rho_mu}
	S_\rho[f] = S_\rho^{\rm UV}[f] + \frac{i N}{8} \int_{-\infty}^0 d\mu \,  \int_{s,x} \int_{k,\epsilon} 
	\left[ \overline{G}^{\,3}(\mu) \partial_k \epsilon_k \times \frac{b'\partial_x b}{b^2} + 
	\overline{G}^{\,4}(\mu) \epsilon_k \partial_k \epsilon_k \times \frac{b'\partial_x b}{b^3}\right].
\end{equation}
Here, the first contribution $S_\rho^{\rm UV}[f]$ arises from the leading-order Moyal expansion. We will analyze it at the end of this subsection. For now, we focus on the second contribution, which involves a spatial gradient $\partial_x b$.
After integration over the auxiliary energy $\mu$, it becomes
\begin{equation}
\Delta S_\rho[f] = -  \frac{i N}{8}  \int_{k,\epsilon}  \int_{s,x}
\left[ {G}^2_{\epsilon,k}(\Lambda/b) \partial_k \epsilon_k \times \frac{b'\partial_x b}{2 b^2} + 
{G}^3_{\epsilon,k}(\Lambda/b) \epsilon_k \partial_k \epsilon_k \times \frac{b'\partial_x b}{3 b^3}\right],
\end{equation}
where the mean-field Green's function ${G}_{\epsilon,k}(\Lambda)$ was defined in Eq.~\eqref{eq:G_mf_Ek}.
At this stage, it is important to rescale the energy integration variable by setting $\epsilon = \widetilde \epsilon/b^2$.
Using the transformation law~\eqref{eq:Green_rescale} for the Green's function, one finds
that energy-momentum and space-time integrations are factorized, yielding
\begin{equation}
\label{eq:dS_tau}
\Delta S_\rho[f] = -  \frac{i N}{8}   \int_{k,\widetilde\epsilon} \partial_k \epsilon_k \left[
\frac 12 G^2_{\widetilde\epsilon,k}(\Lambda)   + \frac 13 G^3_{\widetilde\epsilon,k}(\Lambda) \,\epsilon_k \right] \times 
\int_{s,x} \frac{b'\partial_x b}{b^2}.
\end{equation}
To extract the leading logarithmic expression for the coupling constant from here, one observes that at large energy, 
the Green's function scales as $G_{\epsilon,k}(\Lambda) \sim |\epsilon|^{-1/2}$.
Hence, the first term in Eq.~(\ref{eq:dS_tau}) is a logarithmic integral, while the second one gives a subleading contribution.
We further note that in the relevant range of energies, 
$T_\Lambda \ll |\epsilon| \ll J$, the kinetic energy is negligible as compared to the self-energy, i.e.
$|\epsilon_k| < \Lambda \ll \Sigma_\epsilon^0$, where $\Sigma^{0}_{\epsilon} \propto \sqrt{|\epsilon|}$ 
is the SYK self-energy~\eqref{eq:GS0_SYK_E}.
Approximating the Green's function as $G_{\epsilon,k}(\Lambda) \simeq - (\Sigma^{0}_{\epsilon})^{-1}$, 
one then arrives at 
\begin{equation}
\label{eq:counter_term_S_rho}
\Delta S_\rho[f] \simeq - \frac {i N}{16} \int \frac{d\epsilon(k)}{(2\pi)^2} \!\!\! \int\limits_{T_\Lambda<|\widetilde\epsilon|<J} \frac{d\widetilde\epsilon}{\bigl(\Sigma^0_{\widetilde\epsilon}\bigr)^{2}}  \times \int_{s,x} \frac{b'\partial_x b}{b^2} = \frac{i C}{12 \pi} \int_{s,x} \frac{b'\partial_x b}{b^2},
\end{equation}
where $C$ is the central charge given by Eq.~\eqref{eq:Cu_high_T}. In what follows (see Subsection~\ref{App:second_order_expansion}), 
we will interpret the above contribution as a counter-term, which contributes to the 
imaginary part of the action~$S^{\rm AS}_\pm[b]$.

We now return to the first contribution in Eq.~\eqref{eq:S_rho_mu}. After the integration over $\mu$ it reduces to
the following simple expression,
\begin{equation}
\label{eq:St_int_xs_ek}
S_\rho^{\rm UV}[f] = - \frac {i N }{2} \int_{s,x} \int_{k,\epsilon} \, (b \epsilon)\,  {\cal G}_{\epsilon, k},
\end{equation}
which can be shown to be a $f$-independent constant. The way to demonstrate it follows the analysis of Sec.~\ref{App:Rep_invariance_S_star}.
We start by substituting the exact Wigner symbol ${\cal G}_{\epsilon, k}$ by its leading approximation $G_{\epsilon,k}(\Lambda/b)$.
For this choice, by restricting the energy integration to $|\epsilon|<J/b^2$ and introducing 
$\epsilon = \widetilde \epsilon/b^2$ as before, one employs the scaling property~\eqref{eq:G_ek_rescaling} 
of the Green's function to find:
\begin{equation}
\label{eq:S_rho_int_xs_ek}
S_\rho^{\rm UV}[f] = - \frac {iN}{2}\int \frac{ds dx}{b^2} \times \int_k\int_{{|\widetilde\epsilon|}<J}
\!\!\!\!\widetilde{\epsilon} \,G_{\widetilde{\epsilon}, k}(\Lambda)
\propto - \frac {iN}{2}\int {d\tau dx},
\end{equation}
where the relation $d\tau = F' ds = ds/b^2$ was used. Therefore, the spatio-temporal integral produces a $b$-independent result as it was claimed above.

We next show that a variation of the above result due to gradient corrections~\eqref{eq:delta_G} vanishes, i.e.
\begin{equation}
\label{eq:delta_St_int_xs_ek}
\int_{s,x} \int_{k,\epsilon} \, (b \epsilon)\,  \delta {\cal G}_{\epsilon, k} = 0,
\end{equation}
which follows from the analytic structure of the energy integral. For the sake of illustration, consider the
first term contributing to the gradient correction of the propagator,
\begin{equation}
\label{eq:delta_Wigner_1}
\delta  \overline{\cal G}^{(1)}_{\epsilon,k} =  - \frac{\epsilon_{k}}{2}\left(\frac 1 b\right)''  \overline G_{\epsilon,k}
    \overleftrightarrow{\Box}_\epsilon \overline G_{\epsilon,k}.
\end{equation}
On changing the energy integration variable to $\epsilon = \widetilde \epsilon/b^2$ and integrating by parts a  few times, we can transform~\eqref{eq:delta_St_int_xs_ek} to the following integral,
\begin{equation}
    {\cal J}^{(1)} \propto \int_{s,x} {b'}^2 \int_{\widetilde{\epsilon},k} \epsilon_k \widetilde{\epsilon}\,
    [\partial_{\widetilde{\epsilon}} G_{\widetilde{\epsilon}, k}(\Lambda)]^2,
\end{equation}
which nominally looks like a contribution to the real part of the AS action~\eqref{eq:S_RI_b}. 
We further employ our standard approximation by using the SYK result $\Sigma_\epsilon^0$ for the self-energy, 
and change the integration variable to $\widetilde{\epsilon} = {\rm sgn}(\lambda) \,\lambda^2/J$, which yeilds
\begin{equation}
    {\cal J}^{(1)} \propto \int_{s,x} {b'}^2 \int_{k} \epsilon_k \int_{-\infty}^{+\infty} G^4_{\lambda,k}  \,\lambda d\lambda\,  \rightarrow\, 0 , \qquad 
    -\, G_{\lambda,k} = \frac{1}{\epsilon_k -{i}(\lambda/\pi)}.
\end{equation}
For a given momentum $k$, the integral over $\lambda$ now evaluates to zero due to simple analytic properties of the Green's function $G_{\lambda,k}$. 
The analysis of the second gradient correction to the Wigner symbol $\overline{\cal G}_{\epsilon,k}$,
\begin{equation}
\label{eq:grad_corr_G_second}
 \delta  \overline{\cal G}^{(2)}_{\epsilon,k} =   \frac {\epsilon^2_{k}}{2} {\left(\frac 1 b\right)'}^2
    \left(\partial_\epsilon \overline G{}_{\epsilon,k} \overleftrightarrow{\partial_\epsilon} \,\overline G_{\epsilon,k}^2 \right),
\end{equation}
can be accomplished along the same lines as above, thereby substantiating the validity of Eq.~\eqref{eq:delta_St_int_xs_ek}.

We further explore the action $S_j[f]$ following the same principles. Since the current operator $j^1$ already contains
one spatial derivative, it is sufficient to use the local Moyal approximation. Taking into account the relation
between the two Wigner symbols, $j_{\epsilon,k}^1 = i(\partial_k \epsilon_k \partial_x F) \rho_{\epsilon,k}$, see Eq.~\eqref{eq:Wigner_rho_j}, one arrives at the action
\begin{equation}
\label{eq:S_j_UV}
 S_j^{\rm UV}[f] = \frac {N }{2} \int_{s,x} \int_{k,\epsilon} \,j_{\epsilon,k}^1 \, {G}_{\epsilon, k}(\Lambda/b)
 = \frac {N }{2} \int\limits_{s,x} \partial_x F F' \times \int_k\int_{{|\widetilde\epsilon|}<J}\!\!\!\!
 \widetilde{\epsilon} \partial_k \epsilon_k \,G_{\widetilde{\epsilon}, k}(\Lambda),
\end{equation}
which is a counterpart of Eqs.~\eqref{eq:St_int_xs_ek} and \eqref{eq:S_rho_int_xs_ek}. However, 
this result corresponds neither to any of the contributions to the desired AS action~(\ref{eq:S_RI_b}), nor is a full derivative.
We will demonstrate how it cancels in combination with higher-order terms in the gradient expansion in the following subsection.

\subsection{Second-order gradient terms}
\label{App:second_order_expansion}

In this Appendix, we continue with the second-order expansion over gradients, $\rho$ and $j^1$, in the 
fluctuation action $S_{\rm fl}[f]$, see Eq.~\eqref{eq:S_fl_f}. First, we discuss how this expansion 
can be expressed in the Wigner representation and consider a typical second-order term $S_{\rho\rho}[f]$.
Using the definition of the operator $\rho$, see Eq.~(\ref{eq:bar_dtau}), one can rewrite this piece of the action as
\begin{equation}
\label{eq:S_2_PI}
S_{\rho\rho}[f] = \int \, b_{t_1} \Pi_{t_1 t_2} b_{t_2} \, d^2 t dx, 
\end{equation}
with $\Pi_{t_1 t_2}$ being a generalized polarization operator. Its most general expression takes the following form,
\begin{equation}
\label{eq:Pi_12}
\Pi_{t_1 t_2} = - \frac N 4\int_k \, \overline{\cal G}_{t_1 t_2}(k) \, \overleftrightarrow{\partial_{t_1}} \,\overleftrightarrow{\partial_{t_2}}\,\overline{\cal G}_{t_2 t_1}(k),
\qquad \overleftrightarrow{\partial_t} := \frac 12 ( \overrightarrow{\partial_t} - \overleftarrow{\partial_t} ).
\end{equation} 
Here $\overline{\cal G}_{t_1 t_2}(k)$ refers to the Wigner symbol of the propagator $\overline{\cal G}$ w.r.t. position coordinates, and
the expression~(\ref{eq:S_2_PI}) is written in the leading order Moyal expansion, assuming $b$ to change adiabatically in space.
It is a reasonable approximation since the polarization operator already contains two temporal gradients.
We can simplify (\ref{eq:Pi_12}) further by using the full Wigner symbol of the Green's function, $\overline{\cal G}_{\epsilon,k}(s,x)$, 
evaluated with respect to both temporal- and spatial indices. Then, after some simple algebra, one can check that
the polarization operator $\Pi_{t_1 t_2}$ in the Wigner representation acquires the following form,
\begin{equation}
\label{eq:Pi_ws}
\Pi_{\omega} = -\frac N 4 \int_{k,\epsilon} \overline{\cal G}_{\epsilon_+,k} 
\left(\epsilon^2 + \tfrac 14 \overleftrightarrow{\Box}_s \right) \overline{\cal G}_{\epsilon_-,k} = \Pi_{\omega}^{\rm I} + \Pi_{\omega}^{\rm II}, 
\quad \epsilon_\pm = \epsilon \pm \tfrac 12 \omega,
\end{equation}
where the 1st and 2nd terms above refer, respectively, to $\epsilon^2$ and $s$-derivative parts.
Here, all dependencies on the slow time $s$ and the position $x$ were left implicit for brevity. 
We have also introduced the second-order differential operator $\overleftrightarrow{\Box}_s$ which 
needs both left and right functions to act on and is defined as
\begin{equation}
\label{eq:D_Alembert_operator}
g(s) \overleftrightarrow{\Box}_s\, h(s) : = \frac{d^2 }{dr^2} \Bigl( g(s+ \tfrac 12 r) h(s- \tfrac 12 r )\Bigr)\Bigl|_{r=0} = 
\frac 14 g''_{s^2} h + \frac 14 g  h''_{s^2} - \frac 12 g'_s h'_s. 
\end{equation}
Similarly, one can analyze other terms in the gradient expansion, i.e. $S_{j\rho}$ and $S_{jj}$. 

\subsubsection{Evaluation of the Schwarzian action}
\label{sec:Schwarzian_action_derivation}
In the following two subsections, we demonstrate how to derive the AS action within the leading logarithmic approximation. Specifically, we show that in order to determine coupling constants of the AS action with logarithmic accuracy, it suffices to use the local approximation for the exact Wigner symbol of the Green's functions defining the polarization operator~\eqref{eq:Pi_ws} [see also the comment around Eq.~\eqref{eq:G_ek_rescaling}]. That is, one can substitute 
$\overline{\cal G}_{\epsilon,k} \to G_{\epsilon,k}(\Lambda/b)$, as suggested in Eq.~\eqref{eq:cal_G_Moyal}.

We start from the Schwarzian piece, which constitutes the real part in the action $S_\pm^{\rm AS}[b]$ and originates from $S_{\rho\rho}[f]$, see Eq.~\eqref{eq:S_2_PI}. It is more instructive to discuss {Model II} first. Here the main contribution comes from the flat band region, $|k|<k_0$, where the Green's function $\overline G_{\epsilon,k}(\Lambda/b)$ simplifies to the SYK solution
$G^{0}_\epsilon$, see Eq.~\eqref{eq:GS0_SYK_E}, and becomes $k$ (momentum) and $s$ (slow time) independent. 
In Fourier space one obtains,
\begin{equation}
\label{eq:S_tau_tau_exp}
	S_{\rho\rho}[f] = \int dx  \sum_\omega b_\omega \Pi_\omega b_{-\omega}, \quad \Pi_\omega =  - \frac{N}{4} \int_{k,\epsilon}  G^0_{\epsilon_+} G^0_{\epsilon_-} \epsilon^2 = \Pi_0 + \frac{\omega^2} 2 \Pi''_0 + \dots 
\end{equation}
The $\Pi_0$-contribution here is UV-divergent and we use the parametrization dependent cut-off scheme discussed in Appendix~\ref{App:Rep_invariance_S_star} to handle it, which gives
\begin{equation}
	\Pi_0  \propto \int_0^{J/b^2} \epsilon d\epsilon \overset{\widetilde\epsilon = \epsilon b^2}{=} \frac{1}{b^4} \int_0^J  
    \widetilde{\epsilon} d\widetilde{\epsilon} \propto \frac{J^2}{b^4}. 
\end{equation}
We then see that it contributes a constant to the action
\begin{equation}
\label{eq:S0_rr}
		S^{0}_{\rho\rho} \propto \int\frac{ dx ds }{b^2} = \int { dx ds } F' =  \int { dx d\tau } = \beta L. 
\end{equation}
On other hand, the second-order order frequency expansion of the polarization operator evaluates to
\begin{equation}
\label{eq:Pi_2}
	\frac{1} 2 \Pi''_0  =  - \frac N 8  \int_{k,\epsilon} (  G^0_{\epsilon,k}\, \overleftrightarrow{\Box}_{\!\omega}\,  G^0_{\epsilon,k} ) \epsilon^2 \overset{|k|<k_0}{=}
   - \frac {N}{32}  \int_{k,\epsilon} (\Sigma^0_{\epsilon})^{-2} = \frac{N k_0}{32 J}  \int\limits_{\sim T}^J \frac{d\epsilon}{\epsilon} =  \frac{N k_0}{32 J} \ln\frac{J}{T},
\end{equation} 
where the logarithmic infra-red singularity was cut by the temperature $T$. The above frequency expansion brings the Schwarzian action,
\begin{equation}
\label{eq:S_rho2}
S_{\rho\rho}[f] = \frac 12 \Pi''_0 \int dx  \sum_\omega b_\omega \omega^2 b_{-\omega} =  
\frac 12 \Pi''_0 \int_{s,x} b'^2 =  
- \frac{N k_0}{64 J} \ln\frac{J}{T} \int_{x,\tau} \{f,\tau\}.
\end{equation}
In the last transformation results of Sec.~\ref{sec:gradients_GSigma} were used, see Eq.~\eqref{eq:S_RI_b}.
Comparing $S_{\rho\rho}[f]$ to the real part of the AS action, one derives the relation
\begin{equation}
\label{eq:C_to_u_ratio}
\frac{C}{u(T)} =  \left(\frac{3\pi}{8}\right) \frac{N k_0}{J}  \ln\frac{J}{T},
\end{equation}
which fixes the ratio of a central charge $C$ to the temperature dependent velocity $u(T)$.
One can also easily analyze the contributions stemming from the small range of momenta $k_0<|k|<\pi/a$,
where $\epsilon_k$ is nonvanishing. Here, the term proportional to $\epsilon^2$ in the polarization operator~\eqref{eq:Pi_ws}
produces a logarithmic correction to $S_{\rho\rho}[f]$, which scales as $({N k_1}/{J}) \ln ({J}/{T_\Lambda})$.
It is negligible compared to the main contribution~\eqref{eq:S_rho2}. The contributions originating from the 
$\overleftrightarrow{\Box}_s$~-~term in Eq.~\eqref{eq:Pi_ws} are even smaller. 
It can be shown that their functional dependence on the reparametrizations reduces to the Schwarzian form, 
as discussed above, and at the same time corresponding diagrams defining the prefactor are all IR convergent and free from UV logarithmic 
divergencies (see also additional comments in the end of this subsection).

We now turn to the analysis of {Model I}. Similar to the discussion above, the primary contribution to the Schwarzian action arises from the first part of the polarization operator. Here, we begin with the local approximation for the propagators, employing Eq.~\eqref{eq:cal_G_Moyal}, which yields
\begin{equation}
\label{eq:Pi_ws_Model_I}
\Pi^{\rm I}_{\omega} = -\frac N 4 \int_{k,\epsilon} \epsilon^2 \, {G}_{\epsilon_+,k}(\Lambda/b) {G}_{\epsilon_-,k}(\Lambda/b) = 
\Pi^{\rm I}_0 + \frac{\omega^2} 2 (\Pi^{\rm I}){''}_{\!\!\!\!\!0} + \dots 
\end{equation} 
The zero frequency piece, after rescaling $\epsilon = \widetilde{\epsilon}/b^2$ and using the transformation law~\eqref{eq:Green_rescale},
evaluates to
\begin{equation}
    \Pi^{\rm I}_0 \propto  -\frac {N}{b^4} \int_{k} \int_{|\widetilde{\epsilon}|<J} \widetilde{\epsilon}^{\,2} \, {G}^2_{\widetilde{\epsilon},k}(\Lambda) 
    \propto \frac{\rm Const}{b^4}.
\end{equation}
It contributes only an inessential constant to the overall action, see Eq.~\eqref{eq:S0_rr}. In turn, for the frequency-dependent piece, we obtain with the logarithmic accuracy
\begin{eqnarray}
    \frac 12 (\Pi^{\rm I}){''}_{\!\!\!\!\!0} &=& - \frac N 8  \int_{k,\epsilon} \left[  G_{\epsilon,k}(\Lambda/b)\, \overleftrightarrow{\Box}_{\!\epsilon}\,  G_{\epsilon,k} (\Lambda/b) \right] \epsilon^2 \overset{\widetilde{\epsilon}= \epsilon b^2}{=}  \\
    \label{eq:Schw_log_Model_I}
    &-& \frac N 8  \int_{k,\widetilde{\epsilon}} \left[  G_{\widetilde{\epsilon},k}(\Lambda)\, 
    \overleftrightarrow{\Box}_{\!\widetilde{\epsilon}}\,  G_{\widetilde{\epsilon},k} (\Lambda) \right] \widetilde{\epsilon}^{\,2} \simeq 
    - \frac {N k_0}{32\pi^2}  \int_{\sim T_\Lambda}^J \frac{d\widetilde{\epsilon}}{(\Sigma^0_{\widetilde{\epsilon}})^2} = 
     \frac{N k_0}{32 J} \ln\frac{J}{T_\Lambda}. \nonumber
\end{eqnarray}
To derive this result, we employed~\eqref{eq:Green_rescale} to eliminate the $b$-dependence from the integral and substituted $G_{\epsilon,k}^{-1} \to - \Sigma^0_\epsilon$, which
is valid in the high-energy limit $T_\Lambda \ll \epsilon \ll J$.
We then relied on the fact that the resulting infra-red logarithmic divergence in
this case is regularized by the finite kinetic energy $\epsilon_k$ and thus is effectively cut-off at the scale $T_\Lambda$.
The latter implies the relation
\begin{equation}
\label{eq:C_to_u_ratio_I}
\frac{C}{u} =  \left(\frac{3\pi}{8}\right) \frac{N k_0}{J}  \ln\frac{J}{T_\Lambda},
\end{equation}
where the velocity $u$ turns out to be temperature independent at variance with our previous result~\eqref{eq:C_to_u_ratio} for {Model II}.

To proceed, we demonstrate that the gradient corrections~\eqref{eq:delta_G} to the Wigner symbol ${\cal G}_{\epsilon,k}$ of the Green's function do not alter the derived result. For concreteness, we illustrate this by considering the second correction~\eqref{eq:grad_corr_G_second}. Since the latter already includes temporal gradients in terms of $b'^2$, it suffies
to take the zero frequency limit of the polarization operator, $\Pi^{\rm I}_{\omega \to 0}$, see~\eqref{eq:Pi_ws},
to obtain a correction to the action $S_{\rho\rho}[f]$. With the integration variable $\epsilon$ rescaled as usual, 
we obtain an intermediate result in the following form:
\begin{eqnarray}
    \delta S_{\rho\rho}^{(2)}[f] &=& - \frac{N}{4}\int_{s,x}\left(\frac{b'}{b}\right)^2 \int_{\epsilon,k}  \epsilon^2 
    \epsilon_k^2 \left(\partial_\epsilon \overline G{}_{\epsilon,k} \overleftrightarrow{\partial_\epsilon} \,\overline G_{\epsilon,k}^2 \right) \overline G{}_{\epsilon,k} \nonumber \\
    \label{eq:delta_S_rho_rho}
    & \overset{\widetilde{\epsilon} = \epsilon b^2}{=}& 
    - \frac{N}{4}\int_{s,x} b'^2  \times \int_{{\widetilde{\epsilon}},k}  \widetilde{\epsilon}^2 
    \epsilon_k^2 \left[\partial_{\widetilde{\epsilon}}\, G_{{\widetilde{\epsilon}},k}(\Lambda) \overleftrightarrow{\partial_{\widetilde{\epsilon}}} \,G_{{\widetilde{\epsilon}},k}^2(\Lambda) \right] G{}_{{\widetilde{\epsilon}},k}(\Lambda),
\end{eqnarray}
where the relation~\eqref{eq:Green_rescale} was used to transition from the first to the second line. Referring to Eq.~\eqref{eq:S_RI_b},
it becomes evident that the above expression is that of the Schwarzian.
To further estimate the energy-momentum integral, we use the
UV behavior of the Green's function: $G_{\epsilon,k} \sim |\epsilon J|^{-1/2}$ in the range $T_\Lambda \ll \epsilon \ll J$, and find
\begin{equation}
\label{eq:subleading_Schwarzian}
     \delta S_{\rho\rho}^{(2)}[f] \sim N \int_{s,x} b'^2 \int_{|k|<k_0}  \epsilon_k^2 \int_{\epsilon_k^2/J}^{+\infty} 
     \frac{d\epsilon}{(J \epsilon)^2} \sim \frac{N k_0}{J} \int_{x,\tau}\{f,\tau\}.    
\end{equation}
Note that the resulting contribution to the Schwarzian action is subleading compared to the logarithmic one~\eqref{eq:Schw_log_Model_I}, as stated earlier. An exact evaluation of the integral~\eqref{eq:delta_S_rho_rho}, which we do not reproduce here, confirms this estimate. Additionally, considering another gradient correction to the Wigner symbol, as given by Eq.~\eqref{eq:delta_St_int_xs_ek}, 
yields a similar result. 

The evaluation of the second part of the polarization operator, $\Pi_\omega^{\rm II}$, see Eq.~\eqref{eq:Pi_ws},
which contributes to the Schwarzian action, proceeds along the same lines as above and does not provide any new insights.
Due to the presence of the second-order differential operator $\overleftrightarrow{\Box}_s$, acting on slow times, one can, as before,
take a zero frequency limit, $\omega \to 0$, in the above polarization operator. 
The analytical structure of resulting one-loop momentum-energy integrals
happens to be analogous to that in Eq.~\eqref{eq:Schw_log_Model_I}. They are free from both UV and IR divergencies and
provide an additional contribution of order $N k_0/J$ to the coupling constant in front of the Schwarzian, 
cf. Eq.~\eqref{eq:subleading_Schwarzian}. Consequently, the relation~\eqref{eq:C_to_u_ratio_I} remains unchanged when 
understood with logarithmic accuracy.

\subsubsection{Central charge and the kinetic term in the AS action}
We next proceed with a derivation of the central charge $C$ by analyzing the action $S_{j\rho }[f]$, which
brings about the imaginary contribution to the AS action. Following the steps around Eqs.~(\ref{eq:S_2_PI}--\ref{eq:Pi_ws}),
one can rewrite this action in terms of a corresponding polarization operator,
\begin{equation}
\label{eq:S_tau_x_def}
S_{j\rho}[f] = \int \, \partial_x F_1 b_{t_1} P_{t_1 t_2} b_{t_2} \, d^2 t dx,  \qquad F_1=F(t_1, x),
\end{equation}
where the Wigner symbol of $P_{t_1 t_2}$ takes the following form:
\begin{equation}
P_{\omega} =  -\frac{i N}{2} \int_{\epsilon,k} \partial_k \epsilon_k\,  \overline{\cal G}_{\epsilon_+,k} 
\left(\epsilon^2 + \tfrac 14 \overleftrightarrow{\Box}_s \right)  \overline{\cal G}_{\epsilon_+,k}  = 
P_{\omega}^{\rm I} + P_{\omega}^{\rm II}, \qquad \epsilon_\pm = \epsilon \pm \omega/2,
\end{equation}
with the first and the second terms referring to $\epsilon^2$- and $s$-derivative parts of the above polarization operator, respectively.
Following our detailed discussion in the previous subsection~\ref{sec:Schwarzian_action_derivation}, we may disregard $P_{\omega}^{\rm II}$
and use the local approximation to the Wigner symbols, $\overline{\cal G}_{\epsilon,k} \to G_{\epsilon,k}(\Lambda/b)$, 
when deriving the central charge $C$ with logarithmic accuracy.

The first part of the polarization operator can be expanded as $P_{\omega}^{\rm I} = P_{0}^{\rm I}+\frac{\omega^2}{2} P''_0 + \dots$,
with the leading zero-frequency term being UV dominated. As before, we evaluate it by changing the energy integration variable to 
$\widetilde{\epsilon} = \epsilon b^2$, which yields
\begin{equation}
P_0^{\rm I} = - \frac{i}{b^4} \times \frac N 2\int_k\int_{|\widetilde{\epsilon}|<J}  \!\!\!\!\
 \widetilde{\epsilon}^{\,2}\, \partial_k \epsilon_k \, G^2_{\widetilde{\epsilon},k} (\Lambda). 
\end{equation}
On making use of the general relation~(\ref{eq:S_tau_x_def}), the UV contribution to the action  $S_{j\rho}[f]$ then reads
\begin{equation}
\label{eq:S_j_UV_2}
S_{j\rho}^{\rm UV}[f] =  - \frac {iN}{2}\int_{s,x} \partial_x F F'  \times
\int_k\int_{|\widetilde{\epsilon}|<J} 
\!\!\!\! \widetilde{\epsilon}^{\,2}\, \partial_k \epsilon_k\,  G^2_{\widetilde{\epsilon},k} (\Lambda).
\end{equation} 
It needs to be analyzed further in combination with the analogous contribution~(\ref{eq:S_j_UV}) arising from the first-order gradient expansion, which was derived previously. At the end of this subsection, we will demonstrate how a mutual cancellation of these terms, and similar ones, can be achieved.

For now, we proceed by discussing the $P_0''$-term in the frequency expansion, which contributes to the AS action, as will become clear shortly. Following the derivation leading to Eq.~(\ref{eq:Schw_log_Model_I}), we evaluate the constant $P_0''$ as
\begin{eqnarray}
    \frac 12 (P^{\rm I}){''}_{\!\!\!\!\!0} &=& - \frac{iN}{4}  \int_{k,\epsilon}  \epsilon^2 \partial_k \epsilon_k
    \left[  G_{\epsilon,k}(\Lambda/b)\, \overleftrightarrow{\Box}_{\!\epsilon}\,  G_{\epsilon,k} (\Lambda/b) \right]\overset{\widetilde{\epsilon}= \epsilon b^2}{=}  \\
    \label{eq:Kinematic_log_Model_I}
    &-& \frac{i N}{32\pi^2} \int_k \partial_k \epsilon_k  \int_{\sim T_\Lambda}^J \frac{d\widetilde{\epsilon}}{(\Sigma^0_{\widetilde{\epsilon}})^2} = 
     \frac{i N \Lambda}{16 J} \ln\frac{J}{T_\Lambda}. \nonumber
\end{eqnarray}
To obtain the above result, which is found with logarithmic accuracy, it is sufficient to use the 
high-energy asymptotics for the Green's function $G^{-1}_{\epsilon,k} \to - \Sigma^{0}_\epsilon$, obtained by neglecting 
its exact momentum dependence, and to cut off the energy integral by the scale $T_\Lambda$ in the IR limit.
This second-order frequency expansion produces the following contribution to the action,
\begin{equation}
\label{eq:S_tau_x_Fourier}
	S_{j\rho}[f] = \frac 12 P_0'' \int dx \sum_\omega (\partial_x F b_s)_{-\omega} \,\omega^2 b_\omega = 
	\frac{i C}{12\pi}  \int dx  \sum_\omega (\partial_x F b)_{-\omega} \,\omega^2 b_\omega,
\end{equation}
where we have introduced the central charge,
\begin{equation}
\label{eq:C_value}
	C =  \left(\frac{3\pi}{4}\right) \frac{N \Lambda}{J}  \ln\frac{J}{T_\Lambda}.
\end{equation}
By transforming the result~(\ref{eq:S_tau_x_Fourier}) to the time domain and using integration by parts, 
the action $S_{j\rho}[f]$ reduces to
\begin{equation}
	\label{eq:S_tx_R}
	S_{j\rho}[f] = - \frac{i C}{12\pi}  \int_{s,x} \partial_x F\, b b'' =  
    - \frac{i C}{24\pi}  \int_{s,x} \partial_x F (b b'' -  b'^2 )  
	  - \frac{i C}{12\pi}  \int_{s,x} \frac{(\partial_x b) b'}{b^2},
\end{equation}
where, to obtain the final form of the action, the relation $\partial_x F' = - 2\partial_x b/b^3$ has been employed
(it follows from $F'=1/b^2$).
Using Eq.~(\ref{eq:S_RI_b}) for the imaginary part of the AS action, we see that Eq.~(\ref{eq:S_tx_R}) 
indeed reproduces its expected form up to a residual contribution, i.e.
\begin{equation}
\label{eq:S_tau_x_plus_dS}
		S_{j\rho}[f]  = - \frac{i C}{48\pi}  \int_{x,\tau}\frac{f''\partial_x f'}{f'^2} - 
        \frac{i C}{12\pi}  \int_{s,x} \frac{(\partial_x b) b'}{b^2}
		\equiv i \,{\rm Im }\, S^{\rm AS}_{-}[b] + \Delta S_{j\rho}[b].
\end{equation} 
(Note that the negative index in the action $S^{\rm AS}_{-}[b]$ corresponds to right moving Majoranas). 
On comparing the result with the counter term $\Delta S_\rho[f]$~\eqref{eq:counter_term_S_rho}, 
we see that the latter exactly cancels against $\Delta S_{j\rho}[b]$.
In this way we have achieved our goal: To show that the sum of two actions, $S_{\rho}[f] + S_{j\rho}[f]$,
reproduces the kinematic (or imaginary) part of the AS action. By using the
previously found ratios of the coupling constants $C/u$ for both models, see Eqs.~\eqref{eq:C_to_u_ratio} and
\eqref{eq:C_to_u_ratio_I}, together with the value of the central charge~\eqref{eq:C_value}, the velocity $u(T)$ can 
now be easily recovered as listed in the beginning of Sec.~\ref{sec:gradients_GSigma}.

\subsubsection{Analysis of the UV terms}
Here we discuss why the UV contributions~\eqref{eq:S_j_UV} and \eqref{eq:S_j_UV_2}, which were previously identified, can be disregarded. 
We start by noting that similar UV contributions arise when one analyses higher-order terms of the type ${\rm tr}(j^1 \overline{\cal G} [\rho\overline{\cal G}]^n)$
in the action of fluctuations~\eqref{eq:S_fl_expansion}. It is not hard to verify that the UV part of such $n$-th order term reads
\begin{equation}
\label{eq:S_j_UV_n}
S_{j\rho^n}^{\rm UV}[f] =  \frac {N}{2}\int_{s,x} \partial_x F F'  \times
\int_k\int_{|\widetilde{\epsilon}|<J} 
\!\!\!\! \widetilde{\epsilon}\, \partial_k \epsilon_k\, G_{\widetilde{\epsilon},k} (\Lambda)\, [-i\widetilde{\epsilon} G_{\widetilde{\epsilon},k} (\Lambda)]^n.
\end{equation}
As a result, the following geometric series can be summed up,
\begin{equation}
\label{eq:S_j_UV_all}
    {\cal S}^{\rm UV}[f] =  \sum_{n=0}^{+\infty} S_{j\rho^n}^{\rm UV}[f] =  \frac {N}{2}\int_{s,x} \partial_x F F'  \times
    \int_k\int_{|\widetilde{\epsilon}|<J} \!\!\!\! \widetilde{\epsilon}\, \partial_k \epsilon_k\, {\bf G}_{\widetilde{\epsilon},k} (\Lambda),
\end{equation}
where the full Green's function is given by the standard relation
\begin{equation}
    {\bf G}^{-1}_{\widetilde{\epsilon},k}(\Lambda) = i \widetilde{\epsilon} + G^{-1}_{\widetilde{\epsilon},k} (\Lambda) = 
    i \widetilde{\epsilon} - \epsilon_k - \Sigma(\widetilde{\epsilon}, \Lambda),
\end{equation}
with $\Sigma(\widetilde{\epsilon}, \Lambda)$ being the self-energy found in the mean-field approximation. At this stage, we observe that, 
up to terms of order ${\cal O}(a_x^2)$, the result~\eqref{eq:S_j_UV_all} can be equivalently represented as a difference of two actions, 
\begin{equation}
     {\cal I}^{\rm UV}_{a_x}[f] - {\cal I}^{\rm UV}_{a_x = 0}[f] = {\cal S}^{\rm UV}[f] + {\cal O}(a_x^2), 
     \qquad a_x = \partial_x F.
\end{equation}
Here, the newly introduced action ${\cal I}^{\rm UV}_{a_x}[f]$ is defined as
\begin{equation}
\label{eq:I_action_ax}
    {\cal I}^{\rm UV}_{a_x}[f] =   - \frac {N}{2}\int_{s,x}  F' \int_k\int_{|\widetilde{\epsilon}|<J} \!\!\!\! 
    \ln\left[  i \widetilde{\epsilon} - \epsilon_{k + \widetilde{\epsilon} a_x } - \Sigma(\widetilde{\epsilon}, \Lambda) \right],
\end{equation}
with the momentum $k$ in the dispersion relation being shifted by the 'vector potential' 
$a_x$ and the energy $\widetilde{\epsilon}$ playing the role of an effective charge.

At this stage, the Green's function ${\bf G}_{\widetilde{\epsilon},k}(\Lambda)$ can be regarded as an operator acting 
in momentum space and parametrically depending on the energy $\widetilde{\epsilon}$. Following this logic, 
the UV action can be expressed via the determinant,
\begin{equation}
     {\cal I}^{\rm UV}_{a_x}[f] =   \frac {N}{2}\int_{s,x}  F'  \int_{|\widetilde{\epsilon}|<J} \!\!\!\! 
     \ln {\rm det}\, {\bf G}_{\widetilde{\epsilon},\hat k + \widetilde{\epsilon} a_x}(\Lambda).
\end{equation}
Taking into account the commutation relation $[\hat{x}, \hat{k}] = i$, one can further write:
\begin{equation}
\label{eq:gauge_G_k}
    {\bf G}_{\widetilde{\epsilon},\hat k + \widetilde{\epsilon} a_x }(\Lambda) = 
    e^{- i\widetilde{\epsilon}\, F(s,x) }\, {\bf G}_{\widetilde{\epsilon},\hat k}(\Lambda) \,  e^{ i\widetilde{\epsilon} \,F(s, x) },
\end{equation}
which shows that the Green's functions above are related by a local gauge transformation in space.
If the dispersion relation $\epsilon_k$ did not describe chiral edge modes, it would be possible to identify the determinants of the two Green's functions 
--- one with $a_x = \partial_x F$ and the other with $a_x = 0$ --- and the action $ {\cal S}^{\rm UV}[f]$ would vanish. 
Consequently, the non-zero UV contribution \eqref{eq:S_j_UV_all} should be interpreted as the chiral anomaly.
We attribute the appearance of this anomaly to the transformation ${\cal M}_{3/2}$, which was employed to obtain the regularized action~\eqref{eq:S_reg_bar}
from its defining expression~\eqref{eq:S_reg_f}. Recall that the determinant of the differential operator ${\cal D}$ in that action 
was regularized by dividing it by ${\rm det}\, \widetilde{\Sigma}^0$, which contains no kinetic part. 
As a result, any possible chiral anomaly remained uncompensated.

Bearing these considerations in mind, we conclude that the UV contributions~\eqref{eq:S_j_UV} and \eqref{eq:S_j_UV_2} 
are absent in the starting $G\Sigma$-action~\eqref{eq:S_reg_bar} and therefore must be disregarded.

\section{Aspects of (semi)-classical Liouville field theory} \label{App:Liouville}
This appendix deals with three aspects of (semi)-classical Liouville field theory. Firstly, we calculate the one-loop partition function of the Liouville action~\eqref{eq:SL} and demonstrate its agreement with the result~\eqref{eq:Z-vacuum}. Secondly, we detail on how solutions to the disk equations of motion~\eqref{eq:Liouville-General-solution} can be transformed to the strip. At last, we extend the Liouville action to the presence of sources and show that we obtain a finite on-shell action, which allows us to calculate correlation functions.

\subsection{One-loop partition function} \label{App:Liouville-one-loop}
The one-loop partition function is obtained by considering Gaussian fluctuations around the field $\phi_0$ given in~\eqref{eq:phi-0-strip} evaluated on the action~\eqref{eq:SL}. We expand $\phi(\tau,x)=\phi_0(\tau)+\ep(\tau,x)$ and impose $\ep(0,x)=\ep(\beta/2,x)=0$ since the ZZ-boundary conditions are fulfilled by the saddle-point solution. We then obtain
\begin{align}
    S_2[\epsilon]=&S_0+\frac{C}{96 \pi  u} \int_0^{L} \int_0^{\beta/2} \dd{x} \dd{\tau} \ep(\tau,x)\qty(-u^2\partial_x^2 -\partial_{\tau}^2+\qty(\frac{2\pi}{\beta})^2 2\csc(\frac{2\pi \tau}{\beta})^2) \ep(\tau,x)\,, \label{eq:SL-2}
\end{align}
where all source terms integrate to boundary terms. Note that the on-shell action coincides with $S_0$ calculated below~\eqref{eq:LiouvSemicl} and is finite since all divergencies of $\phi_0$ close to the boundary of the strip at $\tau=0, \beta/2$ cancel. The fluctuation determinant is evaluated as
\begin{align}
    Z_2= \int_{\ep(0,x)=\ep(\beta/2,x)=0} \mathcal{D}[\ep] e^{-S_2[\ep]}=  \frac{\tilde{N}}{\det(D)^{\frac{1}{2}}} e^{-S_0},
\end{align}
where $\det(D)$ is the determinant of the differential operator in the quadratic action~\eqref{eq:SL-2} acting on functions with prescribed boundary conditions. To obtain its eigenvalues, we first factorize the part acting in the $x$-direction by a Fourier expansion
\begin{align}
    \ep(\tau,x)=\frac{1}{L}\sum_m e^{2\pi i m x/L} \ep_m(\tau).
\end{align}
 Ignoring the spatial dependence for now, we are left with the Schroedinger problem
\begin{align}
    \qty(-\partial_{\tau}^2+2\qty(\frac{2\pi}{\beta})^2\csc(\frac{2\pi \tau}{\beta})^2) \ep(\tau)=\lambda^2 \ep(\tau)\,.
\end{align}
Using the substitution
\begin{align}
    u=\cot(\frac{2\pi \tau}{\beta}), \qquad  (u^2+1)=\csc(\frac{2\pi \tau}{\beta})^2\,,
\end{align}
with $u(0)\rightarrow\infty$ and $ u(\beta/2) \rightarrow -\infty$, the problem is brought into the form
\begin{align}
    \qty(\frac{2\pi}{\beta})^2(u^2+1)\qty[-(u^2+1)\partial_u^2-2u\partial_u+2]\ep(\tau)=\lambda^2 \ep(\tau)\,.
\end{align}
This is the Legendre differential equation with two independent solutions~\cite{Cohl2021}
\begin{align}
    \ep(u)=c_1\,P_1^{n}(iu)+c_2\,Q_1^{n}(iu),   \qquad n=\frac{\beta \lambda}{2\pi},
\end{align}
where $Q_m^n(u)$ and $P_m^n(u)$ are the associated Legendre functions. Our goal is now to find a restriction on the allowed values of $\lambda$. We start by noting that $P_1^n(i u)$ grows as $\abs{u}$ for $\abs{u}\rightarrow\infty$ for all $n$ and hence is not normalizable, i.e. it does not fulfill the given boundary conditions. More interesting is the solution
\begin{align}
    Q_1^n(u) = \frac{\pi}{2} \qty( -\sin(\frac{(1+n)\pi}{2})w_1(1,n,u)+\cos(\frac{(1+n)\pi}{2}) w_2(1,n,x)),
\end{align}
with 
\begin{align}
    w_1(1,n,u)=&\frac{2^n \Gamma(1+\frac{n}{2})}{\Gamma(\frac{3}{2}-\frac{n}{2})}(1-u^2)^{-n/2} \pFq{1}{2}{-\frac{1+n}{2},\frac{2-n}{2},\frac{1}{2}}{u^2}, \nonumber \\
     w_2(1,n,u)=&\frac{2^n \Gamma(1+\frac{n}{2})}{\Gamma(1-\frac{n}{2})}u(1-u^2)^{-n/2} \pFq{1}{2}{-\frac{n}{2},\frac{3-n}{2},\frac{3}{2}}{u^2} ,
\end{align}
where $\pFq{1}{2}{a,b,c}{d}$ denote the hypergeometric functions and the representation of $ Q_1^n(u)$ can be continued to $u\in \mathbb{C}\backslash ((-\infty,-1] \cup [1,\infty)) $ for $n \in \mathbb{C}$ and $1+n \notin -\mathbb{N}$, c.f. \cite{Cohl2021} Theorem 4.3. Importantly, all branch cuts of the function lie in the excluded region of the complex plane. Then, using the expansion of the hypergeometric function for large arguments\footnote{Denoting the first two arguments of the hypergeometric function by $a$ and $b$, the expansion is valid if $a-b \notin \mathbb{N}$, which is the case for us.}, we obtain
\begin{align}
    Q_1^n(iu) \approx \frac{\alpha u}{\Gamma(2-n)}+\frac{\beta}{\Gamma(2-n) u}+ \frac{\gamma}{u^2}+O\qty(\frac{1}{u^3}),  \qquad \abs{u} \to \infty
\end{align}
with $\alpha,\beta, \gamma \in \mathbb{C}$. In order for the diverging term to vanish, we need to demand $2-n$ to lie on the poles of the $\Gamma$-function. This allows us to conclude that $n$  must take values in $\mathbb{N}_{>1}$, which is the desired quantization condition on $n$.

Including the spatial modes, we are now in a position to evaluate the one-loop determinant as
\begin{align}
     &\operatorname{Det}\qty[\qty(-\partial_{\tau}^2-u^2\partial_{x}^2+2\qty(\frac{2\pi}{\beta})^2\csc(\frac{2\pi \tau}{\beta})^2)] \nonumber \\
     &=\prod_{m=-\infty}^\infty\prod_{n>1}^\infty\qty(\qty(\frac{2\pi m   u}{L})^2+\qty(\frac{2\pi n}{\beta})^2)=N\prod_{m=-\infty}^\infty\prod_{n>1}^\infty\qty(m^2+\qty(\frac{L n}{ u\beta})^2) \nonumber \\
     &=N\prod_{m=-\infty}^\infty\prod_{n>1}^\infty\qty(m^2-\frac{n^2}{(-u \mop)^2})=N\prod_{m=-\infty}^\infty\prod_{n>1}^\infty\qty(m-\frac{n}{u \mop})\qty(m+\frac{n}{u \mop})=N\prod_{m=-\infty}^\infty\prod_{\abs{n}>1}^\infty\qty(m+ \frac{n}{u \mop})\,.
\end{align}
Up to the global prefactor $N$ and the transformation $\mop \mapsto - 1/\mop$, this determinant is equal to the one obtained in (5.19) of \cite{Cotler2018}. Following the $\zeta$-function regularization outlined in their work, one obtains exactly the AS partition function  $Z_2=Z(\mop u )$ given in~\eqref{eq:Z-vacuum} with renormalized central charge $c=C+13$.

\subsection{Liouville field on the strip}
\label{sec:L_field_strip}
The purpose of this subsection is to derive the general solution for the Liouville field 
$\phi(w, \overline{w})$ on the strip, see Eq.~\eqref{eq:Strip-general-solution}, from the corresponding 
solution~\eqref{eq:Liouville-General-solution} on the disk. 
To simplify calculations, 
we introduce an additional pair of coordinates $(z, \overline z)$ that parametrize the full complex
plane and consider two conformal transformations:
\begin{equation}
\label{eq:conf_maps_z_xi}
     z(w) = e^{\frac{2 \pi}{\beta} w},  \qquad  \xi(z) = \frac{z-i}{z+i}.
\end{equation}
The first transformation here maps the half-strip $S_+$ on the upper-half complex plane $C_+$, i.e. $z: S_+ \mapsto C_+$,
while the second one maps $C_+$ onto the unit disc $D$, meaning $\xi: C_+ \mapsto D$.
The combination of these two mappings, $(\xi \circ z)(w)$, gives the Cayley map~\eqref{eq:Cayley-map}.

Our approach will be first to obtain the general solution $\widetilde{\phi}(z, \overline{z})$
on the upper-half complex plane by relating it to $\phi'(\xi, \bar \xi)$, and then to transform
$\widetilde{\phi}(z, \overline{z})$ into the desired solution $\phi(w, \overline{w})$ on the strip.
To this end, given a 'conformal reparametrization' $f'(\xi)$ of the disk, 
we consider a related reparametrization $h(z)$ on the upper-half plane, which is defined by the relation
\begin{equation}
\label{eq:f_prime_h}
 f'(\xi(z)) \equiv \xi(h(z)).
\end{equation}
The logic behind this construction can be gained from the following commutative diagram:
\begin{center}
\begin{tikzcd}[row sep=2em]
f': D \arrow{r}{f'}  & D \\
h : C_+ \arrow{u}{\xi} \arrow{r}{h} & \arrow{u}{\xi} C_+
\end{tikzcd}
\end{center}
In turn, the Liouville fields in domains $C_+$ and $D$ are related by the transformation law~\eqref{eq:Liouville-Trafo},
which for the case under consideration becomes
\begin{equation}
    e^{\widetilde{\phi}(z, \overline{z})} = e^{\phi'(\xi, \bar \xi)} \abs{\pdv{\xi(z)}{z}}^2. 
\end{equation}
Taking into account the analytical form of the solution~\eqref{eq:Liouville-General-solution} for 
the field $\phi'(\xi, \bar \xi)$, it is instructive to consider the chain of relations:
\begin{equation}
    \partial_\xi f'(\xi)\Bigl|_{\xi(z)} \pdv{\xi(z)}{z} \overset{\eqref{eq:f_prime_h}}{=} 
    \pdv{\xi(z)}{z}\Bigl|_{h(z)} \partial_z h(z) \equiv  \partial_z   \xi(h(z)).
\end{equation}
The latter enables us to express the field $\widetilde{\phi}(z, \overline{z})$ solely in terms of 
the conformal reparametrization $h(z)$ on the upper-half plane:
\begin{equation}
\label{eq:exp_tilde_phi}
    e^{\widetilde{\phi}(z, \overline{z})} = \frac{4 \abs{\partial_z\xi(h(z))}^2}{(1-\abs{ \xi(h(z))}^2)^2} 
    \overset{\eqref{eq:conf_maps_z_xi}}{=}
    - \frac{4 \partial_z h \, \partial_{\overline z} \overline{h}}
      {\bigl(h(z) - \overline{h(z)}\,\bigr)^2},
\end{equation}
where the explicit form of the conformal transformation $\xi(z)$, see Eq.~\eqref{eq:conf_maps_z_xi}, was used
to derive the very last relation.

At this stage, one can rely on the Schwarz reflection principle to state that $\overline{h(z)} = h (\overline{z})$ 
and thereby to simplify the denominator in Eq.~\eqref{eq:exp_tilde_phi}.
Indeed, the reparametrization on the disk satisfies the  
asymptotic relation $|f'(\xi)| \to 1$ for $|\xi| \to 1$. On the other hand, the (invertible) conformal
transformation $\xi(z)$ maps the real axis onto the unit circle, i.e. $\xi : \mathds{R} \mapsto  \partial D$,
where $\partial D$ denotes a boundary of the disk. Then from the defining relation~\eqref{eq:f_prime_h} it follows
that the function $h(z)$ is real valued on the real axis. This latter property justifies 
the usage of the Schwarz principle, as it was suggested above.

In the second step, following the same procedure as above, 
one can transform $\widetilde{\phi}$ to the
desired solution $\phi$ on the half-strip. These two Liouville fields are related as
\begin{align}
\label{eq:Liouville_S_C}
	 \qty(\frac{2 \pi}{\beta})^2 e^{\phi(w, \overline{w})}=e^{\widetilde{\phi}(z, \overline{z})}
     \abs{\pdv{z(w)}{w}}^2.
\end{align}
With the help of the commutative diagram,
\begin{center}
\begin{tikzcd}[row sep=2em]
h: C_+ \arrow{r}{h}  & C_+ \\
f : S_+ \arrow{u}{z} \arrow{r}{f} & \arrow{u}{z} S_+
\end{tikzcd}
\end{center}
we define the classical reparametrization $f(w)$ on the half-strip $S_+$ by the relation
\begin{equation}
\label{eq:h_f}
 h(z(w)) \equiv z(f(w)),
\end{equation}
which in somewhat more explicit terms reads as
\begin{equation}
    f(w) = \frac{\beta}{2\pi} \ln \left[ h(e^{\frac{2\pi}{\beta} w}) \right].
\end{equation}
In particular, one can see from the above analytical formula that the Schwarz reflection principle 
for the reparametrization $h(z)$ on the disk guarantees that the same relation, 
$\overline{f(w)} = f(\overline{w})$, holds on the strip.

To further resolve the Liouville field $\phi(w, \overline{w})$ in terms of $f(z)$, one
again manipulates with derivatives,
\begin{equation}
    \partial_z h(z)\Bigl|_{z(w)} \pdv{z(w)}{w} \overset{\eqref{eq:h_f}}{=} \partial_w \, z(f(w)).
\end{equation}
Then from relation~\eqref{eq:Liouville_S_C} we finally find
\begin{equation}
    e^{\phi(w, \overline{w})} =  - \qty(\frac{\beta}{\pi})^2 \frac{\abs{\partial_w \, z(f(w))}^2 }
    {\bigl( z(f(w)) - z(f(\overline{w}))\bigr)^2} \overset{\eqref{eq:conf_maps_z_xi}}{=}
    \frac{\abs{\partial_w f(w)}^2}{\sin^2(i\pi(f(\overline{w})-f(w))/\beta)},
\end{equation}
thereby recovering the result~\eqref{eq:Strip-general-solution} from Sec.~\ref{sec:light}.

\subsection{Action in presence of a source} \label{App:Liouville-one-source}
In this subsection, we generalize the action~\eqref{eq:SL} to the presence of a source of weight $l$ at position $w_0$ and also calculate its on-shell value. Inspired by the geometric construction of Ref.\cite{Menotti2006}, 
we set
\begin{equation}
\label{eq:S_l_w0_tilde}
    S(l, w_0)= \widetilde S(l, w_0) + 2 l \delta\ln\frac{2\pi \tuv}{\beta},
\end{equation}
where $\tuv \sim 1/J$ is the short-time cutoff and
\begin{align} 
    \widetilde S(l,w_0)=\lim_{\tuv \to 0}\Bigg[ &\frac{C}{24 \pi u} \qty( \int_{S \backslash \Delta_{\tuv}} \dd{x} \dd{\tau}\qty( \frac{(\partial_\tau \phi)^2}{4}+\frac{(u\partial_x \phi)^2}{4}+ \qty(\frac{2\pi}{\beta})^2 e^{\phi})- 
    \oint_{\partial S} d\vec{l} \cdot \nabla_{\vec{n}} \phi ) \nonumber \\
    -& \frac{l}{4\pi i} \oint_{\partial \Delta_{\tuv}} \phi\left(\frac{\dd{w}}{w-w_0}-\frac{\dd{\wb}}{\wb-\wb_0}\right)
    - 2 l \delta \ln \frac{2\pi \tuv}{\beta}\Bigg],\label{eq:SL-one-source}
\end{align}
is the regularized action, constructed to remain well-defined in the limit ${\tuv} \to 0$. 
In the above definition, $\Delta_{\tuv}$ denotes a disk of radius $\tuv$ around the source, which is omitted in the bulk term,
$\vec{n}$ represents the outward pointing normal vector at the boundaries of the strip $S$, and $\delta=6l/C$. 

We now comment on the terms in the second line of~\eqref{eq:SL-one-source}. To this end, we inspect potential singularities in the integrand and show that the subtracted counter-term, $2 l \delta \ln ({\tuv}/{\beta})$, ensures that the regularized 
action $\widetilde S(l, w_0)$ is finite.
Starting with the behavior near the boundary as $\tau \to 0,\beta/2$, only the first line of~\eqref{eq:SL-one-source} contributes. Notably, in this limit, for the zero-source solution $\phi_0$, the integrand remains finite even though the field itself diverges, as discussed in the previous subsection. By inspection, this result can be generalized to solutions $\phi$  with insertions at arbitrary positions: In the Liouville equation with ZZ boundary conditions~\eqref{eq:LiouvSemicl}, the source term becomes negligible close to the boundary compared to the diverging exponential. As a result, it does not alter the leading divergence but only introduces a correction linear in the distance to the boundary, ensuring that the action remains finite.

Moving to the the second line of Eq.~\eqref{eq:SL-one-source}, the first term is the source included in the action expressed as a contour integral via Cauchy's theorem (it is a regularized form of the $\delta$-function in two dimensions).
The second one, which we add to~\eqref{eq:S_l_w0_tilde} and subtract here, 
is a counter-term making the expression in square brackets regular and allowing to take the $\tuv \to 0$ limit there.  This can be seen the following way: By~\eqref{eq:Liouville-Trafo} and~\eqref{eq:LivOneSourceZero}, the one-source solution close to the source in any geometry behaves as  $\phi = -4 \delta \ln(r/\beta) +O(1)$, where $r$ is the distance to the source in $w$-coordinates. 
The only divergence in the first line of \eqref{eq:SL-one-source} comes from the kinetic term, 
whose leading divergence is given by
\begin{equation}
   \frac{C}{24 \pi} \int_{S \backslash \Delta_{\tuv}} \!\!\!\!\! d^2 x \frac 1 4 \partial_\mu \phi \partial^\mu \phi
    \sim 2l \delta \int_{\tuv}^\beta \dd{r} r (\partial_r \ln(r/ \beta))^2= -2 l \delta \ln( \tuv/\beta)+O(1).
\end{equation}
On other hand, the source term diverges as $(-l \phi)= 4l\delta \ln(r/\beta)+O(1)$ and both terms together cancel the divergence of the counter term, which is also logarithmic in $\tuv$. Again, due to conformal covariance of $\phi$, this result holds for any source insertion on any geometry. 
By construction, the full cut-off dependence is shifted onto the last term of~\eqref{eq:SL-one-source}, yielding a normalization constant in the correlation function~\eqref{eq:Heavy-correlator}.

In a next step, we want to find the on-shell value of $S(l, w_0)$ which is required to calculate the heavy one-point function~\eqref{eq:Heavy-correlator}, and on the way also confirm the conformal transformation behavior used in its derivation.
To this end, we define the Liouville action on the $\xi$-disk

\begin{align} 
    \tilde{S}'(l,\xi_0)=\lim_{\tuv \to 0}\Bigg[ &\frac{C}{24 \pi} \qty( \int_{D \backslash \Delta_{\tuv'}} \dd{\xi} \dd{\xib}\qty( \partial_\xi \phi' \partial_{\xib} \phi'+ e^{\phi'})- 
    \oint_{\partial D} d\vec{l} \cdot \nabla_{\vec{n}} \phi' )
    \nonumber \\
    -& \frac{l}{4\pi i} \oint_{\partial \Delta_{\tuv'}} \phi'\left(\frac{\dd{\xi}}{\xi-\xi_0}-\frac{\dd{\xib}}{\xib-\xib_0}\right)
    - 2 l \delta \ln \tuv' \Bigg],\label{eq:SL-one-source-xi}
\end{align} 
where the solution $\phi'$ is related to the strip solution $\phi$ via the transformation law~\eqref{eq:Liouville-Trafo}, 
$D$~denotes the $\xi$-disk, and $ \Delta_{\tuv'}$ is again a small disk of radius 
\begin{equation}
    \tuv'=\abs{\xi(w_0+\tuv e^{i\theta})-\xi(w_0)} \simeq\tuv \abs{\xi'(w_0)}
\end{equation}
around the source. 
The difference of the actions on the strip and on the disk then reads
\begin{align} \label{eq:S-difference}
     ( \tilde{S}(l, w_0)-S(0, w_0))-(\tilde{S}'(l, \xi_0)-S'(0, \xi_0))= - l(1-\delta)\ln \abs{ \frac{\beta}{2\pi} \xi'(w_0)}^2,
\end{align}
 where we also subtracted the vacuum-actions $S_0=\tilde{S}_0$ and $S'_0=\tilde{S}_0'$. Now, the right-hand side of~\eqref{eq:S-difference} stems purely from the second line of~\eqref{eq:SL-one-source} and~\eqref{eq:SL-one-source-xi}: The contribution $\sim l$ comes from sending $\phi \mapsto \phi'$ using~\eqref{eq:Liouville-Trafo} in the source term and the term $\sim l \delta$ is obtained by substituting $\tuv \mapsto \tuv'$ in the counter term. Together, both terms give the classical scaling dimension of the vertex operator. To conclude the argument, one must show that there are no other contributions, or equivalently, the first lines of the vacuum-normalized actions~\eqref{eq:SL-one-source} and~\eqref{eq:SL-one-source-xi} are equivalent. Since on both geometries the boundary term cancels between the one-source- and the vacuum action and the potential $e^{\phi'}$ is manifestly covariant, the only non-trivial terms are the kinetic ones. For these one can invoke Green's theorem to show that
 \begin{align}
     \int \dd^2{\xi} \qty(\partial_\xi \phi' \partial_{\xib} \phi'-\partial_\xi \phi_0' \partial_{\xib} \phi_0' )=  \int \dd^2{\xi} \qty(\partial_\xi \phi \partial_{\xib} \phi-\partial_\xi \phi_0 \partial_{\xib} \phi_0),
 \end{align}
i.e. the logarithms appearing in the transformation law from $\phi$ to $\phi' $ can be discarded.
From here, one applies a change of variables to recover exactly $\tilde{S}(l,w_0)-S(0,w_0)$, confirming the assertion.
 
To further evaluate the expression~\eqref{eq:S-difference}, we focus on the action difference on the $\xi$-disk. It can be calculated to be finite as \cite{Menotti2006}
\begin{align} \label{eq:xi-change}
    \tilde{S}'(l, \xi_0)- S(0, \xi_0)&=(\tilde{S}'(l, 0)-S'(0, \xi_0))+(\tilde{S}'(l, \xi_0)-\tilde{S}'(l, 0)) \nonumber\\
    &=  - \ln U(l)-l(1-\delta)\ln(\abs{h_\xi'(\xi,\xi_0)}^2_{\xi=\xi_0}).
\end{align}
Here the second term in the final result follows from the same arguments as above and the first term
\begin{align}  \label{eq:one-point-coeff}
   \ln U(l)= -\frac{C}{6}\qty(2\delta(1-\ln2)+(1-2\delta)\ln(1-2\delta))
\end{align}
contains all the information independent of the position. To find it, we introduce a cut-off $\eta \ll 1$ close to the boundary $\abs{\xi}=1$ and evaluate the on-shell action $\tilde{S}[l,0]$ without the boundary term on the solution 
$\phi'(r) \equiv \phi'(r;0,l)$ with $|\xi|=r$ (c.f.~\eqref{eq:LivOneSourceZero} for its definition). 
Expanding this solution close to the origin, one finds 
that the source term at $\tuv' \ll 1$ evaluates to
\begin{equation}
    S_{*}(\delta) =   - \frac{l}{4\pi i} \oint\limits_{|\xi|={\tuv'}} \!\!\!\!\phi'\left(\frac{\dd{\xi}}{\xi-\xi_0} - {\rm c.c.} \right) 
    = - l \phi'(\tuv')  = \frac{C \delta}{3}(2\delta \ln \tuv' - \ln(2-4\delta)).
\end{equation}
Then invoking polar coordinates, the on-shell action becomes (we use $\delta = 6l/C$)
\begin{align}
    I(\delta)&= \frac{C}{12} \int_{\tuv'}^{1-\eta} \qty(\frac{1}{4} \left[\partial_r \phi'(r)\right]^2 +  e^{\phi'(r)}) r\dd{r}+ S_{*}(\delta) - \frac{C \delta^2}{3}\ln \tuv'
    \nonumber\\
    &\overset{\tuv' \to 0}{=}\frac{C}{6}  \left(2 \delta(1-\ln2) + (1-2 \delta ) \ln (1-2 \delta )+
    \eta^{-1} + \ln(2 \eta) -3/2\right) + {\cal O}(\eta),
\end{align}
 and we obtain $I(0)-I(\delta)\overset{\eta \to 0}{=}\ln U(l)$, as given by Eq.~\eqref{eq:one-point-coeff}.
 The constant $ U(l)$ can be identified with the semi-classical limit of the ZZ one-point coefficient \cite{Zamolodchikov2001}, modifying the prefactor of the correlation function, while the second term depending on~\eqref{eq:h-def} contains the coordinate-dependence of the correlation function. Putting everything together, using $l=C\delta/6$, and fixing the UV-scale as $\tuv=1/J$ we finally obtain the one-source action
\begin{equation} \label{eq:SL-one-source-explicit}
    \!\!\!\! S(l, w_0)=S_0 - \ln U(l)  - \frac{ C \delta  (1-\delta) }{6} \ln( \frac{\beta^2}{4\pi^2}\abs{\xi'(w_0)}^2\, \abs{h_\xi'(\xi,\xi_0)}^2_{\xi=\xi_0})+\frac{C \delta^ 2}{6}\ln \abs{\frac{2\pi }{\beta J}}^2, 
\end{equation}
which leads to Eq.~\eqref{eq:SL-one-source-explicit-main} from the main text after application of the chain rule for derivatives.

\section{One-source solution to Liouville equation} \label{App:One-source}
In this appendix, we derive the one-source solution~\eqref{eq:LivOneSourceZero} to the Liouville equation with heavy insertion at $\xi_0=0$, which can be derived either from~\eqref{eq:LiouvSemicl} by applying the transformation law~\eqref{eq:Liouville-Trafo}, or directly from the action~\eqref{eq:SL-one-source-xi}. The strategy is to solve the equation without source insertion and to later implement the corresponding boundary conditions close to the source and close to the boundary. Since the solution will be radially symmetric around the origin, we move to polar coordinates $\xi=r e^{i \theta}, \xib=r e^{-i \theta}$ and neglect the $\theta$-dependence of the equation. We also write the equations in terms of $\Phi= e^\phi$, leading to 
\begin{align}
   -2 \Phi^2+\frac{\Phi' }{r}-\frac{\Phi'^2}{\Phi}+\Phi''=0,
\end{align}
where $\Phi=\Phi(r)$. The general solution reads
\begin{align}
    \Phi(r)=\frac{4 c_1^2}{r^2\qty(e^{c_2}r^{-c_1}-e^{-c_2}r^{c_1})^2}, \label{eq:liv-1-source-1}
\end{align}
in terms of two real constants $c_1$ and $c_2$. The behavior close to the source at $r \to 0$ can be determined as $\Phi \sim \abs{r}^{-4\delta}$ by neglecting the exponential term  and using $\delta^{(2)}(\xi)=\pi^{-1}\partialb\xi^{-1}$ , with $\delta=6 l_1/C <1/2$. Comparing this result to the solution \eqref{eq:liv-1-source-1}, we obtain $c_1=\pm(1-2\delta)$, both choices leading to the same result. For the behavior close to the boundary, as proven in slightly different settings in Appendix A of \cite{Menotti2006} and Appendix B of \cite{Hulik2016}, for $r\to 1$, the Liouville field should diverge as $\Phi \sim 4(1-r^2)^{-2}+O(1)$, i.e. no  $O((1-r^2)^{-1})$ term is present. The function $\Phi$ fulfilling this condition is obtained by setting $c_2=0$, uniquely recovering the solution~\eqref{eq:LivOneSourceZero}.
\end{appendix}
\bibliography{library.bib}

\begin{thebibliography}{10}
\providecommand{\url}[1]{\texttt{#1}}
\providecommand{\urlprefix}{URL }
\expandafter\ifx\csname urlstyle\endcsname\relax
  \providecommand{\doi}[1]{doi:\discretionary{}{}{}#1}\else
  \providecommand{\doi}{doi:\discretionary{}{}{}\begingroup
  \urlstyle{rm}\Url}\fi
\providecommand{\eprint}[2][]{\url{#2}}

\bibitem{1993thooft}
G.~{'t Hooft},
\newblock \emph{{Dimensional Reduction in Quantum Gravity}},
\newblock arXiv e-prints gr-qc/9310026 (1993),
\newblock \doi{10.48550/arXiv.gr-qc/9310026},
\newblock \eprint{gr-qc/9310026}.

\bibitem{Susskind:1994vu}
L.~Susskind,
\newblock \emph{{The World as a hologram}},
\newblock J. Math. Phys. \textbf{36}, 6377 (1995),
\newblock \doi{10.1063/1.531249},
\newblock \eprint{hep-th/9409089}.

\bibitem{Maldacena1999}
J.~Maldacena,
\newblock \emph{The large-{N} limit of superconformal field theories and
  supergravity},
\newblock International Journal of Theoretical Physics \textbf{38}(4), 1113
  (1999),
\newblock \doi{10.1023/A:1026654312961}.

\bibitem{Gubser1998}
S.~S. {Gubser}, I.~R. {Klebanov} and A.~M. {Polyakov},
\newblock \emph{{Gauge theory correlators from non-critical string theory}},
\newblock Physics Letters B \textbf{428}(1-2), 105 (1998),
\newblock \doi{10.1016/S0370-2693(98)00377-3},
\newblock \eprint{hep-th/9802109}.

\bibitem{Witten1998}
E.~{Witten},
\newblock \emph{{Anti-de Sitter space and holography}},
\newblock Advances in Theoretical and Mathematical Physics \textbf{2}, 253
  (1998),
\newblock \doi{10.48550/arXiv.hep-th/9802150},
\newblock \eprint{hep-th/9802150}.

\bibitem{AHARONY2000183}
O.~Aharony, S.~S. Gubser, J.~Maldacena, H.~Ooguri and Y.~Oz,
\newblock \emph{Large {N} field theories, string theory and gravity},
\newblock Physics Reports \textbf{323}(3), 183 (2000),
\newblock \doi{https://doi.org/10.1016/S0370-1573(99)00083-6}.

\bibitem{1983Teitelboim}
C.~Teitelboim,
\newblock \emph{Gravitation and hamiltonian structure in two spacetime
  dimensions},
\newblock Physics Letters B \textbf{126}(1-2), 41 (1983),
\newblock \doi{10.1016/0370-2693(83)90012-6}.

\bibitem{JACKIW1985343}
R.~Jackiw,
\newblock \emph{Lower dimensional gravity},
\newblock Nuclear Physics B \textbf{252}, 343 (1985),
\newblock \doi{https://doi.org/10.1016/0550-3213(85)90448-1}.

\bibitem{1993Sachdev}
S.~{Sachdev} and J.~{Ye},
\newblock \emph{Gapless spin-fluid ground state in a random quantum
  {H}eisenberg magnet},
\newblock {PRL} \textbf{70}(21), 3339 (1993),
\newblock \doi{10.1103/PhysRevLett.70.3339},
\newblock \eprint{cond-mat/9212030}.

\bibitem{Kitaev:2018}
A.~Kitaev and S.~J. Suh,
\newblock \emph{The soft mode in the {S}achdev-{Y}e-{K}itaev model and its
  gravity dual},
\newblock Journal of High Energy Physics \textbf{2018}(5), 183 (2018),
\newblock \doi{10.1007/JHEP05(2018)183}.

\bibitem{Almheiri2015}
A.~Almheiri and J.~Polchinski,
\newblock \emph{Models of {A}d{S}2 backreaction and holography},
\newblock Journal of High Energy Physics \textbf{2015}(11), 14 (2015),
\newblock \doi{10.1007/JHEP11(2015)014}.

\bibitem{Maldacena2016}
J.~Maldacena, D.~Stanford and Z.~Yang,
\newblock \emph{Conformal symmetry and its breaking in two-dimensional nearly
  anti-de {S}itter space},
\newblock Progress of Theoretical and Experimental Physics \textbf{2016}(12),
  12C104 (2016),
\newblock \doi{10.1093/ptep/ptw124}.

\bibitem{Maldacena2016Remarks}
J.~Maldacena and D.~Stanford,
\newblock \emph{Remarks on the {S}achdev-{Y}e-{K}itaev model},
\newblock Phys. Rev. D \textbf{94}, 106002 (2016),
\newblock \doi{10.1103/PhysRevD.94.106002}.

\bibitem{Bagrets2016}
D.~Bagrets, A.~Altland and A.~Kamenev,
\newblock \emph{Sachdev-{Y}e-{K}itaev model as {L}iouville quantum mechanics},
\newblock Nuclear Physics B \textbf{911}, 191 (2016),
\newblock \doi{https://doi.org/10.1016/j.nuclphysb.2016.08.002}.

\bibitem{Bagrets2017}
D.~Bagrets, A.~Altland and A.~Kamenev,
\newblock \emph{Power-law out of time order correlation functions in the {SYK}
  model},
\newblock Nuclear Physics B \textbf{921}, 727 (2017),
\newblock \doi{https://doi.org/10.1016/j.nuclphysb.2017.06.012}.

\bibitem{Polchinski2016}
J.~Polchinski and V.~Rosenhaus,
\newblock \emph{The spectrum in the {S}achdev-{Y}e-{K}itaev model},
\newblock Journal of High Energy Physics \textbf{2016}(4), 1 (2016),
\newblock \doi{10.1007/JHEP04(2016)001}.

\bibitem{Mertens162}
J.~Engelsöy, T.~G. Mertens and H.~Verlinde,
\newblock \emph{An investigation of {A}d{S}$_2$ backreaction and holography},
\newblock Journal of High Energy Physics  (2016),
\newblock \doi{10.1007/JHEP07(2016)139}.

\bibitem{Saad2019}
P.~{Saad}, S.~H. {Shenker} and D.~{Stanford},
\newblock \emph{{JT gravity as a matrix integral}},
\newblock arXiv e-prints arXiv:1903.11115 (2019),
\newblock \doi{10.48550/arXiv.1903.11115},
\newblock \eprint{1903.11115}.

\bibitem{Cotler2017}
J.~S. Cotler, G.~Gur-Ari, M.~Hanada, J.~Polchinski, P.~Saad, S.~H. Shenker,
  D.~Stanford, A.~Streicher and M.~Tezuka,
\newblock \emph{Black holes and random matrices},
\newblock Journal of High Energy Physics \textbf{2017}(5), 118 (2017),
\newblock \doi{10.1007/JHEP05(2017)118}.

\bibitem{SSS2018}
P.~{Saad}, S.~H. {Shenker} and D.~{Stanford},
\newblock \emph{{A semiclassical ramp in SYK and in gravity}},
\newblock arXiv:1806.06840  (2018),
\newblock \doi{10.48550/arXiv.1806.06840}.

\bibitem{Altland2018}
A.~Altland and D.~Bagrets,
\newblock \emph{Quantum ergodicity in the {S}{Y}{K} model},
\newblock Nuclear Physics B \textbf{930}, 45 (2018),
\newblock \doi{10.1016/j.nuclphysb.2018.02.015}.

\bibitem{Altland21}
A.~Altland, D.~Bagrets, P.~Nayak, J.~Sonner and M.~Vielma,
\newblock \emph{From operator statistics to wormholes},
\newblock Phys. Rev. Res. \textbf{3}, 033259 (2021),
\newblock \doi{10.1103/PhysRevResearch.3.033259}.

\bibitem{Jensen:2016pah}
K.~Jensen,
\newblock \emph{{Chaos in AdS$_2$ Holography}},
\newblock Phys. Rev. Lett. \textbf{117}(11), 111601 (2016),
\newblock \doi{10.1103/PhysRevLett.117.111601},
\newblock \eprint{1605.06098}.

\bibitem{Alekseev1989}
A.~Alekseev and S.~Shatashvili,
\newblock \emph{Path integral quantization of the coadjoint orbits of the
  {V}irasoro group and 2-d gravity},
\newblock Nuclear Physics B \textbf{323}, 719 (1989),
\newblock \doi{10.1016/0550-3213(89)90130-2}.

\bibitem{Alekseev1990}
A.~Alekseev and S.~Shatashvili,
\newblock \emph{From geometric quantization to conformal field theory},
\newblock Communications in Mathematical Physics \textbf{128}, 197 (1990),
\newblock \doi{10.1007/BF02097053}.

\bibitem{Cotler2018}
J.~Cotler and K.~Jensen,
\newblock \emph{A theory of reparameterizations for {AdS3} gravity},
\newblock Journal of High Energy Physics \textbf{2019}(2), 79 (2019),
\newblock \doi{10.1007/JHEP02(2019)079}.

\bibitem{Mertens2022}
T.~G. Mertens, J.~Sim{\'o}n and G.~Wong,
\newblock \emph{A proposal for 3d quantum gravity and its bulk factorization},
\newblock Journal of High Energy Physics \textbf{2023}(6), 134 (2023),
\newblock \doi{10.1007/JHEP06(2023)134}.

\bibitem{Banados1992}
M.~Bañados, C.~Teitelboim and J.~Zanelli,
\newblock \emph{Black hole in three-dimensional spacetime},
\newblock Phys. Rev. Lett. \textbf{69}, 1849 (1992),
\newblock \doi{10.1103/PhysRevLett.69.1849}.

\bibitem{Banados1993}
M.~Bañados, M.~Henneaux, C.~Teitelboim and J.~Zanelli,
\newblock \emph{Geometry of the 2+1 black hole},
\newblock Phys. Rev. D \textbf{48}, 1506 (1993),
\newblock \doi{10.1103/PhysRevD.48.1506}.

\bibitem{Altland2023}
A.~Altland and B.~Simons,
\newblock \emph{Condensed {Matter} {Field} {Theory}},
\newblock Cambridge University Press,
\newblock ISBN 978-1-108-49460-1,
\newblock \doi{10.1017/CBO9780511789984} (2023).

\bibitem{wuHelicalLiquidEdge2006}
C.~Wu, B.~A. Bernevig and S.-C. Zhang,
\newblock \emph{Helical {Liquid} and the {Edge} of {Quantum} {Spin} {Hall}
  {Systems}},
\newblock Physical Review Letters \textbf{96}(10), 106401 (2006),
\newblock \doi{10.1103/PhysRevLett.96.106401},
\newblock Publisher: American Physical Society.

\bibitem{Alhassid99}
Y.~Alhassid, P.~Jacquod and A.~Wobst,
\newblock \emph{Random matrix model for quantum dots with interactions and the
  conductance peak spacing distribution},
\newblock Phys. Rev. B \textbf{61}, R13357 (2000),
\newblock \doi{10.1103/PhysRevB.61.R13357}.

\bibitem{Lian2019}
B.~Lian, S.~L. Sondhi and Z.~Yang,
\newblock \emph{The chiral {SYK} model},
\newblock Journal of High Energy Physics \textbf{2019} (2019),
\newblock \doi{10.1007/JHEP09(2019)067}.

\bibitem{berkooz2017}
M.~Berkooz, P.~Narayan, M.~Rozali and J.~Sim{\'o}n,
\newblock \emph{Comments on the random {T}hirring model},
\newblock Journal of High Energy Physics \textbf{2017}(9), 57 (2017),
\newblock \doi{10.1007/JHEP09(2017)057}.

\bibitem{Turiaci2017}
G.~J. Turiaci and H.~Verlinde,
\newblock \emph{Towards a 2d {QFT} analog of the {SYK} model},
\newblock Journal of High Energy Physics \textbf{2017}(10), 167 (2017),
\newblock \doi{10.1007/JHEP10(2017)167}.

\bibitem{Pasterski2022}
S.~Pasterski and H.~Verlinde,
\newblock \emph{Mapping {SYK} to the sky},
\newblock Journal of High Energy Physics \textbf{2022}(9), 47 (2022),
\newblock \doi{10.1007/JHEP09(2022)047}.

\bibitem{Achucarro:1993fd}
A.~Achucarro and M.~E. Ortiz,
\newblock \emph{{Relating black holes in two-dimensions and three-dimensions}},
\newblock Phys. Rev. D \textbf{48}, 3600 (1993),
\newblock \doi{10.1103/PhysRevD.48.3600},
\newblock \eprint{hep-th/9304068}.

\bibitem{Callebaut2023}
N.~Callebaut,
\newblock \emph{Entanglement in Conformal Field Theory and Holography}, pp.
  239--271,
\newblock Springer Nature Switzerland, Cham,
\newblock ISBN 978-3-031-42096-2,
\newblock \doi{10.1007/978-3-031-42096-2_10} (2023).

\bibitem{Henneaux2020}
M.~Henneaux, W.~Merbis and A.~Ranjbar,
\newblock \emph{Asymptotic dynamics of {AdS3} gravity with two asymptotic
  regions},
\newblock Journal of High Energy Physics \textbf{2020}, 64 (2020),
\newblock \doi{10.1007/JHEP03(2020)064}.

\bibitem{Chua2023}
W.~Z. Chua and Y.~Jiang,
\newblock \emph{Hartle-{H}awking state and its factorization in 3d gravity},
\newblock Journal of High Energy Physics \textbf{2024}(3), 135 (2024),
\newblock \doi{10.1007/JHEP03(2024)135}.

\bibitem{Kraus2021}
P.~Kraus, R.~Monten and R.~M. Myers,
\newblock \emph{3d gravity in a box},
\newblock SciPost Physics \textbf{11}, 070 (2021),
\newblock \doi{10.21468/SciPostPhys.11.3.070}.

\bibitem{Ebert2022}
S.~Ebert, E.~Hijano, P.~Kraus, R.~Monten and R.~M. Myers,
\newblock \emph{Field theory of interacting boundary gravitons},
\newblock SciPost Phys. \textbf{13}, 038 (2022),
\newblock \doi{10.21468/SciPostPhys.13.2.038}.

\bibitem{Banerjee2022}
S.~Banerjee, M.~Dorband, J.~Erdmenger, R.~Meyer and A.-L. Weigel,
\newblock \emph{Berry phases, wormholes and factorization in {AdS/CFT}},
\newblock Journal of High Energy Physics \textbf{2022}, 162 (2022),
\newblock \doi{10.1007/JHEP08(2022)162}.

\bibitem{Cotler2021}
J.~Cotler and K.~Jensen,
\newblock \emph{Ad{S3} gravity and random {CFT}},
\newblock Journal of High Energy Physics \textbf{2021}, 33 (2021),
\newblock \doi{10.1007/JHEP04(2021)033}.

\bibitem{luttingerGroundStateEnergyManyFermion1960}
J.~M. Luttinger and J.~C. Ward,
\newblock \emph{Ground-state energy of a many-fermion system. {II}},
\newblock Phys. Rev. \textbf{118}, 1417 (1960),
\newblock \doi{10.1103/PhysRev.118.1417}.

\bibitem{Geo:2022}
G.~Jose, K.~Seo and B.~Uchoa,
\newblock \emph{Non-{F}ermi liquid behavior in the {S}achdev-{Y}e-{K}itaev
  model for a one-dimensional incoherent semimetal},
\newblock Phys. Rev. Res. \textbf{4}, 013145 (2022),
\newblock \doi{10.1103/PhysRevResearch.4.013145}.

\bibitem{Balents:2017}
X.-Y. Song, C.-M. Jian and L.~Balents,
\newblock \emph{Strongly correlated metal built from {S}achdev-{Y}e-{K}itaev
  models},
\newblock Phys. Rev. Lett. \textbf{119}, 216601 (2017),
\newblock \doi{10.1103/PhysRevLett.119.216601}.

\bibitem{Altland:2019}
A.~Altland, D.~Bagrets and A.~Kamenev,
\newblock \emph{Quantum criticality of granular {S}achdev-{Y}e-{K}itaev
  matter},
\newblock Phys. Rev. Lett. \textbf{123}, 106601 (2019),
\newblock \doi{10.1103/PhysRevLett.123.106601}.

\bibitem{Chklovskii:1992}
D.~B. Chklovskii, B.~I. Shklovskii and L.~I. Glazman,
\newblock \emph{Electrostatics of edge channels},
\newblock Phys. Rev. B \textbf{46}, 4026 (1992),
\newblock \doi{10.1103/PhysRevB.46.4026}.

\bibitem{Chklovskii:1993}
D.~B. Chklovskii, K.~A. Matveev and B.~I. Shklovskii,
\newblock \emph{Ballistic conductance of interacting electrons in the quantum
  hall regime},
\newblock Phys. Rev. B \textbf{47}, 12605 (1993),
\newblock \doi{10.1103/PhysRevB.47.12605}.

\bibitem{Rammer:1986}
J.~Rammer and H.~Smith,
\newblock \emph{Quantum field-theoretical methods in transport theory of
  metals},
\newblock Rev. Mod. Phys. \textbf{58}, 323 (1986),
\newblock \doi{10.1103/RevModPhys.58.323}.

\bibitem{Kamenev:2011}
A.~Kamenev,
\newblock \emph{Field Theory of Non-Equilibrium Systems},
\newblock Cambridge University Press,
\newblock \doi{10.1017/9781108769266} (2023).

\bibitem{Carlip:1998uc}
S.~Carlip,
\newblock \emph{{Quantum gravity in 2+1 dimensions}},
\newblock Cambridge Monographs on Mathematical Physics. Cambridge University
  Press,
\newblock ISBN 978-0-521-54588-4, 978-0-511-82229-2,
\newblock \doi{10.1017/CBO9780511564192} (2003).

\bibitem{Banados1999}
M.~Bañados,
\newblock \emph{Three-dimensional quantum geometry and black holes},
\newblock AIP Conference Proceedings \textbf{484}(1), 147 (1999),
\newblock \doi{10.1063/1.59661}.

\bibitem{Roberts:2012aq}
M.~M. Roberts,
\newblock \emph{{Time evolution of entanglement entropy from a pulse}},
\newblock JHEP \textbf{12}, 027 (2012),
\newblock \doi{10.1007/JHEP12(2012)027},
\newblock \eprint{1204.1982}.

\bibitem{Brown1986}
J.~D. Brown and M.~Henneaux,
\newblock \emph{Central charges in the canonical realization of asymptotic
  symmetries: An example from three dimensional gravity},
\newblock Communications in Mathematical Physics \textbf{104}, 207 (1986),
\newblock \doi{10.1007/BF01211590}.

\bibitem{Cotler:2020ugk}
J.~Cotler and K.~Jensen,
\newblock \emph{{AdS$_{3}$ gravity and random CFT}},
\newblock JHEP \textbf{04}, 033 (2021),
\newblock \doi{10.1007/JHEP04(2021)033},
\newblock \eprint{2006.08648}.

\bibitem{Witten1988}
E.~Witten,
\newblock \emph{2 + 1 dimensional gravity as an exactly soluble system},
\newblock Nuclear Physics B \textbf{311}, 46 (1988),
\newblock \doi{10.1016/0550-3213(88)90143-5}.

\bibitem{ATownsend}
A.~Achúcarro and P.~Townsend,
\newblock \emph{A {C}hern-{S}imons action for three-dimensional anti-de
  {S}itter supergravity theories},
\newblock Physics Letters B \textbf{180}, 89 (1986),
\newblock \doi{10.1016/0370-2693(86)90140-1}.

\bibitem{Donnay2016}
L.~Donnay,
\newblock \emph{{Asymptotic dynamics of three-dimensional gravity}},
\newblock PoS \textbf{Modave2015}, 001 (2016),
\newblock \doi{10.22323/1.271.0001}.

\bibitem{Witten:1989ip}
E.~Witten,
\newblock \emph{{Quantization of {Chern-Simons} Gauge Theory With Complex Gauge
  Group}},
\newblock Commun. Math. Phys. \textbf{137}, 29 (1991),
\newblock \doi{10.1007/BF02099116}.

\bibitem{Bunster2014}
C.~Bunster, M.~Henneaux, A.~P{\'e}rez, D.~Tempo and R.~Troncoso,
\newblock \emph{Generalized black holes in three-dimensional spacetime},
\newblock Journal of High Energy Physics \textbf{2014}(5), 31 (2014),
\newblock \doi{10.1007/JHEP05(2014)031}.

\bibitem{Elitzur89}
S.~Elitzur, G.~Moore, A.~Schwimmer and N.~Seiberg,
\newblock \emph{Remarks on the canonical quantization of the
  {C}hern-{S}imons-{W}itten theory},
\newblock Nuclear Physics B \textbf{326}, 108 (1989),
\newblock \doi{10.1016/0550-3213(89)90436-7}.

\bibitem{Banados1998}
M.~Bañados, T.~Brotz and M.~E. Ortiz,
\newblock \emph{Boundary dynamics and the statistical mechanics of the 2 +
  1-dimensional black hole},
\newblock Nuclear Physics B \textbf{545}(1), 340 (1999),
\newblock \doi{https://doi.org/10.1016/S0550-3213(99)00069-3}.

\bibitem{Banados:2016zim}
M.~Ba\~nados and I.~A. Reyes,
\newblock \emph{A short review on {N}oethers theorems, gauge symmetries and
  boundary terms},
\newblock Int. J. Mod. Phys. D \textbf{25}(10), 1630021 (2016),
\newblock \doi{10.1142/S0218271816300214},
\newblock \eprint{1601.03616}.

\bibitem{Mertens2018}
T.~G. Mertens,
\newblock \emph{The {S}chwarzian theory — origins},
\newblock Journal of High Energy Physics \textbf{2018}, 36 (2018),
\newblock \doi{10.1007/JHEP05(2018)036}.

\bibitem{Coussaert1995}
O.~Coussaert, M.~Henneaux and P.~van Driel,
\newblock \emph{The asymptotic dynamics of three-dimensional {E}instein gravity
  with a negative cosmological constant},
\newblock Classical and Quantum Gravity \textbf{12}, 2961 (1995),
\newblock \doi{10.1088/0264-9381/12/12/012}.

\bibitem{Fitzpatrick2014}
A.~L. Fitzpatrick, J.~Kaplan and M.~T. Walters,
\newblock \emph{Universality of long-distance {AdS} physics from the {CFT}
  bootstrap},
\newblock Journal of High Energy Physics \textbf{2014}(8), 145 (2014),
\newblock \doi{10.1007/JHEP08(2014)145}.

\bibitem{Fitzpatrick2015}
A.~L. Fitzpatrick, J.~Kaplan and M.~T. Walters,
\newblock \emph{Virasoro conformal blocks and thermality from classical
  background fields},
\newblock Journal of High Energy Physics \textbf{2015}(11), 200 (2015),
\newblock \doi{10.1007/JHEP11(2015)200}.

\bibitem{Fitzpatrick2016}
A.~L. Fitzpatrick and J.~Kaplan,
\newblock \emph{A quantum correction to chaos},
\newblock Journal of High Energy Physics \textbf{2016}(5), 70 (2016),
\newblock \doi{10.1007/JHEP05(2016)070}.

\bibitem{Hulik2016}
O.~Hul{\'i}k, T.~Proch{\'a}zka and J.~Raeymaekers,
\newblock \emph{Multi-centered {AdS3} solutions from {V}irasoro conformal
  blocks},
\newblock Journal of High Energy Physics \textbf{2017}(3), 129 (2017),
\newblock \doi{10.1007/JHEP03(2017)129}.

\bibitem{Haehl2018}
F.~M. Haehl and M.~Rozali,
\newblock \emph{Effective field theory for chaotic {CFTs}},
\newblock Journal of High Energy Physics \textbf{2018}(10), 118 (2018),
\newblock \doi{10.1007/JHEP10(2018)118}.

\bibitem{Mertens2017}
T.~G. Mertens, G.~J. Turiaci and H.~L. Verlinde,
\newblock \emph{Solving the {S}chwarzian via the conformal bootstrap},
\newblock Journal of High Energy Physics \textbf{2017}, 136 (2017),
\newblock \doi{10.1007/JHEP08(2017)136}.

\bibitem{Menotti2004}
P.~Menotti and E.~Tonni,
\newblock \emph{Standard and geometric approaches to quantum {L}iouville theory
  on the pseudosphere},
\newblock Nuclear Physics B \textbf{707}(3), 321 (2005),
\newblock \doi{https://doi.org/10.1016/j.nuclphysb.2004.11.003}.

\bibitem{Zamolodchikov1995}
A.~Zamolodchikov and A.~Zamolodchikov,
\newblock \emph{Conformal bootstrap in {L}iouville field theory},
\newblock Nuclear Physics B \textbf{477}(2), 577 (1996),
\newblock \doi{https://doi.org/10.1016/0550-3213(96)00351-3}.

\bibitem{Seiberg1990}
N.~Seiberg,
\newblock \emph{Notes on quantum {L}iouville theory and quantum gravity},
\newblock Progress of Theoretical Physics Supplement \textbf{102}, 319 (1990),
\newblock \doi{10.1143/PTPS.102.319}.

\bibitem{Ginsparg1993}
P.~Ginsparg and G.~Moore,
\newblock \emph{Lectures on 2d gravity and 2d string theory (tasi 1992)},
\newblock arXiv:hep-th/9304011  (1993),
\newblock \doi{10.48550/arXiv.hep-th/9304011}.

\bibitem{Menotti2006}
P.~Menotti and E.~Tonni,
\newblock \emph{Liouville field theory with heavy charges, {I}. {T}he
  pseudosphere},
\newblock Journal of High Energy Physics \textbf{2006}(06), 020 (2006),
\newblock \doi{10.1088/1126-6708/2006/06/020}.

\bibitem{Teschner2000}
J.~Teschner,
\newblock \emph{Remarks on {L}iouville theory with boundary},
\newblock In \emph{Non-perturbative Quantum Effects 2000}, vol.~6, p. 041.
  SISSA Medialab,
\newblock \doi{10.48550/arXiv.hep-th/0009138} (2000).

\bibitem{Teschner2001}
J.~Teschner,
\newblock \emph{Liouville theory revisited},
\newblock Classical and Quantum Gravity \textbf{18}, R153 (2001),
\newblock \doi{10.1088/0264-9381/18/23/201}.

\bibitem{Schomerus2005}
V.~Schomerus,
\newblock \emph{Non-compact string backgrounds and non-rational {CFT}},
\newblock Physics Reports \textbf{431}(2), 39 (2006),
\newblock \doi{https://doi.org/10.1016/j.physrep.2006.05.001}.

\bibitem{Nazarov_Blanter:2009}
Y.~V. Nazarov and Y.~M. Blanter,
\newblock \emph{Quantum Transport: Introduction to Nanoscience},
\newblock Cambridge University Press,
\newblock \doi{10.1017/CBO9780511626906} (2009).

\bibitem{Nazarov:1989}
Y.~V. Nazarov,
\newblock \emph{Anomalous current-voltage characteristics of tunnel junctions},
\newblock Journal of Experimental and Theoretical Physics \textbf{68}(3), 561
  (1989).

\bibitem{Kamenev:1999}
A.~Kamenev and A.~Andreev,
\newblock \emph{Electron-electron interactions in disordered metals: {K}eldysh
  formalism},
\newblock Phys. Rev. B \textbf{60}, 2218 (1999),
\newblock \doi{10.1103/PhysRevB.60.2218}.

\bibitem{Mora:2007}
C.~Mora, R.~Egger and A.~Altland,
\newblock \emph{From {L}uttinger liquid to {A}ltshuler-{A}ronov anomaly in
  multichannel quantum wires},
\newblock Phys. Rev. B \textbf{75}, 035310 (2007),
\newblock \doi{10.1103/PhysRevB.75.035310}.

\bibitem{Ngo_Dinh:2012}
S.~{Ngo Dinh}, D.~A. Bagrets and A.~D. Mirlin,
\newblock \emph{Nonequilibrium functional bosonization of quantum wire
  networks},
\newblock Annals of Physics \textbf{327}(11), 2794 (2012),
\newblock \doi{10.1016/j.aop.2012.06.004}.

\bibitem{Levitov:2001}
L.~S. Levitov, A.~V. Shytov and B.~I. Halperin,
\newblock \emph{Effective action of a compressible quantum {H}all state edge:
  {A}pplication to tunneling},
\newblock Phys. Rev. B \textbf{64}, 075322 (2001),
\newblock \doi{10.1103/PhysRevB.64.075322}.

\bibitem{Gu2017}
Y.~Gu, X.-L. Qi and D.~Stanford,
\newblock \emph{Local criticality, diffusion and chaos in generalized
  {S}achdev-{Y}e-{K}itaev models},
\newblock Journal of High Energy Physics \textbf{2017}(5), 125 (2017),
\newblock \doi{10.1007/JHEP05(2017)125}.

\bibitem{JianYao17}
S.-K. Jian and H.~Yao,
\newblock \emph{Solvable {S}achdev-{Y}e-{K}itaev models in higher dimensions:
  From diffusion to many-body localization},
\newblock Phys. Rev. Lett. \textbf{119}, 206602 (2017),
\newblock \doi{10.1103/PhysRevLett.119.206602}.

\bibitem{Sachdev17}
R.~A. Davison, W.~Fu, A.~Georges, Y.~Gu, K.~Jensen and S.~Sachdev,
\newblock \emph{Thermoelectric transport in disordered metals without
  quasiparticles: The {S}achdev-{Y}e-{K}itaev models and holography},
\newblock Phys. Rev. B \textbf{95}, 155131 (2017),
\newblock \doi{10.1103/PhysRevB.95.155131}.

\bibitem{Cai2018}
W.~Cai, X.-H. Ge and G.-H. Yang,
\newblock \emph{Diffusion in higher dimensional {SYK} model with complex
  fermions},
\newblock Journal of High Energy Physics \textbf{2018}(1), 76 (2018),
\newblock \doi{10.1007/JHEP01(2018)076}.

\bibitem{Blake2018}
M.~Blake, H.~Lee and H.~Liu,
\newblock \emph{A quantum hydrodynamical description for scrambling and
  many-body chaos},
\newblock Journal of High Energy Physics \textbf{2018}(10), 127 (2018),
\newblock \doi{10.1007/JHEP10(2018)127}.

\bibitem{Blake20182}
M.~Blake, R.~A. Davison, S.~Grozdanov and H.~Liu,
\newblock \emph{Many-body chaos and energy dynamics in holography},
\newblock Journal of High Energy Physics \textbf{2018}(10), 35 (2018),
\newblock \doi{10.1007/JHEP10(2018)035}.

\bibitem{Maldacena2015}
J.~Maldacena, S.~H. Shenker and D.~Stanford,
\newblock \emph{A bound on chaos},
\newblock Journal of High Energy Physics \textbf{2016}(8), 106 (2016),
\newblock \doi{10.1007/JHEP08(2016)106}.

\bibitem{Lam2018}
H.~T. Lam, T.~G. Mertens, G.~J. Turiaci and H.~Verlinde,
\newblock \emph{Shockwave {S}-matrix from {S}chwarzian quantum mechanics},
\newblock Journal of High Energy Physics \textbf{2018}(11), 182 (2018),
\newblock \doi{10.1007/JHEP11(2018)182}.

\bibitem{Roberts2014}
D.~A. Roberts and D.~Stanford,
\newblock \emph{Diagnosing chaos using four-point functions in two-dimensional
  conformal field theory},
\newblock Phys. Rev. Lett. \textbf{115}, 131603 (2015),
\newblock \doi{10.1103/PhysRevLett.115.131603}.

\bibitem{Cohl2021}
H.~S. Cohl, J.~Park and H.~Volkmer,
\newblock \emph{Gauss hypergeometric representations of the {F}errers function
  of the second kind},
\newblock Symmetry, Integrability and Geometry: Methods and Applications
  (2021),
\newblock \doi{10.3842/SIGMA.2021.053}.

\bibitem{Zamolodchikov2001}
A.~Zamolodchikov and A.~Zamolodchikov,
\newblock \emph{Liouville field theory on a pseudosphere},
\newblock arXiv:hep-th/0101152  (2001),
\newblock \doi{10.48550/arXiv.hep-th/0101152}.

\end{thebibliography}
\end{document}